# WHAT ARE THE HIDDEN QUANTUM PROCESSES IN EINSTEIN'S WEAK PRINCIPLE OF EQUIVALENCE?

**Tom Ostoma and Mike Trushyk**

48 O'HARA PLACE, Brampton, Ontario, L6Y 3R8
emqg@rogerswave.ca

Monday April 12, 2000

**ACKNOWLEDGMENTS**

We wish to thank R. Mongrain (P.Eng) for our lengthy conversations on the nature of space, time, light, matter, and CA theory.

## ABSTRACT

***We provide a quantum derivation of Einstein's Weak Equivalence Principle (WEP) of general relativity using a new quantum gravity theory proposed by the authors called Electro-Magnetic Quantum Gravity or EMQG (ref. 1). EMQG is manifestly compatible with Cellular Automata (CA) theory (ref. 2 and 4), and is also based on a new theory of inertia (ref. 5) proposed by R. Haisch, A. Rueda, and H. Puthoff (which we modified and called Quantum Inertia, QI). QI states that classical Newtonian Inertia is a property of matter due to the strictly local electrical force interactions contributed by each of the (electrically charged) elementary particles of the mass with the surrounding (electrically charged) virtual particles (virtual masseons) of the quantum vacuum. The sum of all the tiny electrical forces (photon exchanges with the vacuum particles) originating in each charged elementary particle of the accelerated mass is the source of the total inertial force of a mass which opposes accelerated motion in Newton's law 'F = MA'. The well known paradoxes that arise from considerations of accelerated motion (Mach's principle) are resolved, and Newton's laws of motion are now understood at the deeper quantum level.***

***We found that gravity also involves the same 'inertial' electromagnetic force component that exists in inertial mass. We propose that Einstein's general relativistic Weak Equivalence Principle (WEP) originates from common 'lower level' quantum vacuum processes occurring in both inertial mass and gravitational mass, in accordance with the principles of quantum field theory. Gravitational mass results from the quantum activities of both the electrical force (photon exchanges) component and the pure gravitational force (graviton exchanges) component, acting simultaneously on the elementary particles that make up a mass. However, inertial mass is strictly the result of the electrical force process only, as given by quantum inertia principle (with negligible graviton processes present).***

***Under gravitation, the elementary particles of a test mass near the earth exchanges gravitons with the earth. However what is frequently neglected is that the surrounding (electrically charged) virtual fermion particles also exchange gravitons with the earth, causing the virtual fermion particles to accelerate (fall) towards the earth. A test mass does move under the influence of these direct graviton exchanges, but more importantly, moves under the influence of the falling (electrically charged) virtual particles of the quantum vacuum, which dominates the total quantum force process. Therefore a test mass under gravity 'sees' that the quantum vacuum accelerate in the same way as it did when the test mass was subjected to acceleration alone. Thus, equivalence arises from the reversal of the acceleration vectors of the quantum vacuum with respect to a test mass undergoing acceleration, as compared to a test mass subjected to a gravitational field. In accelerated frames, it is the mass that accelerates. Inside gravitational fields it is the virtual particles of the quantum vacuum that accelerates.***

***A consequence of EMQG is that all elementary mass particles must consist of combinations of just one fundamental matter particle type, which we call the 'masseon' particle. The masseon has one, fixed (smallest) quanta of mass (similar to a quanta of electric charge), which we call 'mass charge'. The masseon also carries either a positive or negative quanta of electric charge. The masseon particle generates a fixed flux of graviton exchange particles, with a flux rate being completely unaffected by high speed motion (like electric charge). The graviton is the vector boson exchange particle of the pure gravitational force interaction. The physics of graviton exchanges is nearly identical to the photon exchanges of QED, with the same concept of positive and negative gravitational 'mass charge' carried by masseons and anti-masseons respectively. The ratio of the strength of graviton to photon exchange force coupling is about*** $10^{-40}$ ***for an electron. In QED, the quantum vacuum consists of virtual electrons, virtual anti-electrons, and virtual photons. In EMQG, the quantum vacuum consists of virtual masseons, virtual anti-masseons, and virtual gravitons, which also posses both positive and negative electrical charge and positive and negative 'mass charge'. There are almost equal numbers of virtual masseon and anti-masseon particles existing in the quantum vacuum everywhere, and at any given time. This is why the cosmological constant is very close to zero.***

# TABLE OF CONTENTS

## 1. INTRODUCTION

***"I have never been able to understand this principle (principle of equivalence) ... I suggest that it be now buried with appropriate honors."***

**- Synge: Relativity- The General Theory**

Imagine that you are *stationary*, and standing on the surface of the earth. Gravity feels like a force that is holding your gravitational mass to the earth's surface. Yet when you are standing in a rocket undergoing *accelerated motion* (far from gravitational fields, moving with an acceleration of 1 g), the principle of equivalence tells us that there is an identical force exerted against the rocket floor by your inertial mass. However, this force is now caused by your dynamic accelerated motion through empty space. While standing on the earth, however, you were definitely ***not*** in motion. Why should there be such a deep connection between what *appears* to be two completely different physical phenomena: a static force of gravity with no apparent motion, and a dynamic force due to the accelerated motion of your mass? In other words, why should you weigh the same in the rocket as you do on the earth?

The principle of equivalence encompasses this *strange 'coincidence'*, and is one of the founding postulates of general relativity theory. In it's stronger form, it states that ***all the laws of physics are the same*** in the above thought experiment. When stated in it's weaker form, it implies that objects of different mass fall at the same rate of acceleration in a uniform gravity field, and that ***only laws of motion of physics are the same***.

Equivalence also means that the inertial mass, i.e. the mass defined by Newton's law of motion: $m_i = \frac{F_i}{g}$ is ***exactly*** equal to the gravitational mass, which is mass defined by ***a completely different law*** given by Newton's universal gravitational law: $m_g = \frac{F_g r^2}{GM}$ (where $m_i$ and $m_g$ are the inertial and gravitational mass of an object, $F_i$ and $F_g$ are the inertial and gravitational forces exerted on the object, g is the acceleration of the object, M is the mass of the earth, r is the distance from earth's center, and G is Newton's universal gravitational constant). Newton was well aware of this coincidence of mass equivalence, which we refer to here as the Newtonian mass equivalence principle.

The equivalence principle demands that that $m_i = m_g$. Why should this be true in our universe? It has been about 85 years since the discovery of the Einstein equivalence principle, and hundreds of years since the discovery of Newton's mass equivalence principle. Yet, it is still *not* understood ***why*** inertial mass exists in the first place, or why a mass opposes acceleration with a back acting inertial force. More importantly, it is also not known why there are two totally ***different physical definitions*** for inertial and gravitational mass (instead of just one).

Furthermore, masses can have different temperatures (and therefore different energy content), and may be composed of different materials like lead, wood, water. Again, why should $m_i = m_g$, no matter what the material composition and the energy content that a mass may contain? In short, the principle of equivalence is one of the deepest, unsolved mysteries that exists in fundamental physics today!

Lacking any deeper understanding of this question, most physicists prefer to accept equivalence as the fundamental way in which the universe operates. Other physicists maintain that the origin of the principle of equivalence is one of the deepest, unsolved mystery of modern physics, and deserves an explanation. In this paper we reveal for the first time, the low-level quantum processes that are hidden from our view, and provide a quantum field theoretic solution to the principle of equivalence. In other words, we show how the principle of equivalence turns out to be purely a quantum process. Equivalence results from the activities of quantum particles interacting with quantum particles, while obeying the general laws of quantum field theory. This paper is also an invitation to explore a new theory of gravity called ElectroMagnetic Quantum Gravity, or EMQG (ref. 1). EMQG is based on a new understanding of both inertia and the principle of equivalence, which exists on the distance scales of quantum particles.

How is the Einstein principle of equivalence precisely defined? Physicists recognize two main formulations of the principle of equivalence (with some minor variations): the Weak Equivalence Principle (WEP), and the Strong Equivalence Principle (SEP). The strong equivalence principle states that the results of *any given physical experiment* will be precisely identical for an accelerated observer in free space as it is for a non-accelerated observer in a perfectly uniform gravitational field. This is the form of the equivalence principle used by Einstein to formulate his theory of general relativity in 1915.

A weaker form of equivalence principle (WEP) restricts itself only to the laws of motion of masses in these two situations. In other words, the weak principle of equivalence states that only the laws of motion of a mass on the earth is identical to the laws of motion of the same mass inside an accelerated rocket (at 1g). Technically, when comparing the equivalence of a mass in a rocket to a mass on the earth, we assume that the motion is restricted to short distances (mathematically, at a point) on the earth, where gravity does not vary with height. The WEP implies that objects of different mass, different material composition, and different energy content all fall at the same rate of acceleration towards the earth (as they do in a rocket accelerating at 1g). How do we know that the WEP holds true in general?

Newton was well aware that inertial mass was equivalent to gravitational mass (which we call Newtonian equivalence). In fact even before Newton, there was an early experimental demonstration of the equivalence principle at work. This historical experiment was performed by Galileo in Pisa, where two objects of significantly different mass were dropped off the leaning tower of Pisa. Galileo observed that the two masses arrived on the ground roughly at the same time. The weak equivalence principle implies that these two different masses should fall at exactly the same rate (as they obviously do inside an accelerated rocket). Since this early experiment, the equivalence of inertia and gravitational mass has been verified to a phenomenal accuracy of about one part in about $10^{-15}$ (ref 24). Einstein is generally credited with the elevation of the equivalence principle to a fundamental symmetry of nature in 1915.

Conventional wisdom in physics assumes that the strong principle of equivalence is ***exact***, and somehow reflects a fundamental aspect of nature. It is assumed to be applicable under any physical circumstance. It is believed to hold true at the elementary particle level, and under enormously large gravitational fields such as on a neutron star. As a consequence, Einstein's general relativity theory (which is based on this principle) is also assumed to hold true under any physical condition. The principle of equivalence has been tested under a wide variety of gravitational field strengths and distance scales. It has been tested with different material types (ref. 6 and 7). It has been tested to an extremely high precision for laboratory bodies (up to 3 parts in $10^{-12}$). It also has been checked to 1 part in $10^{-12}$ for the acceleration of the moon and earth towards the sun. It has even been tested for individual elementary particles, such as the neutron. To our knowledge, no confirmed violation of the equivalence principle has been reported to date.

Yet after 80 years of close scrutiny, the principle of equivalence has still remained only a postulate of general relativity. It cannot be proven from more fundamental principles. Some of the better literature on general relativity have drawn attention to this fact, and admit that no explanation can be found as to; "why our universe has a deep and mysterious connection between acceleration and gravity" (ref. 8). One must bear in mind that mass is really nothing more than a vast collection of quantum particles, which interact with each other through forces. Forces are also ultimately the result of quantum particles called bosons, which act like the exchange particles that transmit momentum from one particle to another. Therefore, it is essential that equivalence be understood at a quantum particle level in order for a deeper understanding to emerge.

We will show why there is equivalence, and also demonstrate that the principle of equivalence is just an approximation, albeit an extremely close approximation. We will also show that there is actually a tiny imbalance in the equivalence of inertial and gravitational mass, with the gravitational mass of an object being (every so slightly) larger than the inertial mass. This effect is magnified when comparing the free fall times of very large mass to that of an extremely tiny mass on the earth. Furthermore this effect may be measured experimentally in the near future. We will also show that in certain extremely rare physical circumstances, the equivalence principle does not hold at all (for example, anti-matter; described in section 8)!

In order to understand the equivalence principle (and inertia) on the quantum level, we must have an understanding of the basic concepts of a new theory of gravity called ElectroMagnetic Quantum Gravity (EMQG, ref. 1, 1998). The motivation for the development of EMQG was the consideration that the universe may be a vast Cellular Automaton process (ref. 57). In the process of developing EMQG theory with this goal in mind, we discovered the hidden quantum processes that are responsible for the principle of equivalence. Before we can present our derivation of the equivalence principle, we must first briefly review the important concepts of EMQG theory that are required to understand the principle of equivalence. Reference 1 (1998) contains the entire work.

## 2. INTRODUCTION TO ELECTROMAGNETIC QUANTUM GRAVITY

***"The interpretation of geometry advocated here cannot be directly applied to submolecular spaces … it might turn out that such an extrapolation is just as incorrect as an extension of the concept of temperature to particles of a solid of molecular dimensions"***

***A. Einstein (1921)***

Two years ago we introduced a theory of gravity based solely on the activities of quantum particles, which consist of quantum mass particles (fermions) and quantum force particles (bosons). ElectroMagnetic Quantum Gravity (EMQG, ref. 1) grew out of an attempt to understand not just gravity on a quantum scale, but also inertia and the principle of equivalence. Various attempts at the unification of general relativity with quantum theory have not been entirely successful in the past, because these theories do not grasp the true quantum nature of inertia and the hidden quantum processes behind Einstein's principle of equivalence. In developing a theory of quantum gravity, one might ask which of the existing approaches is more relevant or fundamental; quantum field theory or classical general relativity (with conventional 4D space-time continuum)? Currently it seems that both theories are generally not compatible with each other.

EMQG is based on a postulate that the universe operates like some form of a vast Cellular Automaton (CA) (ref. 2, 3, 4, 34, 57). Figure 21 shows a tentative model for the universal CA. This led us to the position that quantum field theory is actually in closer touch to the true inner workings of our universe, than general relativity. This is because many CA's inherently produce stable, oscillating particle-like patterns that are reminiscent of elementary particles interacting. The CA requires strict locality, and therefore forces can only be possible through the exchange of 'force particles'. On the other hand general relativity is taken as a global, classical description of space-time, gravity and matter. General relativity reveals the large-scale patterns and organizing principles that arise from the hidden, local quantum processes existing on the quantum particle distance scales.

A Cellular Automaton consists of a huge array of 'cells' or memory locations that are capable of storing numerical data. The numeric state of *all the cells,* everywhere, changes at a regular synchronized interval called a 'clock cycle'. The change in numeric state depends on the numeric values of a cell's neighbors and on the same set of fixed rules or algorithm, which are located in each and every cell of the cellular automaton. *It is important to note that the 'clock cycle' marks events related to the CA operation, and is not the same thing as events in our measure of time.*

Unlike the ordinary desktop computer, the cellular automata computer updates *all* it's memory locations in a single 'clock cycle', thus making the CA computer vastly more faster and powerful than any of the common computers systems in use today. It should be noted that there already exists computational machines modeled on the CA architecture, which are being used by physicists for various physical modeling projects. These CA machines are used to model nature, such as problems in fluid flow and turbulence, rather then for computing things like income tax or other business applications. However, it is interesting to note how well the CA computer model is suited to modeling various particle interactions, such as for elementary particle physics problems, fluid mechanics, and turbulence problems. The CA structures used for these type of simulations are known as 'Lattice Gases'.

The cellular automaton computer was discovered theoretically by Konrad Zuse and Stanislav Ulam in the late 1940's, and later was put to use by John von Neumann to model the real world behavior of complex spatially extended structures. The best known example of a CA is the 'game of life' originated by John Horton Conway and published in the famous mathematical games department of Scientific American magazine in the 1970's. A direct consequence of CA theory is that our universe must be *utterly simple* on the smallest scales. Furthermore, the universe must be completely deterministic, or at least be deterministic in principle. The apparent randomness of nature set forth by quantum theory must reflect our general ignorance of the exact numeric state of a system on the CA. If we know the structure of the universal CA, the exact mathematical rules that govern the cells of the CA, and the exact numeric state of all the cells in the universe at a given instant, then we have all the information needed to predict the exact future state of the system. Of course this is completely impossible to do in practice, for a variety of reasons. Even if we can somehow gather all this information, we would need a computer just as powerful as the universal CA to process the data in a reasonable period of time!

We believe that Quantum Field Theory (QFT) is currently our best mathematical description of the inner fundamental workings of the universe. QFT tells us that all forces originate from a quantum particle exchange process. These particle exchanges transfer momentum from one quantum particle to another, and the huge numbers of particles exchanged produces generates what appears to be a smooth force interaction. The exchange process is universal, and applies to electromagnetic, weak and strong nuclear forces, and also (as we hope to show here) for the gravitational force. The generic name given to the force exchange particle is the 'vector boson' particle.

ElectroMagnetic Quantum Gravity (EMQG) is basically a quantum field theory model of gravity. Figure 17 shows a block diagram of the relationship of EMQG with the rest of physics. Basically EMQG supports the view that gravity is based on particle exchange processes in accordance with the general principles of quantum field theory. EMQG is based on the graviton (the exchange particle for the *pure* gravitational force), and *also* on the photon (the exchange particle for electromagnetic force) exchange particles. What is unique about EMQG theory is that gravitation involves ***both*** the photon and graviton exchange particles operating at the same time, where now the photon plays a very important role in gravity! In fact, the photon exchange process dominates all gravitational interactions and is, in most part, responsible for the principle of equivalence. The photon particle is also responsible for another important property that all matter possesses; *Newtonian inertia*.

When formulating a new theory of quantum gravity, a mechanism must be found that produces the gravitational force, while somehow being linked to the principle of equivalence. In addition, this mechanism should naturally lead to 4D space-time curvature and should be compatible with the principles of general relativity theory. Nature has another long-range force called electromagnetism, which is extremely prominent in all physical interactions, and has been described quite successfully by the principles of quantum field theory. This well-known theory is called Quantum ElectroDynamics (QED), and this theory has been tested for electromagnetic phenomena to an unprecedented accuracy. Electromagnetism plays the essential role of atomic and molecular interactions, and is responsible for most observable phenomena, including virtually all of chemistry and biology.

Electromagnetic force is a long range force. Gravitational force is also a long range force. Therefore we believe that it is reasonable to assume that gravitational force should, somehow, result from a similar process as electromagnetism. The similarity between the two basic forces is also evident from classical physics, which describe the Coulomb electrical force and Newton's law for the gravitational force. However a few obstacles lie in the way for this line of reasoning.

First, gravitational force is observed to be <u>always</u> attractive, and never repulsive! In QED, electrical forces can be both attractive or repulsive. One result of this is that there are equal numbers of positive and negative charged virtual particles in the quantum vacuum at any given instant of time. This is because virtual particles are always created in equal and opposite charged particle pairs, according to quantum field theory. Thus, there is a balance of attractive and repulsive electrical forces in the quantum vacuum, and the quantum vacuum is electrically neutral, overall. If this were not the case, the electrically charged virtual charged particles of one charge type in the vacuum would dominate over all other physical interactions involving real matter, due to the enormous number of vacuum particles involved.

Secondly, QED is formulated in a relativistic, flat 4D space-time with no curvature. In QED, electrical charge is defined as the fixed rate of emission of photons (strictly speaking, the fixed probability of emitting a photon) from a given charged particle. Electromagnetic forces are caused by the exchange of photons, which propagate between the charged particles. The photons transfer momentum from the source charge to the destination charge, and travel in flat 4D space-time (assuming no gravity). From these basic considerations, a successful theory of quantum gravity should have an exchange particle or graviton, which is emitted from a mass particle at a fixed rate as in QED. This is called 'mass charge', and is analogous to electrical charge in QED. Therefore, the graviton transfers momentum from one mass to another, which is the root cause of gravitational force. Yet, the graviton exchanges must somehow produce disturbances of the normally flat space and time, when originating from a very large mass.

Since mass is known to vary with velocity (special relativity), one might expect that 'mass charge' of a particle might also vary with velocity. In QED the electromagnetic force exchange process is governed by a fixed, universal constant ($\alpha$) which is not affected by anything like motion (more will be said about this point later). Should this not be true for graviton exchange process in quantum gravity as well? As we hinted, gravity is a long-range force, governed by a similar mathematical law as found in Coulomb's Electrical Force law. Coulomb's Electric Force law states: $F = KQ_1Q_2/R^2$ , and Newton's Gravitational Force law: $F=GM_1M_2/R^2$. Both are inverse square laws, both depend linearly on the product of the two masses (or charges). Both obey the principle of superposition.

We believe that this suggests that there is a deep connection between gravity and electromagnetism. Yet, gravity supposedly has no counterpart to negative electrical charge. Thus might leads one to believe that there is no such thing as negative 'mass charge' for gravity. Furthermore, QED also has no analogous phenomena as the principle of equivalence. Why should gravity be connected with the principle of equivalence, and thus inertia, and yet no analogy of this phenomena exists for electromagnetic phenomena?

To answer the question of negative 'mass charge', EMQG postulates the existence of negative 'mass charge' for gravity, in close analogy to electromagnetism. Furthermore, we claim that this property of matter is possessed by all anti-particles that carry mass. Therefore anti-particles, which are opposite in electrical charge to ordinary particles, are also opposite in 'mass charge'. In fact, negative 'mass charge' is not only abundant in nature, it comprises nearly half of all the mass particles in the form of 'virtual' particles in the universe! The other half exists as positive 'mass charge', also in the form of virtual particles. Furthermore, all familiar ordinary (real) matter comprises only a tiny fraction of the total 'mass charge' in the universe, and is experimentally found to be almost all positive! Real anti-matter seems to be very scarce in nature, and no search for it in the cosmos has revealed any to date.

Both positive and negative 'mass charge' appear in huge numbers in the form of virtual particles, which quickly pop in and out of existence in the quantum vacuum (section 3), everywhere in empty space. We suggest that the existence of negative 'mass charge' is the key to the solution to the famous problem of the cosmological constant (ref. 46), which is one of the great unresolved mysteries of modern physics. Finally, we propose that the negative energy, or the antimatter solution of the famous Dirac equation of quantum field theory is also the ***negative 'mass charge' solution for fermions***.

Previous attempts at quantizing the gravitational field have been made using the principles of quantum field theory. They focused on using the graviton force exchange particle alone, which is proposed as the quanta of the gravitational field in direct analogy with the quantization of electromagnetic fields with photons. The graviton particle is chosen with the right mathematical characteristic to quantize gravity in accordance with quantum field theory and general relativity. These attempts however, fail to account for the origin of space-time curvature. Specifically, how does a graviton 'produce' curvature when propagating from one mass to another? Does the graviton move in an already existing 4D space-time curvature? If it does, how is the space-time produced by the graviton? If not, how is 4D space-time curvature produced? In other words, if the 4D space-time curvature is not caused by the graviton exchanges, then what is the cause?

Furthermore, why do the virtual particles of the quantum vacuum ***not*** contribute a nearly infinite amount of curvature to the whole universe? After all, the force of gravity is universally attractive. According to quantum field theory, virtual gravitons should exist in huge numbers in the quantum vacuum, and should therefore contribute huge amounts of attractive forces and a large amount of space-time curvature. This infamous question is known as the problem of the cosmological constant.

Does graviton exchange processes get affected by high velocity motion (with respect to some other reference frame)? In other words, do the number of gravitons exchanged increase as the velocity of the mass increases, as seems to be required by the special relativistic mass increase with velocity formula? Why does the state of motion of an observer near a gravitational field affect his local 4D space-time curvature? For example, why does an observer in free fall near the earth affect his local space-time conditions in such a way as to match an observer in far space (who lives in flat space-time)? EMQG provides answers for many of these questions. Interested readers should refer to reference 1 (this is quite lengthy) for the answers to these and other questions to do with EMQG. In the interest of being concise, this work presents only the important results from EMQG that are needed to understand the principle of equivalence. One of the key players of gravity in EQMG is the quantum vacuum. Knowledge of this important medium is essential to the understanding of the principle of equivalence. Therefore we now briefly review this subject.

## 3. THE QUANTUM VACUUM AND IT’S RELATIONSHIP EMQG

***Philosophers:*** ***“Nature abhors a vacuum.”***

On pondering the vacuum, one might think that the vacuum is completely devoid of everything. In fact, the physical vacuum is really far from being empty. The recipe for creating a complete vacuum is to remove all matter from an enclosure. However, one would find that this is still not good enough. One must also lower the temperature down to absolute zero in order to remove all thermal electromagnetic radiation. However, Nernst correctly deduced in 1916 (ref. 32) that empty space is still not completely devoid of all radiation after this is done. He predicted that the vacuum is still permanently filled with an electromagnetic field propagating at the speed of light, called the zero-point fluctuations (sometimes called vacuum fluctuations). This was later confirmed by the full quantum field theory developed in the 1920’s and 30’s. Later, with the development of QED, it was realized that all quantum fields should contribute to the vacuum state, like virtual electrons and positron particles, for example.

In order to make a complete vacuum, one must remove all matter from an enclosure. However one would find that this is still not good enough. One must also lower the temperature inside the closure to absolute zero in order to remove all thermal electromagnetic radiation. Nernst correctly deduced in 1916 (ref. 32) that empty space is still not completely devoid of all radiation after this is done. He predicted that the vacuum is still permanently filled with an electromagnetic field propagating at the speed of light, called the zero-point fluctuations (or sometimes called by the generic name 'vacuum fluctuations'). This result was later confirmed theoretically by the newly developed quantum field theory that was developed in the 1920's and 30's.

Later with the development of QED (the quantum theory of electrons and photons), it was realized that all quantum fields should contribute to the vacuum state. This means that virtual electrons and positron particles should not be excluded from consideration. These particles possess mass and have multiples of half integer spin (such as the electron), and therefore belong to the generic class of particles known as fermions. We refer to virtual electrons and virtual anti-electrons (positron) particles as virtual fermions. We believe that ultimately all fermions can be broken down to a fundamental entity that is also electrically charged, as well as having half integer spin and mass.

According to modern quantum field theory, the perfect vacuum is teeming with activity as all types of quantum virtual particles (and virtual bosons or the force carrying particles) from the various quantum fields appear and disappear spontaneously. These particles are called 'virtual' particles because they result from quantum processes that generally have short lifetimes, and are mostly undetectable. One way to look at the existence of the quantum vacuum is to consider that quantum theory forbids the complete absence of propagating fields. This is in accordance with the famous Heisenberg uncertainty principle. In general, it is known that all the possible real particles types (for example electrons, quarks, etc.) will also be present in the quantum vacuum in their virtual particle form.

In the QED vacuum, the quantum fermion vacuum is produced from the virtual particle pair creation and annihilation processes that create enormous numbers of virtual electron and virtual positron pairs. We also have in QED the creation of the zero-point-fluctuation (ZPF) of the vacuum consisting of the electromagnetic field or virtual photon particles. Indeed in the standard model, we also find in the vacuum every possible boson particle, such as the gluons, gravitons, etc., and also every possible fermion particle, such as virtual quarks, virtual neutrinos, etc.

Some estimates of the quantum vacuum particle density (ref. 11) maintain that the number of vacuum particle is a staggering $10^{90}$ particles per cubic meter of empty space! The densest places in the universe, for example neutron stars, contain nowhere near this many particles. For comparison purposes, a neutron star contains about $10^{45}$ nucleons per cubic meter, and therefore the quantum vacuum particles has on the order of $10^{45}$ times more density than a neutron star!! We believe that the quantum vacuum plays a much more prominent role in physics than is generally believed by most physicists. We maintain that the effects of the quantum vacuum are present in virtually all physical activity. In fact, Newton’s three laws of motion can be understood to originate directly from the effects of the quantum vacuum (ref. 48).

One of the most important experimental evidence for the vacuum particles is the Casimir effect, which we discuss in detail.

## 3.1 INTRODUCTION TO THE CASIMIR FORCE EFFECT

The existence of virtual particles of the quantum vacuum reveals itself in the famous Casimir effect (ref. 49), which is an effect predicted theoretically by the Dutch scientist Hendrik Casimir in 1948. The Casimir effect refers to the tiny attractive force that occurs between two neutral metal plates suspended in a vacuum. He predicted theoretically that the force ‘F’ per unit area ‘A’ for plate separation D is given by:

$$\frac{F}{A} = \frac{-p^2 hc}{240 D^4} \quad \ldots \quad \text{Newton's per square meter (Casimir Force 'F')} \qquad (3.1)$$

Casimir obtained this formula by calculating the sum of the quantum-mechanical zero-point energies of the normal modes of the electromagnetic field (virtual photons) between two conductive plates.

The origin of this minute force can be traced to the disruption of the normal quantum vacuum virtual photon distribution between two nearby metallic plates as compared to the vacuum state outside the plates. Certain virtual photon wavelengths (and therefore energies) are forbidden to exist between the plates, because these waves do not ‘fit’ between the two plates (which are both at a relative classical electrical potential of zero). This creates a negative pressure due to the unequal energy distribution of virtual particles inside the plates as compared to those outside the plate region. The pressure imbalance can be visualized as causing the two plates to be drawn together by radiation pressure. Note: Even in the vacuum state, the virtual photon particles ***do*** carry energy and momentum while they exist.

Although the Casimir effect has been attributed to the zero-point fluctuations (ZPF) in the EM field inside the plates, Schwinger showed in the late 70’s that the Casimir effect can also be derived in terms of his source theory (ref. 50), which has no explicit reference to the ZPF of the EM field between the plates. Recently Milonni and Shih have developed a theory of the Casimir force effect, which is totally within the framework of conventional QED (ref. 51). Therefore it seems that it is only a matter of taste whether we attribute the Casimir force effect to the ZPF fields or to the matter fields in vacuum (ref. 52).

Recently Lamoreaux made accurate experimental measurements for the first time of the Casimir force existing between two gold-coated quartz surfaces that were spaced on the order of a micrometer apart (ref. 53). Lamoreaux found a force value of about 1 billionth of a Newton, agreeing with the Casimir theory to within an accuracy of about 5%. More recently, U. Mohideen and A. Roy have made an even more precise measurement in the 0.1 to 0.9 micrometer plate spacing to an accuracy of about 1% (1998, ref. 54). Therefore the experimental reality of this effect is beyond question.

Can the vacuum state be disrupted by other physical processes besides the Casimir plates? One might ask what happens to the virtual particles of the quantum vacuum that are subjected to a large gravitational field like the earth? Since the quantum vacuum is composed of virtual fermions (as well as virtual bosons), the conclusion is inescapable: ***all the virtual fermions possessing mass must be falling (accelerating) on the average towards the earth during their very brief lifetimes***. This vacuum state is definitely different from the vacuum of far outer space away from gravitational fields. Yet to our knowledge, no previous authors have acknowledged the existence of this effect, or studied the physical consequences that result from this. It turns out that the free fall state of the virtual, electrically charged fermion particles of the vacuum is actually the root cause of 4D space-time curvature and also leads to a full understanding of the principle of equivalence. In EMQG we fully study the consequences of a falling quantum vacuum in quantum gravity, which does lead to new experimentally testable predictions.

The physics of the Casimir force effect implies that the quantum vacuum contains an enormous reservoir of energy (ref. 11). Although in standard quantum field theory the number density of virtual particles is unlimited, some estimates place a high frequency cut-off at the plank scale which is estimated to be a density of $10^{90}$ particles per cubic meter (ref. 11)! Generally this energy-density is not available because the energy-density is uniform and it permeates everything. It's like the situation in the deep ocean, where deep sea fishes easily tolerate the extreme pressures in the abyss, because the pressure inside and outside the fish's body balance. If a human goes into these depths, a great difference in pressure must be maintained to support atmospheric pressure inside the human body. Some physicists are looking at ways in which this vast energy reservoir can be tapped (ref. 11). Is there any other evidence for the existence of the quantum vacuum?

## 3.2 EVIDENCE FOR THE EXISTENCE OF VIRTUAL PARTICLES

Aside from the Casimir force effect, there is other evidence in theoretical and experimental physics for the existence of virtual particles. We present a very brief review of some theoretical and experimental evidence for the existence of the virtual particles of the quantum vacuum. Some knowledgeable readers may wish to skip this section.

(1) The extreme precision in the theoretical calculations of the hyper-fine structure of the energy levels of the hydrogen atom, and the anomalous magnetic moment of the electron and muon are both based on the existence of virtual particles in the framework of QED. These effects have been calculated in QED to a very high precision (approximately 10 decimal places), and these values have also been verified experimentally to an unprecedented accuracy. This indeed is a great achievement for QED, which is essentially a perturbation theory of the electromagnetic quantum vacuum. Indeed, this is one of physics greatest achievements.

(2) Recently, vacuum polarization (the polarization of electron-positron pairs near a real electron particle) has been observed experimentally by a team of physicists led by David Koltick. Vacuum polarization causes a cloud of virtual particles to form around the electron in such a way as to produce an electron charge screening effect. This is because virtual positrons tend to migrate towards the real electron, and the virtual electrons tend to migrate away. A team of physicists fired high-energy particles at electrons, and found that the effect of this cloud of virtual particles was reduced the closer a particle penetrated towards the electron. They reported that the effect of the higher charge for the penetration of the electron cloud with energetic 58 giga-electron volt particles was equivalent to a fine structure constant of 1/129.6. This agreed well with their theoretical prediction of 128.5 of QED. This can be taken as verification of the vacuum polarization effect predicted by QED, and further evidence for the existence of the quantum vacuum.

(3) The quantum vacuum explains why cooling alone will never freeze liquid helium. Unless pressure is applied, vacuum energy fluctuations prevent its atoms from getting close enough to trigger solidification.

(4) For fluorescent strip lamps, the random energy fluctuations of the virtual particles of the quantum vacuum cause the atoms of mercury, which are in their exited state, to spontaneously emit photons by eventually knocking them out of their unstable energy orbital. In this way, spontaneous emission in an atom can be viewed as being directly caused by the state of the surrounding quantum vacuum.

(5) In electronics, there is a limit as to how much a radio signal can be amplified. Random noise signals are always added to the original signal. This is due to the presence of the virtual particles of the quantum vacuum as the real radio photons from the transmitter propagate in space. The vacuum fluctuations add a random noise pattern to the signal by slightly modifying the energy of the propagating radio photons.

(6) Recent theoretical and experimental work done in the field of Cavity Quantum Electrodynamics suggests that the orbital electron transition time for excited atoms can be affected by the state of the virtual particles of the quantum vacuum immediately surrounding the excited atom in a cavity, where the size of the cavity modifies the spectrum of the virtual particles.

In the weight of all this evidence, only a few physicists doubt the existence of the virtual particles of the quantum vacuum. Yet to us, it seems strange that the quantum vacuum should barely reveal it’s presence to us, and that we only know about it’s existence through rather obscure physical effects like the Casimir force effect and Davies-Unruh effect. This is especially odd considering that the real observable particles in our universe constitute a minute fraction of the total population of virtual particles of the quantum vacuum in any given point in time (even when considering in the densest places in the universe).

We believe that the vacuum particle plays a much more prominent role in physics, and is totally responsible for Newtonian Inertia. Furthermore, it plays a major role in the principle of equivalence. The quantum vacuum mechanism responsible for inertia is the common electrical force process originating from photon exchanges. We describe this process for inertia in more detail in the next section.

## 4. INTRODUCTION TO QUANTUM INERTIA THEORY

**"Under the hypothesis that ordinary matter is ultimately made of subelementary constitutive primary charged entities or 'partons' bound in the manner of traditional elementary Plank oscillators, it is shown that a heretofore uninvestigated Lorentz force (specifically, the magnetic component of the Lorentz force) arises in any accelerated reference frame from the interaction of the partons with the vacuum electromagnetic zero-point-field (ZPF). ... The Lorentz force, though originating at the subelementary parton level, appears to produce an opposition to the acceleration of material objects at a macroscopic level having the correct characteristics to account for the property of inertia."**

**- B. Haisch, A. Rueda, H. E. Puthoff (1994)**

It has been recently proposed (1994) that Newtonian Inertia is strictly a quantum vacuum phenomenon (as the quote above suggests)! If this is true, then the existence of the quantum vacuum actually reveals it's presence to us in all daily activities! Unlike the hard-to-measure quantum vacuum effects such as the Casimir force, the presence of the inertial force is universal and it's presence prevails throughout all of physics. For example, when you step on the gas pedal of your car you feel an inertial force pushing you against the seat. If inertia is truly a quantum vacuum effect, then every time you are accelerating you are witnessing the effects of the quantum vacuum! This is a far cry from the rather obscure and exotic status that the quantum vacuum now holds in current physical thought.

In 1994 three physicists, R. Haisch, A. Rueda, and H. Puthoff (ref. 5), were the first to propose such a vacuum theory of inertia (known here as HRP Inertia), in which the quantum vacuum played a central role in acceleration and inertial mass. They proposed that inertia is due to the strictly local electrical force interactions of charged matter particles with the immediate background virtual particles of the quantum vacuum (in particular the virtual photons, or ZPF as the authors called it). They found that inertia is caused by the magnetic component of the Lorentz force, which arises between what the author's call the charged 'parton' particles in an accelerated reference frame interacting with the background quantum vacuum virtual particles. The sum of all these tiny forces in this process is the source of the resistance force opposing accelerated motion in Newton's F=MA. The 'parton' is a term that Richard Feynman coined for the constituents of the nuclear particles such as the proton and neutron. Partons are now generally called quarks.

We have found it necessary to make a small modification to HRP Inertia theory as a result of our investigation into the principle of equivalence. Our modified version of HRP inertia is called "Quantum Inertia" (or QI). This theory also resolves the long outstanding problems and paradoxes of accelerated motion introduced by Mach's principle, by suggesting that the vacuum particles themselves serve as Mach's universal reference frame (for <u>acceleration</u> only), without violating the principle of relativity of constant velocity motion.

In other words, our universe offers no observable reference frame to gauge inertial frames (a non-accelerated frame, where Newton's laws of inertia is valid), because the quantum vacuum offers no means to determine absolute constant velocity motion. However for accelerated motion, the quantum vacuum plays a very important role by offering a resistance to the acceleration of a mass, which results in an opposing inertial force. Thus, the very existence of inertial force reveals the absolute value of the acceleration with respect to the net statistical average acceleration of the virtual particles of the quantum vacuum.

In the past there have been various clues to the importance of the state of the virtual particles of the quantum vacuum, with respect to the accelerated motion of real charged particles. One example is the so-called Davies-Unruh effect (ref. 15), where uniform and linearly accelerated charged particles in the vacuum are immersed in a heat bath, with a temperature proportional to acceleration (with the scale of the quantum heat effects being very low). However, the work of reference 5 is the first place we have clearly seen the identification of inertial forces as the direct consequence of the interactions of real matter particles with the virtual particles of the quantum vacuum.

It has even been suggested that the virtual particles of the quantum vacuum are somehow involved in gravitational interactions, a central concept of EMQG. The prominent Russian physicist A. Sakharov proposed in 1968 (ref. 16) that Newtonian gravity could be interpreted as a van der Waals type of force induced by the electromagnetic fluctuations of the virtual particles of the quantum vacuum. Sakharov visualized ordinary neutral matter as a collection of electromagnetically, interacting polarizable particles made of charged point-mass subparticles (partons). He associated the Newtonian gravitational field with the Van Der Waals force present in neutral matter, where the long-range radiation fields are generated by the parton 'Zitterbewegung'. Sakharov went on to develop what he called the 'metric elasticity' concept, where space-time is somehow identified with the 'hydrodynamic elasticity' of the vacuum. However, he did not understand the important clues offered in understanding the equivalence principle, nor the important role that the quantum vacuum plays in inertia and in Mach's principle. We will see that the quantum vacuum also make it's presence felt in a very important way in all gravitational interactions.

After Sakharov there has been further hints that the quantum vacuum is involved in gravitational physics. In 1974 Hawkings (ref. 17) announced that black holes are not completely black. Black holes emit an outgoing thermal flux of radiation due to gravitational interactions of the black hole with the virtual particle pairs created in the quantum vacuum near the event horizon. At first sight, the emission of thermal radiation from a black hole seems paradoxical (since nothing can escape from the event horizon). However, the spontaneous creation of virtual particle and anti-particle pairs in the quantum vacuum near the event horizon can be used to explain this effect (ref. 18). Heuristically, one can imagine that the virtual particle pairs (created with wavelength $\lambda$ are approximately equal to the size of the black hole) 'tunnel' out of the event horizon. For a virtual particle with a wavelength comparable to the size of the hole, strong tidal forces operate to prevent re-annihilation. One virtual particle escapes to infinity with positive energy to contribute to the Hawking radiation, while the corresponding antiparticle enters the black hole to be trapped forever by the deep gravitational potential. Thus, the quantum vacuum is important in order to properly understand the Hawking radiation.

According to Quantum Inertia theory, the property which Newton called the inertial mass of an object, is caused by the vacuum resistance to acceleration of all the individual, electrically charged masseon particles (section 6) that make up the mass. This resistance force is caused by the electromagnetic force interaction (where the details of this process are still unknown at this time) occurring between the electrically charged virtual masseon/anti-masseon particle pairs created in the surrounding quantum vacuum (section 6.1), and all the real masseons particles making up the accelerated mass. Therefore inertia originates in the photon exchanges with the electrically charged virtual masseon particles of the quantum vacuum. The total inertial force $F_i$ of a mass is simply the sum of all the little forces $f_p$ contributed by each of the individual masseons, where the sum is: $F_i = (\Sigma f_p) = MA$ (Newton's law of inertia). Figure 1 shows a schematic diagram of this process for a mass accelerated by a rocket motor..

Any physicist that believes in the existence of the virtual particles of the quantum vacuum and accepts the fact that many virtual particles carry mass (in the form of virtual fermions such as virtual electrons and virtual quarks), will have no trouble in believing that virtual fermion particles are falling in the presence of a large gravitational mass like the earth. Figures 4 and 5 show the falling state of the quantum vacuum, and it's effect on test masses. Yet no one has examined the full ramifications of this phenomena, which is extremely important to EQMG theory.

We believe the existence of the downward accelerating virtual particles (which are accelerating during their very brief lifetimes) under the action of a large gravitational field turns out to be the *missing link* between inertia and gravity. It leads us more or less directly towards a full understanding of the principle of equivalence. Although the quantum vacuum has been studied in detail in the past, to our knowledge no one has examined the direct consequences of a quantum vacuum in a state of free-fall near the earth. Before we continue studying these concepts, we present a rather brief review of the important results from general relativity.

## 5. GENERAL RELATIVITY, ACCELERATION, GRAVITY AND CA

**"The general laws of physics (and gravitation) are to be expressed by equations which hold good for all systems of coordinates."**

**- Albert Einstein**

From the perspective of EMQG, Einstein's gravitational field equations are a set of observer dependent equations, for observers that are subjected to gravity and/or to acceleration. These equations are based on *measurable* relativistic 4D space-time. The core of Einstein's theory is the principle of equivalence and the principle of general covariance, which allows an observer in any state of motion (and coordinate system) to describe gravity and acceleration. We very briefly review these basic postulates, and important concepts of Einstein's general theory of relativity.

### POSTULATES OF GENERAL RELATIVITY

General relativity is a classical field theory founded on all the postulates and results of special relativity. It is also based on Newtonian inertia and Newton's laws of motion, as well as on the following new postulates introduced by Einstein:

**(1) PRINCIPLE OF EQUIVALENCE (STRONG) - The results of any given physical experiment will be precisely identical for an accelerated observer in free space as it is for a non-accelerated observer in a perfectly uniform gravitational field. A weaker form of this postulate states that: objects of different mass fall at the same rate of acceleration in a uniform gravity field.**

**(2) PRINCIPLE OF COVARIANCE - The general laws of physics can be expressed in a form that is independent of the choice of space-time coordinates and the state of motion of an observer.**

As a consequence of postulate 1, the inertial mass of an object is equivalent to it's gravitational mass. Einstein uses this principle to encompass gravity and inertia into his single framework of general relativity in the form of a metric theory of acceleration and gravity, based on quasi-Riemann geometry (where time also enters as a coordinate, unlike pure Riemann geometry).

These postulates, and the additional assumption that when gravitational fields are present nearby, space-time takes the form of a quasi-Riemannian manifold endowed with a metric curvature of the form $ds^2 = g_{ik}\, dx^i\, dx^j$, led Einstein to discover his famous gravitational field equations given below:

$$R_{ik} - \frac{1}{2} g_{ik} R = \frac{8pG}{c^2} T_{ik} \quad \ldots \text{ Einstein's Gravitational Field Equations} \qquad (5.1)$$

where, $g_{ik}$ is the metric tensor, $R_{ik}$ is the covariant Riemann curvature tensor. The left-hand side of the above equation is called the Einstein tensor or $G_{ik}$, which is the mathematical statement of space-time curvature that is reference frame independent, and generally covariant. The right hand side $T_{ik}$ is the stress-energy tensor which is the mathematical statement of the special relativistic treatment of mass-energy density, G is Newton's gravitational constant, and c the velocity of light.

Einstein's law of gravitation (eq 5.1) cannot be derived from any 'rigorous' proof. The famous physicist S. Chandrasekhar writes in regards to this (ref 37):

*"... It seems to this writer that in fact no such derivation exists and that, at the present time, no such can be given. ... It is the object of this paper to show how a mixture of physical reasonableness, mathematical simplicity, and aesthetic sensibility lead one, more or less uniquely, to Einstein's field equations."*

In Einstein's field equation (eq. 5.1) the principle of equivalence (in its strong form) is incorporated in the above framework by the assertion that all accelerations are caused by either gravitational or inertial forces, and are ***metrical*** in nature. In other words, the space-time is equivalent in inertial frame of a rocket at 1g and gravitational field on the earth (in a small vicinity). More precisely, the presence of acceleration caused by either an inertial force or a gravitational field modifies the geometry of space-time such that it is a quasi-Riemannian manifold endowed with a metric.

Furthermore, point particles move in gravitational fields along ***geodesic paths*** governed by the equation:

$$\frac{d^2x^i}{ds^2} + \Gamma^i_{jk}\frac{dx^j}{ds}\frac{dx^k}{ds} = 0 \qquad \text{... Equation for the geodesics} \qquad (5.2)$$

The most striking consequence of general relativity is the existence of curved 4D space-time specified by the metric tensor $g_{ik}$. We will see that in EMQG theory, the meaning of the geodesic is very simple; it is the path taken by light or (force free) point particles through the falling, electrically charged, virtual particles undergoing acceleration in the absence of any other external forces, that maintains a relative, average acceleration of zero with the vacuum particles. In other words for light traveling through the falling quantum vacuum, the photons frequently scatter with the electrically charged virtual particles of the quantum vacuum causing a deflected path. We will see that curvature can be completely understood at the particle level, as a pure vacuum-particle process. Furthermore, we will see that the principle of equivalence is a pure particle interaction process, and not a fundamental law of nature. Before we can show this, we must carefully review the principle of equivalence from the context of general relativity theory.

## 5.1 INTRODUCTION TO THE PRINCIPLE OF EQUIVALENCE

***"I have never been able to understand this principle (principle of equivalence) ... I suggest that it be now buried with appropriate honors."***

**- Synge: Relativity- The General Theory**

Again we would like to emphasize that Einstein did not explain the origin of inertia in general relativity. Instead he retained the classical Newtonian theory of inertia. Inertia is described by Newton in his famous law: F=MA; which states that an object resists being accelerated by a force (F). In other words, a force is required to accelerate an object of mass (M) to an acceleration (A) in order to overcome the inertial back-acting force. Since acceleration is a form of motion, it would seem that a reference frame is required in order to gauge this motion. But this is not the case in Newtonian physics. All observers agree as to which frame is actually accelerating (without observing the actual motions of other observers), by finding out which frames has a force associated with it. Only non-accelerated frames are relative.

If the observers are placed in three rockets with no windows available, and only one observer has the rocket motor running, it is easy for each of the observer to determine if it is their rocket engine that is turned on. The observer simply looks for the presence of a force on his body. Two observers will feel no forces, and the third will. However, the observers who's rocket engine is turned on might conclude that one of the observers who's rocket engine is turned he is accelerating by measure x and t through relative measurements of space and time.

So constant velocities are always relative. Accelerations may be absolute (because we do *not* need to know x or t to determine $a=d^2x/dt^2$, and can determine acceleration by measuring F and m, where a=F/m), or acceleration may be relative (if we decide to measure a by direct determination of x and t) in special relativity. To us this seems very paradoxical. Einstein did not solve this anomaly (which relates to Mach's principle), nor did he provide a reason why the inertial and gravitation masses of an object are equal. This also still remains as a postulate in his theory. It is widely known that the principle of equivalence has been tested to great accuracy. The equivalence of inertia and gravitational mass has been verified to an accuracy of one part in about $10^{-15}$ (ref 24).

Einstein's general theory of relativity is considered a "classical" theory, because matter, space, and time are treated as continuous classical variables. It is known however, that matter is made of discrete particles, and that forces are caused by particle exchanges as described by quantum field theory. A more complete theory of gravity should encompass a detailed quantum process for gravity involving particle interactions only. In the next section we introduce another important player in EMQG, the masseon particle. This particle is required in order to understand the principle of equivalence. The masseon turns out to be the fundamental particle that makes up all fermions.

## 6. PHYSICAL PROPERTIES OF THE GRAVITON AND THE MASSEON

The theory that best describes the quantization of the electromagnetic force field is called Quantum Electrodynamics (QED). Here the charged particles (electrons, positrons) act upon each other through the exchange of force particles, which are called photons. The photons represent the quantization of the classical electromagnetic field. In classical electromagnetic theory, the force due to two charged particles decreases with the inverse square of their separation distance (Coulomb's inverse square law: $F = kq_1q_2/r^2$, where k is a constant, $q_1$ and $q_2$ are the charges, and r is the distance of separation).

QED accounts for this inverse square law by postulating the exchange of photons between the charged particles. The number of photons emitted and absorbed by a given charge (per unit of time) is fixed and is called the charge of the particle. Thus, if the charge doubles, the force doubles because twice as many photons are exchanged during the force interaction. This force interaction process causes the affected particles to accelerate either towards or away from each other depending on if the charge is positive or negative (because positive and negative charges transmit photons with slightly different wave functions that induce attraction or repulsive accelerations).

The strength of the electromagnetic force varies as the inverse square of the distance of separation between the charges in the following way: each charge sends and receives photons from every direction. But, the number of photons per unit area, emitted or received, decreases by the factor $1/4\pi r^2$ (the surface area of a sphere) at a distance 'r' due to the photon emission pattern spreading in all directions. Thus, if the distance doubles, the number of photons exchanged decreases by a factor of four.

Imagine that an electron particle is at the center of a sphere sending out virtual photons in all directions. Imagine that another electron is on the surface of a sphere at a distance 'r' from the emitter, which absorbs some of these photons. The absorption of these photons causes an outward acceleration, and thus a repulsive force. If the charge is doubled on the central electron, there is twice as many photons appearing at the surface of the sphere, and twice the force acting on the other electron. This accounts for the linear product of charge terms in the numerator of the inverse square law. In QED, photons do not interact with each other (through force exchanges). As a result, in-going and out-going photons do *not* affect each other during the exchange process, thorugh the exchange of other force particles.

For gravitational forces, it is experimentally observed that the force originating from two particles possessing mass decreases with the inverse square of their separation distance, and is given by Newton's inverse square law: $F = Gm_1m_2 / r^2$, where G is the gravitational constant, $m_1$ and $m_2$ are the masses, and r is the distance of separation. We have seen that the two force laws are very similar in form. QED theory accounts for Coulomb's law by the photon exchange process. Following the lead from the highly successful QED, EMQG replaces the concept of electrical 'charges' exchanging 'photons' with the idea that 'mass charges' exchange gravitons. Hence, gravitational mass at a fundamental level is simply the ability to emit or absorb gravitons, and pure low-level gravitational mass is interpreted as ‘mass charge’ of a fermion.

For gravity there are gravitons instead of photons, which are the force exchange particles of gravity. Like charge, it is the property called mass-charge that determines the number of exchange gravitons. The larger the mass, the greater the number of gravitons exchanged. Like electromagnetism, the strength of the gravity force decreases with the inverse square of the distance. This conceptual framework for quantum gravity has been around for some time now, but how are we to merge these simple ideas to be compatible with the framework of general relativity? We must be able to explain the Einstein's Principle of Equivalence and the physical connection between inertia, gravity, and curved space-time all within the general framework of graviton particle exchange. General Relativity is based on the idea that the forces experienced in a gravitational field and the forces due to acceleration are equivalent, and both are due to the space-time curvature.

In classical electromagnetism, if a charged particle is accelerated towards an opposite charged particle, the rate of acceleration depends on the electrical charge value. If the charge is doubled, the force doubles, and the rate of acceleration is doubled. If quantum gravity were to work in the exact same way, we would expect that the rate of acceleration of a mass near the earth would double if the mass doubles. The reason for this expectation is that the exchange process for gravitons should be very similar to electromagnetism. In other words, if the ‘mass-charge’ is doubled, the gravitational force is doubled. The only difference between the two forces is that gravity is a lot weaker by a factor of about $10^{-40}$. The weakness of the gravitational forces might be attributed either to the very small interaction cross-section of the graviton particle as compared to the photon particle, or to a very weak coupling constant (the absorption of a single graviton causing a minute amount of acceleration), or both.

Unfortunately, if the graviton exchange process worked ***exactly*** like QED, it would ***not*** reproduce the known nature of gravity. First, there is the problem of variation of mass with velocity as described by special relativity as $m= m_0 \ (1-v^2/c^2)^{-1/2}$ . At face value, this would mean that the number of gravitons exchanged depends on **velocity** of the gravitational mass, which does not easily fit into the framework of a QED type approach to quantum gravity.

Secondly, if two masses are sitting on a table with mass ‘M’ and mass ‘2M’, the forces against the tabletop varies with the mass, just as you would expect in a QED-like exchange of graviton particles. If the mass doubles, the force on the table doubles. Yet, the rate of acceleration is the same for these two masses in free fall. Why? Since twice the number of gravitons is exchanged under mass ‘2M’ in free, you would expect twice the force, and therefore mass ‘2M’ should arrive early. Matter has inertia, and this complicates everything. In almost all quantum gravity theories inertia appears as a separate process that is ‘tacked’ on in an ad hoc manner. As we said, the principle of equivalence merely raises this relationship between inertia and gravitation to the status of a postulate as in Einstein’s theory of general relativity.

All test masses accelerate at the same rate (g=9.8 m/sec$^2$ on the earth) no matter what the value of the test mass is. This is a direct consequence of the principle of equivalence. Mathematically, this follows from Newton's two **different** force laws: inertia and gravity as follows:

$F_i = ma$ ..... (Inertial force) (6. 1)
$F_g = GmM / r^2$ ..... (Gravitational force) (6. 2)

If in free fall, an object (mass m) in the presence of the earth's pull (mass M) is force free, then $F_i = F_g$ (since no other forces are present). Note that the same mass value ‘m’ appears in the two mass definition formulas for equations 6.1 and 6.2.

Therefore, $ma = GmM/ r^2$ or $a = GM/ r^2$ for the mass, independent of the value of the mass m. Thus we see that the rate of acceleration does not depend on the test mass m. All test masses accelerate at the same rate. Thus, inertia and gravity are intimately connected in a deep way because the measure of mass m is the same for acceleration as for gravity.

What is mass? In EMQG, gravitational mass originates from a low-level graviton exchange process that results from ‘mass charge’, where there the emission rate is constant. In fact, mass is quantized in exactly the same way as electric charge in QED. (There exists a fundamental unit of mass charge that is carried by the masseon particle, the lowest quanta of mass). The graviton ultimately responsible for the gravitational force.

We have seen that the graviton particle is the fundamental boson of the pure gravitational force. In EMQG, the graviton is very similar to the photon in physical characteristics. Both particles move at the speed of light, and have the same spin 1 (contrary to conventional wisdom). Both bosons have rest mass of zero. The following table compares the properties of the graviton with the photon particle:

**TABLE #1 PHYSICAL PROPERTIES OF THE PHOTON AND GRAVITON**

| NAME | BOSON PROPERTIES | | | | |
|---|---|---|---|---|---|
| | Symbol | Spin | Rest Mass | Electrical/Mass Charge | |
| Photon Particle | $\gamma$ | 1 | 0 | 0 | 0 |
| Graviton Particle | G | 1 (***not*** 2) | 0 | 0 | 0 |

Gravitons are emitted and absorbed by the ‘masseon’ particle, which is the most elementary form of matter or anti-matter. The masseon carries the lowest possible quanta of gravitational ‘mass charge’. This elementary particle is called the masseon particle (and also comes in anti-masseon varieties, which is the corresponding anti-particle). The masseon is postulated to be the smallest fundamental mass particle and readily combines with other masseons through a new, hypothetical force coupling, which we call the ‘primal force’. Presumably, the primal force comes in positive and negative ‘primal charge’ types. The proposed mediator of this force is called the ‘primon’ particle.

Since the masseon has not yet been detected by particle collisions, we can safely assume that the primal force is very strong. It is not necessary to understand the exact nature of the primal force to achieve and understanding of the principle of equivalence. Suffice it to say that the primal force binds together masseon particles to make all the known fermion particles of the standard model. The masseon carries the lowest possible quanta of positive gravitational ‘mass charge’.

Gravitational ‘mass charge’ is defined as the fixed rate of emission of graviton particles in close analogy to electric charge in QED (actually the probability of emission/absorption). Recall that the graviton is the vector boson of the pure gravitational force. Gravitational ‘mass charge’ is a fixed constant in EMQG, and is analogous to the fixed electrical charge concept. Gravitational ‘mass charge’ is ***not*** governed by the ordinary physical laws of *observable* mass, which includes Einstein’s energy and mass-velocity variation formulae: $E=mc^2$ and $m = m_0 (1 - v^2/c^2)^{-1/2}$ .

Gravitational ‘mass charge’ is the *low-level* mass of a particle, should not be confused with the ordinary observable inertial or observable gravitational mass properties of a particle. These are not the same. It will be assumed that when we use the term ‘low level mass charge’ in this paper, we are talking about the low level gravitational ‘mass charge’ property of a particle, and the associated graviton exchange process. Gravitational mass is the observable mass property that is measured in actual experiments.

Another important property exhibited by the graviton particle is the ***principle of superposition***. This property works exactly the same way as it does for photons. The action of the gravitons originating from several different sources acts to yield a net acceleration vector sum that determines the state of acceleration for a receiving particle. EMQG treats graviton exchanges by the same successful methods developed for the behavior of photons in QED. The dimensionless coupling constant that governs the graviton exchange process is what we call ‘$\beta$‘ in close analogy with the dimensionless coupling constant ‘$\alpha$‘ in QED, where $\beta \approx 10^{-40} \alpha$, tremendously weaker than electrical forces.

As we said, the masseon particle is the fundamental fermion that carries both electrical charge and mass charge. It is very similar to the electron in physical characteristics, where the electron is a composite of many masseons. The following tables compares the properties of the masseon particle with the electron particle:

## TABLE #2 PHYSICAL PROPERTIES OF THE ELECTRON AND MASSEON

| NAME | | FERMION | PROPERTIES | | |
|---|---|---|---|---|---|
| | Symbol | Spin | Rest Mass | Electrical/Mass Charge | |
| Electron particle | e | 1/2 | $5.11 \times 10^{-4}$ Gev/$c^2$ | -1 | +k |
| Masseon particle | m | 1/2 | unknown | + k or -k * | +1 |

*** The constant k less than 1/3 (quark electrical charge), and who's value is not known at this time.**

Masseons simultaneously carry a positive gravitational 'mass charge', and either a positive or negative electrical charge (defined exactly as in QED). Therefore, masseons also exchange photons with other masseon particles. It is important to note that the graviton exchange process is responsible for the low-level gravitational interaction only, which is not directly accessible to our measurements (as we will see later), and is also masked by the presence of the electromagnetic force component in all gravitational measurements.

Masseons are fermions with half integer spin, which behave according to the rules of quantum field theory. ***Gravitons have a spin of one*** (*not spin two*, as is commonly thought), and travel at the speed of light. This paper addresses the gravitational and electromagnetic force interactions only, and the strong and weak nuclear forces are ignored here. (Masseons also carry the strong and weak 'nuclear charge' as well, which is outside the scope of this work).

Anti-masseons carry the lowest quanta of negative gravitational 'mass charge'. Anti-masseons also carry either positive or negative electrical charge, with electrical charge being defined according to QED. An anti-masseon is always created with ordinary masseon in a particle pair, as required by quantum field theory. The negative energy, or the antimatter solution to the famous Dirac equation for fermions in quantum field theory is also the ***negative 'mass charge' solution*** for that fermion**.** Therefore we propose that the symmetry between matter and anti-matter is perfect in Dirac's equation. ***Every physical parameter is reversed in the anti-matter solution of Diarc's equation,*** where previously mass remained positive, and the other physical quantities such as, energy, electrical charge, spin, etc. were reversed when going from matter to anti-matter.

Ordinarily the standard model requires that the mass of any anti-particle is always positive, in order to comply with the Einstein principle of equivalence, or $M_i = M_g$. In EMQG, the principle of equivalence is not taken to be an absolute law of nature, and is definitely grossly violated for anti-particles (the reasons will become clear in section 8). The anti-particles can have both positive inertia mass and *negative* gravitational mass, or $M_i = -M_g$. This violates equivalence and general relativity, and so is an experimentally testable consequence of EMQG theory (section 10, and ref. 1).

There exists a beautiful symmetry between EMQG and QED for gravitational and electromagnetic forces. The masseon-graviton interaction becomes almost identical to the electron-photon interaction. There are only two differences between these forces. First, the ratio of the strength of the electromagnetic over the gravitational forces is on the order of $10^{40}$ for an electron. Secondly, there exists a difference in the nature of attraction and repulsion between positive and negative gravitational 'mass-charges' (as detailed in the table #3 and 4 below). The reason for this slight asymmetry is still an unresolved problem in EMQG theory. We would normally expect that a perfect symmetry should exist between the photon and the graviton.

In QED, the quantum vacuum, as a whole, is electrically neutral over macroscopic distance scales, because the virtual electron and positron (negative electron) particles are always created in particle pairs with equal numbers of positive and negative electrical charge. In EMQG, ***the quantum vacuum is also gravitationally 'neutral' for the same reason***. At any given instant of time, there is a 50-50 mixture of positive and negative virtual gravitational 'mass charge', which is carried by the virtual masseon and anti-masseon pairs. These masseon / anti-masseon particle pairs are created with equal and opposite gravitational 'mass charge'. This is the reason why the cosmological constant is zero (or very close to zero, ref. 1). Half the graviton exchanges between quantum vacuum particles result in attraction, while the other half result in gravitational repulsion. The result is a neutral masseon vacuum. To see how this works, we will closely examine how masseons and anti-masseons interact.

The following tables compares the fundamental attractive and repulsive characteristics of two electrical charged electrons or anti-electrons, and two masseons or anti-masseons possessing 'mass charge', all undergoing electrical or gravitational force interactions:

**TABLE #3 EMQG MASSEON - ANTI-MASSEON GRAVITON EXCHANGE**

| | **(DESTINATION)** | |
|---|---|---|
| | MASSEON | ANTI-MASSEON |
| **(SOURCE)** | | |
| MASSEON | attract | attract |
| ANTI-MASSEON | repel | repel |

**TABLE #4 QED ELECTRON - ANTI-ELECTRON PHOTON EXCHANGE**

| | **(DESTINATION)** | |
|---|---|---|
| | ELECTRON | ANTI-ELECTRON |
| **(SOURCE)** | | |
| ELECTRON | repel | attract |
| ANTI-ELECTRON | attract | repel |

In QED, if the source particle is an electron, it emits photons whose wave function induce repulsion when absorbed by the destination electron, and induces attraction when absorbed by a destination anti-electron. Similarly, if the source is an anti-electron, it emits photons whose wave function induce attraction when absorbed by the destination electron, and induces repulsion when absorbed by a destination anti-electron.

In EMQG, if the source particle is a masseon, it emits gravitons whose wave function induces attraction when absorbed by a destination masseon, and induces attraction when absorbed by a destination anti-masseon. If the source is an anti-masseon, it emits gravitons whose wave function induces repulsion when absorbed by a destination masseon, and induces repulsion when absorbed by a destination anti-masseon. This subtle difference in the nature of graviton exchange process is responsible for some major differences in the way that low-level gravitational 'mass charge' and the electrical charges operate.

It is convenient to think of the photon as occurring in photon and anti-photon varieties (the photon is its own anti-particle). Similarly, the graviton comes in graviton and anti-graviton varieties. Thus, we can say that the masseons emit gravitons, and anti-masseons emit anti-gravitons. The absorption of a graviton by either a masseon or anti-masseon induces attraction. The absorption of an anti-graviton by either a masseon or anti-masseon induces repulsion. Similarly, we can say the electrons emit photons and anti-electrons emit anti-photons. The absorption of a photon by an electron induces repulsion, and the absorption of a photon by an anti-electron induces attraction. The absorption of an anti-photon by an electron induces attraction, and the absorption of an anti-photon by an anti-electron induces repulsion. Now that we have characterized the masseons and gravitons, let us reexamine the nature of the quantum vacuum from the perspective of EMQG.

## 6.1 THE QUANTUM VACUUM AND VIRTUAL MASSEON PARTICLES

In this paper we talk extensively about virtual particles of the quantum vacuum. Precisely what kinds of virtual particles are present in the quantum vacuum? In QED, we talk about the vacuum being populated with virtual electrons and anti-electrons (and virtual muons and tauons in the second and third generation of elementary particles), along with the associated virtual photons. In the standard model of particle physics the quantum vacuum consists of all varieties of virtual fermion and virtual boson particles representing the known virtual matter and virtual force particles, respectively. These include fermions such as virtual electrons, virtual quarks, virtual neutrinos. This also includes bosons such as the virtual photons, virtual gluons, and virtual W and Z bosons.

In EMQG, we restrict ourselves to the study of gravity and electromagnetism. Therefore, the EMQG quantum vacuum consists of the virtual masseons and virtual anti-masseons, and the associated virtual photons and virtual graviton particles (virtual masseon can combine to form virtual electrons, etc.). Recall that ordinary matter consists only of real masseons bound together in certain combinations to form the familiar elementary particles.

We now ask how the virtual electrons and positrons of the QED vacuum behave in the vicinity of a real electrical charge. We want to compare this behavior with virtual masseon and virtual anti-masseon near a real and very large mass-charge like the earth.

First, we review how the QED quantum vacuum is affected by the introduction of a real negative electrical charge. According to QED, the nearby virtual particle pairs become **polarized** around the central charge. This means that the virtual electrons of the quantum vacuum are repelled away from the central negative charge, while the virtual positrons are crowded towards the central negative charge. For real electrons this process is called ***'vacuum polarization'***, and produces the well known phenomena of electric charge screening. Electric charge screening reduces the apparent electric charge of a real electron, when measured over relatively long distances away from the charge.

According to QED, each electron is surrounded by a cloud of virtual particles that winks in and out of existence in pairs (lasting tiny fractions of a second), and this cloud is always present and acts like an electrical shield against the real charge of the electron. Recently, a team of physicists led by D. Koltick of Purdue University in Indiana reported charge screening for an electron, experimentally, at the KEK collider (ref. 33). They fired high-energy particles at electrons and found that the charge screening effect of this cloud of virtual particles was reduced the closer a high-energy charged particle penetrated towards the electron. They report that the effect of the higher charge for an electron, that was penetrated by particles accelerated to an energy 58 giga-electron volts, was equivalent experimentally to a fine structure constant of 1/129.6. This agreed well with the theoretical prediction of 1/128.5, based on QED.

Now we study how the EMQG quantum vacuum is affected by the introduction of a large mass like the earth. According to EMQG, the quantum vacuum virtual masseon particle pairs are ***not*** polarized near a large mass, as we found for electrons (as can be seen from table #3 above). The virtual masseon and anti-masseon pairs are ***both*** attracted towards the mass. It is this ***lack*** of vacuum polarization that results in the main difference between electromagnetism and gravity. This is an extremely important result of EMQG theory.

The earth does ***not*** produce vacuum polarization of virtual particles (as far as mass charge is concerned). In large gravitational fields, all the virtual masseon and virtual anti-masseon particles of the vacuum have more or less the same net average statistical acceleration directed towards a large mass, and produces a net inward flux of quantum vacuum virtual masseon/anti-masseon particles (acceleration vectors only) that can, and does, affect other masses placed nearby. In contrast to this, a large electrically charged object ***does*** produce vacuum polarization in QED; where the positive and negative electric charges accelerates towards and away, respectively from the charged object. Hence in QED there is no energy contribution to other electrical test charges placed nearby (from the vacuum particles only), because the charged vacuum particles contributes equal amounts of force contributions from both directions, i.e. towards and away from the large charge.

On the earth, the ***lack*** of vacuum polarization leads us towards the weak equivalence principle. This is because both the electrically charged positive and negative virtual masseons and anti-masseon particles can act in unison against the electrically charged particles that make up a test mass dropped on the earth. Had there been vacuum polarization for the masseons as far as mass-charge was concerned, the virtual and anti-virtual vacuum particles would accelerate in the two opposite directions from the earth, and hence no net vacuum force would result against a test mass from the vacuum. We have introduced all the major players in EMQG, and we are now ready to state our basic postulates. At this time, these postulates cannot be derived from more basic principles.

## 7. VIRTUAL PARTICLES NEAR SPHERICAL MASS LIKE THE EARTH

We now have enough background material to determine the quantum nature of the gravitational field for a spherically symmetrical, large mass like the earth. In general relativity, Einstein's field equation has been solved exactly for this special case (ref. 39), and the solution is called the Schwarzchild metric (after the discoverer) and is given by:

$$ds^2 = \frac{dr^2}{1-\frac{2GM}{rc^2}} - c^2 dt^2 (1-\frac{2GM}{rc^2}) + r^2 d\Omega^2 \text{ where } d\Omega^2 = d\theta^2 + \sin^2\theta\, d\phi^2 \qquad (7.\ 1)$$

This is a complete mathematical description of the 4D space-time curvature according to general relativity, and also describes the motion of mass near the earth. The metric is given in spherical coordinates. This equation describes the path that light (or matter with additional considerations) takes through the curved 4 D space-time. We will find that this solution is a very good *approximation* to the quantum gravitational field. There are, however, hidden quantum processes involving the virtual particles in EMQG that are responsible for this curvature, and for the ***very tiny*** inaccuracy of this metric due to a slight violation of the principle of equivalence (section 8). We will also see that the metric has a limited range of applicability, and cannot be used on the individual particle scale as r → 0 (due to quantization).

A large spherical mass turns out to have a very simple motion associated with the electrically charged virtual particles that makes up the surrounding quantum vacuum. The normal vacuum condition of the virtual particles near a massive spherical distribution of matter is disrupted, compared to the similar vacuum state in far empty space. Surrounding a large spherically symmetrical mass like the earth, the virtual particles created in the quantum vacuum have a net average acceleration vector that is directed downward towards the earth's center along radius vectors. (**Note:** *We are ignoring the mutual interactions of the vacuum particles, which are why this statement must remain as a statistical statement of an average virtual particle*).

The cause of this downward acceleration of the vacuum is the ***direct*** graviton exchanges between the real masseons that constitute the earth and the virtual masseon particles of the quantum vacuum, which propagate at the speed of light. At any one instant, the vacuum particles have random velocity vectors, which point in all directions (even including the *up* direction). However, the acceleration vectors are generally pointing in the downward direction, on the whole.

The closer the virtual particles are to the earth, the greater the acceleration, as you would expect from the inverse square law. On the quantum scale is due to the geometric spreading of the graviton flux (ref. 1), which gives less graviton flux at greater distances. We conclude that the net average acceleration of the falling virtual particles is directed towards the earth's center, and the magnitude of the acceleration vector varies with the inverse square of the height 'r'.

This accelerated vacuum 'flow', which permeates all matter on the earth, plays an important role in the dynamics of gravity in EMQG. It also naturally ties in with the problem of inertia and the equivalence principle. We suggest that the average net acceleration vector of the electrically charged vacuum particles, which varies in magnitude and direction at different points near the earth, interacts with test masses and with light in such a way as to be equivalent to Schwarzschild 4D space-time metric concept above. This is because the average net acceleration vector of the charged virtual particles at each point in space surrounding the mass, guides the motion of the real photons electromagnetically, and also guides the motion of real electrically charged masseon particles that make up a mass. From simple considerations of the inverse square law of gravity, we easily obtain the general motion of the falling vacuum particles near the earth, and is given by:

***The magnitude and direction of the acceleration of an average, falling, electrically charged virtual masseon particle of the surrounding quantum vacuum at point 'r' above the earth of mass $M_e$ is given by:***

$$\vec{a} = \frac{-GM_e}{r^3}\vec{r} \qquad (7.2)$$

where $\vec{r}$ is the radius vector and $\vec{a}$ is the acceleration vector. The direction of the virtual particle acceleration is along the earth's radius vectors. It is also possible to calculate this from the basic EMQG field equations for a general, slow mass (ref. 1). However, this complex calculation of the state of motion of the vacuum particles is not required in the case of large spherically symmetrical masses like the earth, because of the symmetry and simplicity of the inverse square law for a spherically large mass like the earth.

When a small test mass moves through the space surrounding the earth, the electrical interactions between the real charged masseon particles in the mass with respect to the virtual charged particles quantum vacuum dominates over the pure graviton exchange process between the test mass and the earth. This electrical component plays ***the major*** role in the dynamics of motion of a nearby test mass. The real masseon particles that constitute the mass of the earth exchanges gravitons with the virtual masseons of the quantum vacuum, causing a downward acceleration of the quantum vacuum with a magnitude of 1g at the earth's surface.

If we now introduce a test mass near the earth, the real masseons making up the test mass will fall at the same average rate as that of the net statistical average of virtual particles of the vacuum. This is due to the vastly stronger electrical forces acting between the electrically charged virtual masseons of the surrounding vacuum and the real, electrically charged masseons of the test mass. In some sense, the vacuum acts like a much stronger constraining force that guides the motion of a masseon so that it follows a 'geodesic' path through space.

**Note:** *We have not proved that equation (7.2) is correct. Instead, it is based on the general observations of the motion of a test mass falling near the earth based on the inverse sqaure law, where the electrical forces from the falling vacuum constrains the fall rate of a point-like test mass.. However, this equation can be derived from first principles using the semi-classical EMQG equations of motion (ref. 1). Because equation 7.2 involves space and time, a small correction (for the earth) needs to be applied to compensate for space-time effects as well, since equation 7.2 involves distance and time (ref. 1).*

To fully understand the gravitational field around a spherically symmetrical massive object like the earth, the motion of light near the earth must also be accounted for in EMQG. We will find that the altered behavior of light near a massive object drastically modifies the nature of space-time in equation. This is because the acceleration a=dv/dt, and the velocity v=dx/dt involves distance and time measurements. We will return to this important issue of the meaning of curved 4D space-time in section 9, after deriving the weak principle of equivalence for the motion of matter, from the fundamental postulates of EMQG.

## 8. DERIVATION OF THE WEAK PRINCIPLE OF EQUIVALENCE

***"The principle of equivalence performed the essential office of midwife at the birth of general relativity, but, as Einstein remarked, the infant would never have got beyond its long clothes had it not been for Minkowski's concept [of space-time geometry]. I suggest that the midwife be now buried with appropriate honors and the facts of absolute space-time faced."* - Synge**

We now derive the principle of equivalence from the motion of the virtual particles in the vacuum surrounding the earth. As we stated, the equality of inertial and gravitational mass is known to be true only through physical observation and through experience. Yet until recently, it has been generally believed that it cannot be derived from more fundamental principles. However is the equivalence principle truly an exact statement of the nature of physical reality?

## 8.1 INTRODUCTION

Again, there are two main formulations of the principle of equivalence. The strong equivalence principle states that the results of any given physical experiment will be precisely identical for an accelerated observer in free space as it is for a non-accelerated observer in a perfectly uniform gravitational field. A weaker form of this postulate restricts itself to the laws of motion of masses only. In other words, the laws of motion of identical masses on the earth are identical to the same situation inside an accelerated rocket (at 1g). We have shown (section 5.1) that objects of different mass all fall at the same rate of acceleration in a uniform gravity field. In regards to the strong equivalence principle, Synge writes:

*"… I never been able to understand this Principle … Does it mean that the effects of a gravitational field are indistinguishable from the effects of an observer's acceleration? If so, it is false. In Einstein's theory, either there is a gravitational field or there's none, according as the Riemann tensor does not or does vanish. This is an absolute property. It has nothing to do with any observer's world line … The principle of equivalence performed the essential office of midwife at the birth of general relativity, but, as Einstein remarked, the infant would never have got beyond its long clothes had it not been for Minkowski's concept [of space-time geometry]. I suggest that the midwife be now buried with appropriate honors and the facts of absolute space-time faced."*

Few physicists would doubt the validity of his statement. Synge has hit on an important point in regards to the nature of the equivalence principle and space-time. He is right to say that "*either there is a gravitational field or there's none, according as the Riemann tensor does not or does vanish. This is an absolute property* (of space near a large mass)".

What he means is that the Riemann tensor describing curvature is there, or is not there, depending on whether or not there is a large mass present to distort space-time. (in other words, whether there exists a global space-time curvature or not). There is nothing relative about this fact. The existence of a ***global*** space-time curvature reveals whether you are in a gravitational field. In an accelerated frame, the space-time curvature is local to your motion only, and is not global property of 4D space-time. For accelerated motion from gravitational fields, the global 4D space-time is flat, but the local 4D space-time for the accelerated observer is curved.

There is no mystery here according to EMQG. If a large mass is present, the mass emits huge numbers of graviton particles, and distorts the surrounding virtual particles of the quantum vacuum. This is an absolute statement about the physical nature of the vacuum there. In an accelerated frame, there are very few gravitons, and the quantum vacuum is not affected. However, an observer in the accelerated frame 'sees' the quantum vacuum accelerating with respect to his frame in the same way, and hence the background vacuum state is the same. However, globally the quantum vacuum *still remains (mostly) undisturbed for accelerated motion.*

In other words, on a global scale, i.e. on a scale larger that the accelerated observer and his local frame, the state of relative acceleration of the quantum vacuum is uniform. Thus we conclude that the equivalence principle can be regarded as being a coincidence, due to local quantum vacuum acceleration state happening to appear the same for accelerated observers as it does for a specific observer on the earth.

Recently, we uncovered theoretical evidence that suggests that the strong equivalence principle does ***not*** hold in certain circumstances. First, if gravitons could be detected experimentally with a new and sensitive graviton detector (which is not likely to be possible in the near future), we would be able to distinguish between an inertial frame and a gravitational frame with this detector. This would be possible because inertial frames would have virtually no graviton particles present, whereas the gravitational fields like the earth have enormous numbers of graviton particles. We would be able to construct a device with an indicator light that reads '**GRAVITATIONAL FRAME**' or '**ACCELERATED FRAME**' by setting a certain threshold on the graviton count, and lighting the appropriate electronic indicator.

Thus, we would have performed a physics experiment that can easily detect whether you are in a gravitational field or an accelerated frame in clear violation of the strong equivalence principle. Secondly, recent theoretical considerations of the emission of electromagnetic waves from a uniformly accelerated charge, and the lack of radiation from the same charge subjected to a static gravitational field leads us to the conclusion that the strong equivalence principle also does *not* hold for radiating charged particles. Stephen Parrott (ref 23) has done an extensive analysis of the electromagnetic energy released from an accelerated charge in Minkowski space and a stationary charge in Schwarzchild space. He writes in his paper on "Radiation from a Uniformly Accelerated Charge and the Equivalence Principle":

*"It is generally accepted that any accelerated charge in Minkowski space radiates energy. It is also accepted that a stationary charge in a static gravitational field does not radiate energy. It would seem that these two facts imply that some forms of Einstein's Equivalence Principle do not apply to charged particles.*

*To put the matter in an easily visualized physical framework, imagine that the acceleration of a charged particle in Minkowski space is produced by a tiny rocket engine attached to the particle. Since the particle is radiating energy, that can be detected and used, conservation of energy suggests that the radiated energy must be furnished by the rocket. We must burn more fuel to produce a given accelerated world line than we would to produce the same world line for a neutral particle of the same mass. Now consider a stationary charge in Schwarzchild space-time, and suppose a rocket holds it stationary relative to the coordinate frame (accelerating with respect to local inertial frames). In this case, since no radiation is produced, the rocket should use the same amount of fuel as would be required to hold stationary a similar neutral particle. This gives an experimental test by which we can determine locally whether we*

*are accelerating in Minkowski space or stationary in a gravitational field - simply observe the rocket's fuel consumption."*

He does a detailed analysis of the energy in Minkowski and Schwarzchild space-time, and concludes that strong principle of equivalence does not hold for charged particles in general. This is because the readout of the fuel consumed by the two rocket could be used to drive an indicator light to reveal whether you are in a gravitational field or in an accelerated frame.

As for the weak equivalence principle, so far we can now only specify the accuracy as to which the two different mass types have been shown *experimentally* to be equal in an inertial and gravitational field. In EMQG, we show that the equivalence principle follows from lower level physical processes, and the basic postulates of EMQG. ***We will see that mass equivalence arises from the equivalence of the electrical force generated between the net statistical average acceleration vectors of the matter particles inside a mass interacting with the surrounding quantum vacuum virtual particles inside an accelerating rocket***. The *same* force occurs between the matter particles and virtual particles for a mass near the earth. We will show that weak equivalence is ***not*** perfect, and breaks down when the accuracy of the measurement approaches forty decimal places of accuracy!

Basically mass equivalence arises from the ***reversal*** of the net statistical average acceleration vectors between the charged matter particles and virtual, electrically charged, particles in the famous Einstein rocket, with the same matter particles and virtual particles process occurring near the earth (figures 2 through 5). To fully understand the hidden quantum processes in the principle of equivalence on the earth, we will detail the behavior of test masses and the propagation of light near the earth. Equivalence is shown to hold for the motion of masses in two cases: stationary test masses on the floor, and for freely falling test masses.

First we derive mass equivalence for stationary and falling masses. Next, we show that a quantum principle of equivalence holds for individual elementary particles. Next, we will demonstrate that equivalence also holds for large spherical masses with considerable self-gravity (and self-energy) such as the earth with a hot molten core, and the moon with a considerably colder core, with respect to the motion of third mass which is the sun. We will see that if both the earth and the moon fall towards the sun, they would arrive at the same time to a high degree of precision in the framework of EMQG. Finally, we examine the principle of equivalence for curved Minkowski 4D space-time curvature in a rocket, and on the earth, which is the 4D space-time equivalence.

## 8.1 MASSES INSIDE AN ACCELERATED ROCKET AT 1g

**Figure #2:** There are two different masses at rest on the floor of a rocket which is accelerated upwards at 1 g far from any gravitational sources. The floor of the rocket

experiences a force under the mass ‘2M’ that is twice as great as for the mass ‘M’. In Newtonian physics, the inertial mass is defined in precisely this way, the force ‘F’ that occurs when a mass ‘M’ is accelerated at rate ‘g’ as given by F=Mg. The quantum inertia explanation for this is that the two masses are accelerated with respect to the net average statistical motion of the virtual particles of the vacuum by the rocket. Since mass ‘2M’ has twice the masseon particle count as mass ‘M’, the sum of all the tiny electrical forces between the virtual vacuum and the masseon particles of mass ‘2M’ is twice as great as compared to mass ‘M’, i.e. for mass ‘M’, $F_1=Mg$ and for mass ‘2M’, $F_2=2Mg=2F_1$. Because the particles that make up the masses do not maintain a net zero acceleration with respect to the virtual particles, a force is always present from the rocket floor.

**Figure #3**: There are two different masses (M and 2M) that have just been released, and are in free fall inside the rocket. According to Newtonian physics, no forces are present on the two masses since the acceleration of both masses is zero (the masses are no longer attached to the rocket frame). The two masses hit the rocket floor at the same time. The quantum inertia explanation for this is trivial. The net acceleration between all the real masseons that make up both masses and virtual masseon particles of the vacuum is a net (statistical average) value of zero. The rocket floor simply reaches up to the two masses at the same time, and thus unequal masses fall at the same rate inside an accelerated rocket and arrive at the floor at the same time.

## 8.2 MASSES INSIDE A GRAVITATIONAL FIELD (THE EARTH)

**Figure #4:** There are the same two masses (2M and M) which are at rest on the surface of the earth. The surface of the earth experiences a force under mass ‘2M’ that is twice as great as for that under mass ‘M’. The reason for this is that the two stationary masses do not maintain a net acceleration of zero with respect to the net statistical average acceleration of the virtual masseons in the neighborhood. This is because the virtual particles are all (falling) accelerating towards the center of the earth ($\mathbf{a}=GM/\mathbf{r}^2$) due to the graviton exchanges between the real masseons consisting of the earth and the virtual masseons of the vacuum. Since mass ‘2M’ has twice the masseon particles as mass ‘M’, the sum of all the tiny electrical forces between the virtual masseon particles of the vacuum and the real masseon particles of mass ‘2M’ is twice as great as that for mass ‘M’. Thus, a force is required from the surface of the earth to maintain these masses at rest, with mass ‘2M’ having twice the force of mass ‘M’. The physics of this force is the same as for figure #3 in the rocket, but now the relative acceleration frames of the virtual charged masseons and the real charged masseon particles of the mass are ***interchanged*** (with the exception of the direct graviton induced forces on the masses, which is negligible and discussed later).

**Figure #5**: There are two different masses (M and 2M) that have just been released, and are in free fall towards the earth (no external forces are present on the two masses). The two masses hit the earth at the same time. The ***relative*** average acceleration of the real masseon particles that make up the two masses with respect to the electrically charged

virtual masseon particles of the vacuum is zero, because both masses and the quantum vacuum are accelerated at the same rate through graviton exchanges with the earth.

The electrical forces between the virtual particles and the matter particles of the test mass dominates the local interactions, because the electrical force is $10^{40}$ times stronger than the graviton component. Although mass '2M' has twice the ***pure*** gravitational force (i.e. forces contributed from *direct* graviton exchanges with the earth) due to twice the number of graviton exchanges, this is totally swamped out by the electrical interaction, and the accelerated virtual particles and the test masses are in a state of electrical force equilibrium, as far as acceleration vectors are concerned. Both masses fall at the same rate (neglecting the slight imbalance of the graviton component). In fact, in both cases (i.e. accelerated and gravitational) there is a net equilibrium between the accelerated state of masseons in the mass and virtual masseons of the quantum vacuum (as far as their relative acceleration is concerned, which is zero). Therefore, the two unequal masses fall to the surface of the earth at the same acceleration, and arrive at the same time.

## NEWTONIAN MASS EQUIVALENCE PRINCIPLE

Newton was aware of the mass equivalence principle, i.e. the inertial mass and the gravitational mass of an object are the same. We now revisit the equivalence principle from Newton's perspective, bearing in mind these hidden quantum processes we just described. Now the elegant picture developed above can even be seen to be hidden in Newton's formulation of accelerated motion and gravity. Equivalence between the inertial mass 'M' on a rocket moving with acceleration 'A', and gravitational mass 'M' under the influence of a gravitational field with acceleration 'A' can be seen to follow from Newton's laws if we look at them in the slightly different way:

$F_i = M\ (A)$ ... Stationary vacuum opposes acceleration A of the mass 'M' in rocket.
$F_g = M\ (GM_e/r^2)$ ... Accelerated vacuum (**$A = GM_e/r^2$**) opposes the stationary mass 'M'.

Viewed in this way, even ***Newton's law of gravity looks a lot like his famous third law of inertia***, i.e. Newton's '$F = GM_eM/r^2$' looks a lot like '$F = M\ A$'. Under gravity, the magnitude of the gravitational acceleration is '$A=GM_e/r^2$', which is the same as the magnitude of the acceleration of the rocket. Switching to the reference frame of an average electrically charged masseon in the mass 'M' in both cases, it 'sees' a typical virtual masseon particle in the vacuum near the earth ***in exactly the same way*** as a typical virtual masseon in the rocket. In other words, ***the accelerated quantum vacuum particle state appears the same from the perspective of the mass in an accelerated frame and the same mass on the earth.*** This is the basis of Newtonian mass equivalence.

This example shows why both the masses of figure 2 and 3 are equivalent to the masses in figure 4 and 5. The force magnitude is the same because the calculation of the force involves the same sum of all the tiny electrical forces between the virtual charged masseon particles and the real masseon particles of the mass. The only difference in the physics of the masses is that the relative motions of all the tiny electrical force vectors are interchanged. The other difference is that large numbers of graviton particles (that originate from the earth's mass) are absorbed by the masses in figures 4 and 5 for the earth case, while for the rocket there are negligible gravitons present.

To summarize, two unequal masses in free fall hit the surface of the earth at the same time. The net statistical average acceleration of the real masseon particles that make up the masses and virtual charged masseon particles of the vacuum is zero, because this process is dominated by the electrical force (where the direct graviton exchanges are negligible). The electrical forces between the virtual particles and the matter particles of the test mass dominates the interactions, because the electrical force is $10^{40}$ times stronger than the graviton component.

Although mass '2M' has twice the ***pure*** gravitational force (i.e. forces contributed from *direct* graviton exchanges only with the earth) due to twice the number of graviton exchanges, this is totally swamped out by the electrical interaction, and the accelerated virtual particles and the test masses are in a state of electrical equilibrium as far as acceleration vectors are concerned. Both masses fall at the same rate. In fact, in both cases (i.e. accelerated and gravitational) there is a net equilibrium between the accelerated state of masseons in the mass and virtual masseons of the quantum vacuum as far as their relative acceleration is concerned.

***IT IS IMPORTANT TO NOTE THAT MASS EQUIVALENCE IS NOT PERFECT IN THE ROCKET AND IN THE EARTH-BOUND EXPERIMENTS ABOVE!***

First, the processes in figures 2 and 3 are physically different then in figure 4 and 5. The most important difference is that gravitons are present on the earth, but not in the rocket (strictly speaking there are negligible numbers of gravitons from the rocket structure). Secondly, the force contributed by the direct graviton exchanges between the mass and the earth slightly imbalance the equivalence of the fall rate for the two unequal masses. This discrepancy in the free fall rate of test masses near the earth is extremely minute in magnitude because there is a ratio of about $10^{40}$ in the field strength existing between the electromagnetic and gravitational forces. In principle it could be measured by extremely sensitive experiments, if two test masses are chosen with a very large mass difference, which would amplify this effect (ref. 1).

Does the weak equivalence principle hold for an individual elementary particle?

## 8.3 MICROSCOPIC EQUIVALENCE PRINCIPLE OF PARTICLES

We have been talking about macroscopic masses until now. What about the case for elementary particles? For example, does a proton and an electron simultaneously dropped on the surface of the earth fall at the same rate? In other words, if both particles are dropped inside a rocket that is accelerating at 1 g and also on the earth, would they arrive on the floor at the same time? The answer is yes. In fact, the equivalence principle has actually been experimentally verified for the case of the neutron (ref. 40).

An astute observer may have questioned our reasoning all along, by asking why ***all*** the virtual particles (for example, virtual neutrons, virtual electrons, virtual quarks, etc., which all consist of different mass values) are falling at the same rate. Certainly inside an accelerated rocket an observer who is stationed on the floor will view *all* the virtual particles of the quantum accelerating with respect to him at the same rate, no matter what the masses of the virtual particles are. This is simply because the floor of the rocket accelerates upwards at 1g to meet them, giving the *illusion* (to an observer on the rocket) that virtual particles of different mass are falling at the same rate.

Since the masses of the different types of virtual particles are ***all different*** according to the standard model of particle physics, why are they all falling at the same rate on the earth? Here the cause of the downward acceleration is due to graviton exchanges with the earth. Shouldn't the virtual particles with with greater mass charge fall at a faster rate? Since we are trying to derive the equivalence principle from fundamental concepts, we cannot resort to this principle itself to state that the virtual particles must be accelerating at the same rate on the earth. ***In order to solve this puzzle, we have to postulate the existence of a fundamental mass charge unit for all fermion particles, which is carried by the masseon particle.***

As we have indicated, all particles with mass (virtual or not) must be composed of combinations of the fundamental "masseon" particle, which carries just one fixed quanta of mass charge (postulate #2). Since all virtual masseon particles exchange the same fixed flux of gravitons with the earth, ***all*** the virtual masseons are all falling at the same rate. However, we must bear in mind that virtual masseons can bind momentarily combine together to form all the familiar virtual particles, such as virtual electrons, virtual positrons, virtual quarks, etc. or even currently unknown species of virtual particles.

Recall that masseons carry both gravitational 'mass charge' and ordinary electrical charge. However, the electrical interactions (photon exchanges) will work to equalize the fall rate of virtual masseons that combine to make virtual particles such as virtual electrons and virtual quarks. If a virtual quark consists of say 100 bound masseons (the actual number is not known), the graviton exchanges would normally be cumulative. We would expect to have 100 times more acceleration imparted to the virtual quark, compared to a single virtual masseon. However, virtual masseons dominate the quantum vacuum since they are the fundamental mass particle, and do not have to bound with other masseons to exist. Therefore the lone, unbound virtual masseon is by far the most common virtual mass particle in the quantum vacuum (this is illustrated in figures 10).

No matter how many virtual masseons combine to give other virtual particles such as quarks, the local electrical interaction between the far more numerous lone virtual masseons and the virtual quark will tend to equalize the fall rate in exactly the same as we discussed above for macroscopic masses. ***This process works like a microscopic version*** ***of the EMQG weak principle of equivalence*** for the wide variety of falling virtual particles. In other words, the same action occurring on the particle level for large falling masses is happening on the microscopic scale for elementary particles.

Figure 10 shows the microscopic equivalence principle at work for free falling virtual particles in a gravitational field. To summarize, the electrical forces from the vastly numerous, free virtual masseons of the quantum vacuum (which all fall at the same rate due to equal mass charge) dominates over the action of the more familiar virtual particles that consist of combinations of virtual masseons. The virtual quark would normally fall faster than the virtual electron and the virtual electron faster than an individual virtual masseon *if there were no quantum vacuum*.

## 8.4 THE INERTION: AN ELEMENTARY QUANTA OF INERTIA

Recall that a real neutron particle with no net electrical charge has inertial mass because it is composed of real electrically charged masseon particles, which interact electromagnetically with the charged virtual masseon particles of the quantum vacuum (recall that the number of positive and negative electrically charged masseons inside a neutron are equal). Each real masseon inside the neutron contributes a fundamental unit or 'quanta' of inertia to the neutron, because of the electrical force interaction with the immediate surrounding virtual masseons of the vacuum. Therefore, the sum of the real masseon force contribution to the inertial mass of the neutron defines the neutron's total inertial mass.

We propose that the electrical force that exists between *one* real masseon particle accelerating at a rate of 1g with respect to the surrounding quantum vacuum be given the status of a new universal constant. We call this new constant the '**inertion**' constant, or "i". The inertion has measurement units of force. Thus the inertion represents the lowest possible quanta of inertia force. In order to determine the numerical value of the inertion in the lab, one must determine the number of masseons inside a neutron for example, and divide this into the observable (inertial) mass of the neutron. These numbers are currently unknown at this time.

How does the background quantum vacuum look like from the frame of reference of the neutron particle sitting on the surface of the earth? The answer is that it is the ***same*** as it appears for a neutron on a rocket (1g). As we have found, the virtual masseon and anti-masseon particles of the quantum vacuum are accelerated by the direct graviton exchanges with the earth. Therefore, we can conclude that the 'inertion' also represents the *lowest possible quanta* of gravitational force on the earth.

## 8.5 EQUIVALENCE PRINCIPLE FOR THE SUN-EARTH-MOON SYSTEM

So far we only considered the weak equivalence principle for medium sized masses and elementary particle. Does the weak equivalence principle hold for the following imaginary scenario: the earth and the moon are simultaneously in free fall towards the sun with an acceleration of gravity equal to '$G_{sun}$? In other words, would the earth and moon arrive at the same time on the surface of the sun? Would they also arrive on the floor at the same time when free falling inside a huge rocket undergoing acceleration $G_{sun}$ in space (far from any other large masses)? This question is important because large astronomical bodies have their own internal gravity, and are able to significantly disrupt the virtual particles around them. In other words, the earth's gravity further modifies the acceleration vectors of the virtual particles in it's neighborhood, and adds to contribution from the sun's gravity.

This question also turns out to be at the heart of the so-called metric theories of gravity. In metric theories such as general relativity, all objects with any kind of internal composition follow the natural curvature of space-time. This includes objects with considerable internal energy sources and self-gravity. Any deviation from perfect equivalence would constitute what is generally called the Nordtvedt effect (ref. 8), after the discoverer in 1968. Because the earth's core is molten and very hot, the earth contains a significant internal energy. The earth is also a fairly large source of gravitational energy as well, which significantly contributes to the acceleration to the nearby virtual particles of the quantum vacuum in the earth's vicinity. Contrast this with the moon, which is relatively cool and less energetic, with considerably less gravitational energy.

To see if the weak equivalence works, we start with the fantastic situation where the earth and moon are 'dropped' simultaneously from a height 'h' inside a huge rocket (with negligible mass) accelerating with the same $G_{sun}$ as exists on the surface of the sun (figure 12). The result of this experiment should be obvious. They both arrive on the floor of the rocket at the same time. Actually, it is the floor of the rocket that accelerates upwards and greets both bodies at the same time!

However, one should be aware that the virtual particles of the quantum vacuum are disturbed near these large bodies (by direct graviton exchanges between these bodies and the nearby vacuum), particularly the acceleration vectors of the virtual particles which are in close proximity with the earth and the moon. The vacuum acceleration vectors now tend to point towards the centers of the two bodies, near these regions. This, however, does not affect the results of this experiment. The results are no different than if two small masses are dropped inside the rocket; again because it is the floor that accelerates up to meet them at the same time. However, one must note that the shape of the nearby virtual particle distribution around the bodies in this case (figure 12).

Now we turn to another fantastic situation near the surface of the sun, where the earth and moon are 'dropped' on the sun from the same height 'h' (figure 11). This situation is far more complex than the same experiment inside the rocket. Now all three bodies disturb the virtual particles of the quantum vacuum! The sun sets up a strong $GM_{sun}/r^2$ acceleration field of virtual particles, which applies over very long distances, and points towards the center of the sun. The earth and moon also produce their own vacuum particle acceleration fields $GM_{earth}/r^2$ and $GM_{moon}/r^2$ , respectively in their vicinity, although much weaker over large distances. The sun's acceleration dominates over the surrounding space, except near the moon and near the earth; where some of the virtual particles actually are moving away from the sun (as happens near the surface of the earth on the night side).

How can equivalence with the rocket possibly hold in this scenario? Recall that the electrical action of the virtual particles of the vacuum on the real particles inside the bodies that determines the motion of the earth and the moon undergoing gravitational acceleration. However, in different regions of the earth, the virtual particles are accelerating in different directions! Part of the answer to this problem comes from an important property exhibited by the masseon-graviton particles, which is the ***principle of superposition***. This is a property that is also shared by the electron-photon particle. The action of the gravitons originating from all three sources on a given virtual particle of the quantum vacuum yields a net acceleration that is the net vector sum of the action of all the gravitons received by that virtual particle.

To explain equivalence, we must first recall that equivalence only holds in a sufficiently small region of space (technically, at a given point above the sun) when compared to the equivalent accelerated reference frame. This is because the acceleration of the sun varies with the distance 'r' from the sun's center, whereas inside an accelerated rocket it does not vary with height. Secondly, we must recall that the motion of the virtual particles in the rocket is also disturbed near the vicinity of the earth and the moon. In fact, inside the rocket the virtual particles are also directed along the radius vectors of both the earth and the moon in their vicinity (figure 12). Yet we believe that the sun and the moon still reach the floor of the rocket at the same time. Therefore, the quantum vacuum can be disturbed in the case of the free fall of the earth and the moon in a rocket, and yet can still cause both bodies to fall at the same rate in this case.

A close examination of figures 11 and 12 reveals that the quantum vacuum pattern *is the same* for the two different experiments, when viewed from the ***correct*** reference frame. In figure 11, the observer is stationed on the surface of the sun, so that we can see the reason why both bodies are attracted to the sun. Recall that the electrical interaction between the acceleration vectors of the falling virtual masseons and the real masseons in the earth and moon is the primary reason for the attraction. The direct graviton action is negligible in comparison. Now in order to compare the two experiments of figure 11 and 12, the reference frame for the sun experiment should be equivalent to the rocket.

In the rocket experiment, the frame chosen for our observer is outside the rocket (the quantum vacuum has a relative acceleration of zero) in order to understand the results. Therefore for the sun, the observer's frame should be in free fall, thus restoring the relative acceleration of the quantum vacuum to zero just as for the observer outside the rocket. When this is done, we have to correct the acceleration vectors of the virtual particles near the earth and moon in figure 12. It is easy to show that the result of this operation gives an identical result as figure 12. ***Therefore, equivalence holds in both experiments, because the vacuum acceleration patterns are the same!***

Next we address the important, and more complex issue of light motion equivalence in accelerated and gravitational frames.

## 8.6 LIGHT MOTION IN A ROCKET: SPACE-TIME EFFECTS

According to general relativity, the strong equivalence principle demands that ***all physical phenomena are equivalent*** in a rocket (1g) and on the earth, for a sufficiently small volume. ***This includes the motion of light***. Furthermore, there ought to be equivalent curvature of 4D space-time.

We now review the Einstein famous thought experiment in regards to the motion of light in accelerated and gravitational frames. First we carefully examine three scenarios for the motion of light in a rocket (figures 6 and 7), which is accelerated upwards at 1 g (far from any gravitational sources) and the same situation on the earth (figure 8 and 9). The 1 g acceleration is purposely chosen to match the earth's acceleration. First we study light moving from the floor of the rocket to the ceiling where it is detected by an observer there. Next we look at light moving from the ceiling of the rocket to the floor where it is detected by an observer there. Finally, we examine more closely to what happens to light moving parallel with the floor of the rocket (figure 14) and the same situation on the earth (figure 13), where it follows a curved path.

### (A) LIGHT MOVING FROM THE FLOOR TO THE CEILING OF THE ROCKET

**FIGURE 6 -** Here the light is positioned on the floor of the rocket which is being accelerating upwards at 1 g, and propagates in a straight line up to the observer on the ceiling. Meanwhile, the rocket has accelerated upwards while the light is in flight. What happens to the light? Experimentally, we observer a red shift in the color of the light. Furthermore, one can also conclude that there is a Doppler shift towards the red end of the spectrum. This conclusion is based on to the motion of the observer on the ceiling, who obtains an additional velocity over the light source during the time of flight of the light, when velocities are compared to a third observer outside the rocket. What happens to the velocity of light? To answer this, we must consider the nature of space and time, since velocity is defined as distance/time..

According to general relativity, all observers measure the light velocity as being the standard vacuum value, where each use their measuring instruments that are identically constructed, and calibrated inside each of their reference frame. In other words, the speed of light does not vary under all these circumstances. Closer examination reveals that the clocks in the ceiling differ from the clocks stationed on the floor. In particular, the clock on the floor of the rocket runs slower than the one on the ceiling. Distances measurements are also affected. General relativity explains all these observations with the 4D space-time curvature existing inside accelerated frames. We will return to this example with our EMQG interpretation of these measurements.

(B) LIGHT MOVING FROM THE CEILING TO THE FLOOR OF THE ROCKET

**FIGURE 6 -** Here the light is positioned on the ceiling of the rocket which is accelerating upwards at 1 g, and propagates in a straight line down to the observer on the floor. Meanwhile, the rocket has accelerated upwards while the light is in flight. What happens to the light? Experimentally, we observer a Doppler blue shift in the color of the light. According to general relativity, all observers measure the light velocity as being the standard vacuum value, where each use their measuring instruments that are identically constructed, and calibrated inside each reference frame. Closer examination reveals that the clocks in the ceiling differ from the clocks stationed on the floor. Distances measurements are also affected. General relativity explains all these observations with the 4D space-time curvature existing inside accelerated frames. We will return to this example with our EMQG interpretation of these measurements.

(C) LIGHT MOVING PARALLEL TO THE FLOOR OF THE ROCKET

**FIGURE 7 -** Here the light leaves the light source on the left wall of the rocket which is accelerating upwards at 1 g, and propagates in a straight line towards the observer on the right wall. Meanwhile, the rocket has accelerated upwards while the light is in flight. Therefore an observer in the rocket observes a curved light path. An observer outside the rocket sees a straight light path. With the introduction of the third observer outside the rocket, the reason for curvature becomes readily apparent. Outside the rocket the observer sees the light path go straight across the rocket. Meanwhile, inside the rocket the light leaves the source and is totally de-coupled from the source and moves straight (as seen outside). However, the observer on the floor is drawn upwards by the accelerated motion of the floor. This causes the inside observer to see a curved path.

According to general relativity, the space-time inside the rocket is curved (in the direction of motion), and light moves along the natural geodesics of curved 4D space-time. Meanwhile, the observer outside the rocket lives in flat 4D space-time, and therefore observes light moving in a perfect straight line, which is the geodesic path in flat 4D space-time. We will return to this example with our EMQG interpretation of these measurements. Next we will examine the same scenarios for light motion on the surface of the earth..

## 8.7 LIGHT MOTION NEAR EARTH'S SURFACE - SPACE-TIME EFFECTS

Now we carefully examine the same three scenarios for the motion of light, but now we are on the surface of the earth. We will ignore the variation of acceleration with height found on the earth, by selecting a room that is sufficiently low height and narrowness as to ignore the slight change in the direction of acceleration caused by acceleration vectors being directed along earth's radius vectors. First, light moves from the floor of the room on the surface of the earth to the ceiling, where it is detected. Next, light is moving from the ceiling of the room to the surface of the earth where it is detected by an observer. Finally, light is moving parallel with the earth's surface from the left side of the room to the right, and follows a curved path.

### (A) LIGHT MOVING FROM THE FLOOR TO THE CEILING ON EARTH

**FIGURE 8 -** Light is positioned on the floor of a room on the surface of the earth, and propagates in a straight line up to the observer on the ceiling. What happens to the light? According to the principle of equivalence, the light behaves the same way as in the rocket, i.e. it is red-shifted. However, one cannot claim a Doppler red shift, since there is no relative motion of the source and the detector according to general relativity. Is there any other physical justification for the red shift, other the principle of equivalence? Since light is moving through a gravitational potential, it ought to lose energy as it gains height. According to $E=h\nu$, if the energy 'E' goes down, then the frequency $\nu$ decreases, and we have a red shift. Therefore, when we look at the physics behind accelerated and gravitational frames for this situation (without resorting to the principle of equivalence), we find two different explanations; i.e. Doppler shift and gravitational energy loss. In EMQG, we find that the virtual particle dynamics alone gives an adequate explanation for this.

What happens to the velocity of light? To answer this, we must again consider the nature of space and time, since velocity is defined as distance/time. According to general relativity, all observers measure the light velocity as being the standard vacuum value, where each use their measuring instruments that are identically constructed, and calibrated inside each reference frame. In other words, the speed of light does not vary under all circumstances. Closer examination reveals that the clocks in the ceiling differ from the clocks stationed on the floor. In particular, the clock on the floor on the earth runs slower than the one that is on the ceiling. Distances measurements are also affected. General relativity explains all these observations with the 4D space-time curvature existing inside accelerated frames. We will return to this example with our EMQG interpretation of these measurements.

(B) LIGHT MOVING FROM THE CEILING TO THE FLOOR ON EARTH

**FIGURE 8 -** Here the light is positioned on the ceiling of the room on the surface of the earth, and propagates straight down to the observer on the floor. What happens to the light? Again to answer this, we must again consider the nature of space and time, since velocity is defined as distance/time. According to general relativity, all observers measure the light velocity as being the standard vacuum value, where each one uses their measuring instruments that are identically constructed, and calibrated inside each reference frame. In other words, the speed of light does not vary under all circumstances. Closer examination reveals that the clocks in the ceiling differ from the clocks stationed on the floor. In particular, the clock on the floor on the earth runs slower than one that is on the ceiling. Distances measurements are also affected. General relativity explains all these observations with the 4D space-time curvature existing inside accelerated frames.

According to general relativity, an observer outside the room in free fall observes the light moving downward at the speed of light in a straight path. Meanwhile, according to general relativity, an observer inside the room stationed on the floor also observes the light moving downwards in a straight line. He also observes that the light is blue-shifted. He makes a measurement of the light velocity with his measuring instruments (which were calibrated within his reference frame) and observes that the velocity of light is the same on the floor as he found when he measured received light speed in his internal reference frame with the same instruments. In other words, the speed of light does not vary in all cases. Closer examination reveals that clocks measured in his reference frame differ from the clocks on the ceiling. In particular, the clock on the ceiling of the room runs faster than the one on the floor. Distances are also affected. In general relativity, all these conclusions follow directly from 4D space-time curvature.

(C) LIGHT MOVING PARALLEL TO THE SURFACE OF THE EARTH

**FIGURE 9 -** Here the light leaves the light source on the left wall of the room on the earth and propagates in a curved path towards the observer on the right wall. Meanwhile, an observer in free fall towards the earth's surface sees light moving in a straight path. According to general relativity, and light moves along the natural geodesics of curved 4D space-time in the room. Meanwhile, an observer in free fall lives in flat 4D space-time, and hence an observer sees straight-line paths for light. In general relativity, all these conclusions are identical as for the observer accelerated in the rocket at 1g in accordance with the principle of equivalence. Now we look at 4D space-time curvature from the perspective of EMQG.

## 8.8 GENERAL CONCLUSIONS FOR THE EQUIVALENCE OF LIGHT MOTION

The three elementary observables discussed above are:

(1) The red-shift for light moving straight up in a rocket, or on the earth.
(2) The blue-shift for light moving straight down in a rocket, or on the earth.
(3) The curved light path when light moves parallel to the earth, or in a rocket.

One can easily conclude that there is a Doppler shift for light moving up or down in the rocket, because the detector obtains an ***additional velocity*** with respect to the source during the travel time of the photons (as seen by the observer outside the rocket).

It is also clear that one ***cannot*** conclude that a Doppler shift is responsible for the shift in color of the light as it climbs or descends in the case of light motion in a *gravitational field*. This conclusion is based on the fact that there is *no* relative motion between the light source and the detector during the light travel time. The Doppler shift requires motion.

However, in general relativity the actual motion of the light is identical on the earth and inside the accelerated rocket. It does not say anything about the low level, quantum processes responsible for the light, and why they give that same results! In other words, general relativity prescribes a common cause for the light motion inside the accelerated rocket on the earth; that is the curvature of space and time in both reference frames. ***We maintain that the actual low level physical causes cannot be the same, however because the physics at the quantum level are different for accelerated and gravitational frames!*** This is a central idea in EMQG, where there actually is no strong equivalence principle. We acknowledge that the physical details of the low level processes are different, and we provide an explanation at the quantum scale in this section.

We can view the wave crests of light moving upwards against a gravitational field as a true representation of the emitted history of the light wave. This is because the light crests are conserved during propagation. No wave crests are created on route. Also, no wave crests are lost during the motion of light. For each wave crest that enters the space between the light source and the detector, there must another wave crest leaving that same space. Einstein himself gave a simple and elegant explanation of this in his 1911 paper.

The question to ask is why does an atom that emits a photon upwards in a gravitational field produce a photon of different frequency then the same atom in far empty space? Furthermore why does the atom in an accelerated rocket (1g) that emits light upwards, produce the same frequency as the same atom in the earth's gravitational field? Einstein's answer to these questions is to resort to the strong principle of equivalence and the metric (although in 1911 he used the considerations of gravitational potential energy and arrived at an answer that was off by a factor of two). The strong equivalence principle states that time is slowed in the same way in the rocket and on the earth, thus slowing all atomic processes (without recourse to lower level quantum phenomena).

EMQG provides a simple answer to these questions. ***The accelerated state of the electrically charged quantum vacuum appears the same from the perspective of both the atom in a rocket and an atom on the earth that emits photons.*** That is, the vacuum background conditions for the atom in an accelerated rocket (1g) that emits photons upwards is identical to the same vacuum background for the same atom that is on the earth's surface (neglecting the graviton background for the case of the earth)! Furthermore, ***the electrically charged virtual particles affect the motion of photons***. We will elaborate on this after reviewing photon scattering theory, and the effects of accelerated electrical charges on the propagation of light.

We would like to point out an important problem with the standard general relativistic interpretation of the ***blue shift*** of light (when light travels straight down on the surface of the earth) that has recently been pointed out to us by V. Petkov (ref. 47, 1999). In this scenario, light propagates downwards from the ceiling to the floor of the gravitational field and light becomes blue-shifted rather than red-shifted (figure 6). Why is this so?

V. Petkov makes a strong case for what he calls the anisotropic light velocity in non-inertial frames. He argues that the ***proper velocity*** of light ***changes*** in a gravitational field. Light moves faster when propagates downwards, and light moves slower when it propagates upwards compared to the vacuum value. The proper light velocity is defined by the proper length and proper time measurements. It turns out that Petkov's approach is very similar to the approach we adopted in EQMG, which we now examine fully in the next section (with the exception of a difference in opinion on the nature of the 4D space-time).

## 9. THE EQUIVALENCE OF 4D SPACE-TIME CURVATURE

Here we will contrast the two very different approaches to the problem of 4D space-time curvature, and the equivalence of 4D space-time in accelerated and gravitational frames:

**General Relativity:** *Light velocity is constant, and we have 4D space-time curvature.*
**EMQG&Scattering**: *Photons scatter with the electrically charged falling vacuum.*

The central theme in general relativity is the constancy of light velocity. The curvature of 4D space-time is the root cause of all the observable aspects of light motion in the rocket and on the earth. A deflected light path in a gravitational field would seem to violate the constancy of light velocity, but it is the curving of 4D space-time that is the culprit.

The central theme in EMQG is that photons scatter with the electrically charged, and falling (accelerated) virtual particles of the quantum vacuum. Photon scattering accounts for the illusion of 4D space-time curvature! Furthermore, matter also reacts to the falling vacuum, and together a 4D space-time curvature phenomena emerges from the low level activities of quantum particles on a kind of absolute space and time background (the CA).

We have already presented a unified picture of the equivalence of real matter motion in an accelerated rocket and on the earth, with the model of the falling electrically charged quantum vacuum. We now wish to extend this model to cover the motion of light under the action of the falling vacuum in an accelerated frame and on the surface of the earth, which we claim encompasses the relativistic standard model of 4D space and time.

First we will derive the gravitational time dilation equation in the conventional way, using the general relativistic solution to Einstein's gravitational field equations for a large spherical, non-rotating mass. This solution is called the Schwarzschild metric. Next, we fully develop the EQMG theory of light scattering and 4D space-time curvature. From this, we calculate the quantity of space-time curvature using EMQG theory, and show that the results are the same as relativity for spherical masses. In EMQG the light velocity is anisotropic as Petkov has suggested, ***but the changes in light velocity (compared to the vacuum value) occur not in proper units of length and time (Petkov, ref. 47), but in the CA absolute units of cells and clock cycles.*** For this work, we will not go into the details of the absolute CA space and time, and reference 1 gives a complete account.

## 9.1 GENERAL RELATIVISTIC 4D SPACE-TIME CURVATURE

It is well known that general relativity accounts for the motion of light near a large spherical mass, using the concept of 4D space-time curvature. General Relativity postulates space-time curvature in order to preserve the constancy of the light velocity (a notion first introduced in special relativity) in an accelerated frame or in a gravitational field. The 4D space-time metric must satisfy Einstein's Gravitational field equations:

$$R_{ik} - \frac{1}{2} g_{ik} R = \frac{8pG}{c^2} T_{ik} \quad \text{… Einstein's Gravitational Field Equations} \qquad (9.11)$$

The solution of this equation for the case of spherical mass distribution was given by the Schwarzchild in spherical coordinates (ref. 39):

$$ds^2 = \frac{dr^2}{1 - \frac{2GM}{rc^2}} - c^2 dt^2 (1 - \frac{2GM}{rc^2}) + r^2 d\Omega^2 \text{ where } d\Omega^2 = d\theta^2 + \sin^2\theta \, d\phi^2 \qquad (9.12)$$

This is a complete mathematical description of the 4D space-time curvature near the large spherical mass in spherical coordinates in differential form called the 4D space-time metric. It is of the form of the element of distance 'ds' between two closely spaced points expressed in terms of the chosen coordinates, where $ds^2 = g_{ik}(x)\, dx^i\, dx^k$, and the time coordinate is $g_{00}$. From this, it is easy to show (ref. 39) that the comparison of time measurements between a clock outside a gravitational field (called proper time t( $\infty$ ) to a clock at distance r from the center of a spherical mass distribution (called the coordinate time t( r ) is given by:

$$t(r) = \frac{t(\infty)}{\sqrt{1 - \frac{2GM}{rc^2}}}$$ … which follows from Schwarzchild metric directly. (9.13)

Using the relationship $(1 - x)^{-1/2} \approx 1 - x/2$ when $x << 1$, and realizing the quantity $2GM/rc^2$ is very small, we can write this as:

$$t(r) \approx (1 - \frac{GM}{rc^2})t(\infty) \tag{9.14}$$

This gives the amount of time dilation between a clock on the earth "t( r )" compared to a clock positioned at infinity "t( ∞ )". From this, we see that clocks on the earth run slower then at infinity. The deeper a clock is in a gravitational potential the slower it runs, as compared to it's identically constructed cousin at infinity. So does the frequency of light that is emitted from an atom on the earth.

Similarly, from this metric, we find that the distance Δr at point s( r ) compared to the same distance Δr at infinity follows as:

$$s(\infty) = \frac{s(r)}{\sqrt{1 - \frac{2GM}{rc^2}}} \tag{9.15}$$

or we can also write this as:

$$s(r) \approx \frac{s(\infty)}{1 - \frac{GM}{rc^2}} \tag{9.16}$$

This gives the amount of space distortion for a ruler of length Δr on the earth "s( r )" compared to the identically constructed ruler of length Δr positioned at infinity "s( ∞ )".

Observationally, equation 9.13 has been verified experimentally by atomic clocks mounted on board an airplane (ref. 6 and 7). In these experiments, an atomic clock had spent many hours at a high altitude inside and was returned to the ground and compared to an atomic clock in the laboratory. After correcting for various background effects, the laboratory clock lagged behind the airplanes atomic clock by the predicted amount.

## 9.2 EMQG, LIGHT MOTION, AND 4D SPACE-TIME CURVATURE

The general principles of EMQG are applied here in order to understand 4D space-time curvature and the principle of equivalence of all light motion in an accelerated rocket (1g) on the surface of the earth. To do this we must examine in detail the effects of the background electrically charged, and falling, virtual particles of the quantum vacuum on the propagation of light. The big question to consider is:

***Is the general downward acceleration of the electrically virtual particles of the quantum vacuum near a large mass responsible for the motion of photons in a gravitational field, or is this motion truly the result of an actual 4D space-time geometric curvature that exists near the earth (and which holds to the absolute tiniest distance scales)?***

The answer to this very important question hinges on whether our universe is truly a curved, geometric Minkowski 4D space-time on the smallest of distance scales, or whether curved 4D space-time results merely from the activities of quantum particles such as bosons and fermions (living in an absolute CA space and time) interacting with other virtual quantum particles in the background quantum vacuum. EMQG takes the second approach, in order to be compatible with quantum inertia and CA theory. The second approach also leads to an in- depth understanding of the principle of equivalence on the quantum scale.

According to EMQG theory, light takes on the same general acceleration component as the net statistical average value of the falling, electrically charged, virtual particles of the quantum vacuum, through an electrical 'Fizeau-like' scattering process. By this we mean that the photons are frequently absorbed and re-emitted by the electrically charged virtual particles, which are (on the average) accelerating towards the center of the large mass. When a virtual particle absorbs the real photon, a new photon is re-emitted after a small time delay in the same ***average*** downward direction that the original photon takes. This process is called photon scattering. Photon scattering is well understood for real matter, and the effects of scattering on light velocity are also well known in field of optics. We will see that photon scattering in an accelerated quantum vacuum (where the vacuum can be considered as a fluid medium) is also central to the understanding of 4D space-time curvature.

One of the most important (and completely undisputed) results of classical optics is that light moves slower in water than it does in air. Furthermore it is recognized that the velocity of light in air is slower than that of light's vacuum velocity. This effect is described by the index of refraction 'n', which is defined as the ratio of light velocities in the two different media. What is also an interesting result from classical physics is that the velocity of light in an *ordinary transparent and* ***moving medium*** (such as moving water) is known to *differ* from its value in the same stationary medium (such as stationary water). Fizeau (1851) is credited for demonstrated this experimentally in the laboratory with light propagating through pipes containing currents of water, flowing with a constant velocity in two opposite directions.

In 1915 Lorentz identified the physics of this phenomena as being due to his microscopic semi-classical electromagnetic theory of photon propagation. Einstein was also aware of these results and attributed this to the special relativistic velocity addition rule for the addition of the light and water velocities, without consideration of the low level quantum processes involved.

In EMQG, we propose that in gravitational fields (and for accelerated motion) the moving water of Fizeau's experiment is now replaced by the accelerated and electrically charged virtual particle flow of the quantum vacuum. Like in the Fizeau experiment, photons scatter with the accelerated virtual particles of the quantum vacuum and vary their motion.

Imagine what would happen if you place a clock inside a stream of moving water in the Fizeau experiment. Would the clock keep time properly, when compared to an identically constructed clock placed in stationary water. Of course not! The very idea of this seems almost ridiculous. Of course one could argue that the nature of the hydrodynamic forces in moving and stationary water would alter the dynamics of how the clock operates. We suggest that you should also not be surprised to find that the accelerated flow of virtual particles affects both clocks and rulers placed in a gravitational field, as compared to the identical measuring instruments in far space where the vacuum is undisturbed.

Another important result from EMQG and scattering theory is that the velocity of light in a vacuum, ***without*** the existence of ***any*** virtual particles of the quantum vacuum, should be ***much greater*** than the observed average light velocity in the vacuum. The electrically charged virtual fermion particles of the quantum vacuum frequently scatter photons, which introduce many tiny delays for the photon propagation. This causes a great reduction in the total average light velocity in the vacuum that is populated by countless numbers of virtual, electrically charged, particles. In other words; ***the low level light velocity (between virtual particle scattering events) is much greater than the measured average light velocity after vacuum scattering in the normal vacuum***. A similar effect is known to occur when light propagates through glass, where photons scatter with electrons in the glass molecules, which subsequently reduces the average light velocity through glass compared to the normal vacuum light velocity. This result is frequently overlooked by physicists.

In our study of special relativity, we have seen the importance of the propagation of light in understanding the nature of space and time measurements. Recall that according to standard text books (for example, Serway ref. 19) the definition of an inertial frame in space is a vast rigid 3D grid of identically constructed clocks placed at regular intervals with a ruler (figure 20). These clocks are viewed by an observer with ordinary light, i.e. one can simply 'look' at a clock and determine the time. Since light has a finite speed, we are forced to conclude in this scheme of things that two clocks cannot be simultaneous! If light is somehow affected by gravitational fields, this is bound to distort this concept of space and time, using the above definition of an inertial frame. Therefore, we will closely examine the behavior of light near the earth and in an accelerated rocket.

We must remind the readers that the constancy of light velocity in gravitational frames still ***remains a <u>postulate</u> of general relativity, in spite of 90 years of close scrutiny***. General relativity incorporates all the results of special relativity, including the second postulate of light constancy for all inertial observers. In other words, it is impossible experimentally to distinguish between whether the light velocity changes when light moves upwards, or whether 4D space-time changes with height. What is lacking is a standard to gauge these alternatives. In other words, what measurement standard is assumed in discussions regarding the rates of clocks, the lengths of rulers, and the speed of light. The measurements of local observers cannot be used to answer this question, because we are considering changes to the actual measurement standards that are used by different observers!

First we must carefully understand what is meant by light velocity. Velocity is *defined* as distance divided by time, or c=d/t. Light has *very few* observable characteristics in this regard: we can measure velocity c (the ratio of d/t); frequency $\nu$; wavelength $\lambda$; and we can also measure velocity by the relationship $c=\nu\lambda$. It is important to note that all these observables are closely interrelated. We know that $\nu = 1/t$ (t is the period of one light cycle) and $\lambda=d$ (the length of one light cycle). Thus, c=d/t and $c=\nu\lambda$ are equivalent expressions of space and time. If we transmit green light to an observer on the ceiling of a room on the earth, and he claims that the light is red shifted, it is impossible for him to tell if the red shift was caused by the light velocity changing, or by space and time distortions which causes the timing and length of each of the light cycles to change.

This is very important, and we illustrate this with an example. Imagine that the frequency of light is halved, or $\nu_f = (1/2)\, \nu_i$ and the wavelength doubles $\lambda_f = 2\lambda_i$ , and that you were not aware of both of these changes. Then you would conclude that the velocity of light remains unchanged ($c=\nu\lambda$). However, if the velocity of light is halved, and you were not aware of it, then you could conclude that the frequency is halved, $\nu_f = (1/2)\, \nu_i$ and the wavelength doubles $\lambda_f = 2\lambda_i$.

To illustrate this point, we will now examine what happens if an observer on the floor feeds a ladder (which represents the wave character of light) with equally spaced rungs to an observer on the ceiling, where each observer cannot see what the other observer does with the ladder. A ladder is used as an analogy for the wave characteristic of light, where the distance between rungs represent the wavelength of light (figure 15).

Imagine a perfect ladder with equally space rungs of known length being passed up to you at a known velocity (figure 15), such that it is impossible to tell the motion of the ladder other than by observing the rungs moving past you. If the rung spacing are made larger, you would conclude that either the ladder is slowing down, or that the spacing of the ladder rungs was increased. But it would be impossible to tell which is which. Let us assume that you make a measurement on the moving rungs, and observe a spacing of 1 meter between any two rungs. Then you observe that two rungs move past you every second. You therefore conclude the velocity of the ladder is 2 m/sec.

Now suppose that the ladder is fed to you at half speed or at 1 m/sec, and that you are not aware of this change in velocity. You could conclude that the velocity halved from your measurements, because you now observe that one rung appears in view for every second that elapses instead of two rungs, and that the velocity was thus reduced to 1 m/sec. However, you could just as well concluded that your space and time was altered, and that the velocity of the ladder is constant or unaffected. Since you observe only one rung in view per second instead of the usual two rungs, you could claim that the rung spacing on the ladder is enlarged (red-shifted) or doubled to 2 m by distortions of your measuring equipment, and that the velocity still remained unaltered. From this, you might conclude that the frequency is halved, and that time measurements that might be based on this ladder are now dilated by a factor of two.

Which of these two approaches to the ladder problem is truly correct? It is impossible to say by measurement, unless you know before hand what trait of the ladder was truly altered. For photons, the same problem exists. No known measurement of photons in an accelerated rocket or on the surface of the earth can reveal whether space and time is affected, or whether the velocity of light has changed. This is why the constancy of light remains a postulate in general relativity.

In EMQG theory, the variable light velocity approach is chosen for several important reasons. First, the ***equivalence of light motion*** *in accelerated and gravitational frames now becomes* ***fully understood as a dynamic process*** having to do with motion rather then a fundamental principle of physics. For gravity, it is the hidden virtual particle motion, and for accelerated frames it is the motion of the frame itself. Furthermore, this model retains the same basic behavior as for ordinary matter in motion in accelerated frames and gravitational frames. The equivalence principle now becomes understood on a quantum scale for both light and matter, using the same background vacuum concept.

Secondly, the *physical basis of the curvature* of Minkowski 4D space-time near a large mass now becomes clear. It arises from the interaction of light and matter with the background accelerated virtual particle processes. This process can be visualized as an accelerated quantum vacuum fluid flow in both frames types. However, in gravitation the vacuum fluid flow is caused by direct graviton exchanges acting between the earth and the electrically charged, virtual fermions of the quantum vacuum. In accelerated frames, the vacuum fluid flow is apparent and caused by the actual acceleration of the observer's frame and his light sources (and the measuring instruments he possesses).

Finally, and most importantly, ***the mysterious physical action that occurs between the earth and the surrounding 4D space-time curvature now becomes very clearly understood***. The earth acts on the virtual particles of the quantum vacuum through graviton exchanges, causing them to accelerate towards the earth. The accelerated virtual particles in turn act on light and matter to produce the effects of curved 4D space-time.

Since photon scattering is essential to our understanding of the details of 4D space-time curvature, we will examine scattering in some detail in the following sections. Readers that are familiar with photon scattering theory may wish to skip ahead to section 9.9. First we review the conventional physics of light scattering in a ***real media*** such as water or glass, introducing the concept of the index of refraction and Snell's Law of refraction. Next we introduce photon scattering for a real media moving at a constant velocity, where the velocity of light varies in the moving media (known as the Fizeau effect). Next we introduce an accelerated motion for the real medium and examine how the photons scatter there. This is an important step towards understanding EMQG theory. Finally we generalize these arguments to examine photon scattering with the electrically charged virtual particles of the quantum vacuum. In the next sub-section we show the relationship between photon scattering and 4D space-time curvature.

## 9.3 CLASSICAL SCATTERING OF PHOTONS IN REAL MATTER

It is a well-known result of classical optics that light moves slower in glass than in air. This effect is described by the index of refraction 'n', which is the ratio of light velocities in the two different media. The Feynman Lectures on Physics gives one of the best accounts of the classical theory for the origin of the refractive index and the slowing of light through a transparent material like glass (ref. 42, chap. 31).

When light passes from a vacuum into glass, with an incident angle of $\theta_0$ it deflects and changes it's direction and moves at a new angle $\theta_1$ , where the angles follow Snell's law:

$$n = \sin\theta_0 / \sin\theta_1 \qquad (9.31)$$

This follows geometrically because the wave crests on both sides of the surface of the glass must have the same spacing, since they must travel together (ref. 42). The shortest distance between crests of the wave is the wavelength divided by the frequency. On the vacuum side of the glass surface it is $\lambda_0 = 2\pi c/\omega$, and on the other side it is given by $\lambda = 2\pi v/\omega$ or $2\pi c / \omega n$ since we define v=c/n. If we accept this, then Snell's law follows geometrically (ref. 42). In some sense, the existence of the index of refraction in Snell's law is confirmation of the change in light speed going from the vacuum to glass.

Snell's law does not tell us why we have a change in light velocity, nor does it give us any insight into the phenomena of dispersion and back scattering of light in refraction. A good classical account of the derivation of the index of refraction is given by Feynman himself in ref. 42. Feynman derives the index of refraction for a transparent medium by accepting that the total electric field in any physical circumstance can be represented by the sum of the fields from all charge sources, and by accepting that the field from a single charge is given by it's acceleration evaluated with a retardation speed 'c' (the propagation speed of the exchanged photons). We only summarize the important points of his argument below, and the full details are available in reference 42:

(1) The incoming source electromagnetic wave (light) consists of an oscillating electric and magnetic field. The glass consists of electrons bound elastically to the atoms, such that if a force is applied to an electron the displacement from its normal position will be proportional to the force.

(2) The oscillating electric field of the light causes the electron to be driven in an oscillating motion, thus acting like a new radiator generating a new electromagnetic wave. This new wave is always delayed, or retarded in phase. These delays result from the time delay required for the bound electron to oscillate to full amplitude. Recall that the electron carries mass and therefore inertia. Therefore some time is required to move the electron.
(3) The total resulting electromagnetic wave is the sum of the source electromagnetic wave plus the new phase-delayed electromagnetic wave, where the total resulting wave is phase-shifted.
(4) The resulting phase delay of the electromagnetic wave is the root cause of the reduced velocity of light observed in the medium.

Feynman goes on to derive the classic formula for the index of refraction for atoms with several different resonant frequency $\omega_k$ which is given by:

$$n = 1 + \left( \frac{q_e^2}{2\epsilon_0 m} \right) \sum_k \frac{N_k}{\omega_k^2 - \omega^2 + i \gamma_k \omega} \qquad (9.32)$$

where n is the index of refraction, $q_e$ is the electron charge, m is the electron mass, $\omega$ is the incoming light frequency, $\gamma_k$ is the damping factor, and $N_k$ is the number of atoms per unit volume. This formula describes the index of refraction for many substances, and also describes the dispersion of light through the medium. Dispersion is the phenomenon where the index of refraction of a media varies with the frequency of the incoming light, and is the reason that a glass prism bends light more in the blue end than the red end of the spectrum.

If the medium consists of free, unbound electrons in the form of a gas such as in a plasma (or as the conduction electrons in a simple metal) then the index of refraction with the conditions $\gamma_k << \omega$ and $\omega_k = 0$ is given by (ref. 42):

$$n \approx 1 - [N_k q_e^2 / (2\epsilon_0 m)] / \omega^2 \approx \{1 - [N_k q_e^2 / (\epsilon_0 m)] / \omega^2 \}^{1/2} \qquad (9.33)$$

where we recall that $(1-x)^{1/2} \approx 1 - x/2$ if x is much less than 1.

The quantity $\Omega = [ N_k q_e^2 / (\epsilon_0 m) ]^{1/2}$ is sometimes called the Plasma frequency $\Omega$, where there is a transition to the transparent state at $\Omega = \omega$.

## 9.4 QUANTUM FIELD THEORY OF PHOTON SCATTERING IN MATTER

Although the classical account of scattering predicts the experimentally confirmed results, the correct account must be a quantum mechanical account.

To quote R. Feynman: " … **yes, but the world is quantum not classical *dam-it***".

The propagation of light through a transparent medium is a very difficult subject in QED. It is impossible to compute the interaction of a collection of atoms with light exactly. In fact, it is impossible to treat even one atom's interaction with light exactly in QED. However the interaction of a real atom with photons can be approximated by a simpler quantum system. Since in many cases only two atomic energy levels play a significant role in the interaction of the electromagnetic field with atoms, the atom can be represented by a quantum system with only two energy eigenstates.

In the text book "Optical Coherence and Quantum Optics" a thorough treatment of the absorption and emission of photons in two-level atoms is given (ref. 43, Chap. 15, pg. 762). When a photon is absorbed, and later a new photon of the same frequency is re-emitted by an electron bound to an atom, there exists a time delay before the photon re-emission. The probabilities for emission and absorption of a photon is given as a function of time $\Delta t$ for an atom frequency of $\omega_0$ and photon frequency of $\omega_1$ :

Probability of Photon Absorption is: $K\ [\sin(0.5(\omega_1 - \omega_0)\Delta t)\ /\ (0.5(\omega_1 - \omega_0))]^2$
Probability of Photon Emission is: $M\ [\sin(0.5(\omega_1 - \omega_0)\Delta t)\ /\ (0.5(\omega_1 - \omega_0))]^2$
(9.41)

(where K and M are complex expressions defined in ref. 43)

The important point we want to make from eq. 16.41 is that the probability of absorption or emission depends on the length of time $\Delta t$, where the probability of the emission is zero, if the time $\Delta t = 0$. In other words according to QED, a ***finite time*** is required before re-emission of the photon. There are other factors that affect the probability, of course. For example, the closer the frequency of the photon matches the atomic frequency, the higher the probability of re-emission in some given time period. We maintain that these delays are the actual route cause of the index of refraction in a medium.

We believe that a similar thing happens when photons propagate through the quantum vacuum. Therefore, we want to address the effect of the virtual particles of the quantum vacuum on the propagation velocity of real (non-virtual) photons, a subject that is largely ignored in the physics literature.

## 9.5 THE SCATTERING OF PHOTONS IN THE QUANTUM VACUUM

In section 9.3 we discussed photon scattering in a real matter medium and in a real negatively charged electron gas. The electron gas model is the closest model we have towards understanding photon scattering of the quantum vacuum. However, there are several important differences between the charged electron gas medium and the electrically charged virtual fermion particles of the quantum vacuum as a medium.

First, and most importantly, virtual particles do not carry any net average energy. Instead an individual virtual particle ‘borrows’ a small amount of energy during it’s brief existence, which is then paid back quickly in accordance to the uncertainty principle. It is because quantum mechanics forbids knowing the value of two complementary variables precisely (in this case energy and existence time) for a virtual particle that virtual particles are allowed to exist at all. Therefore unlike the electron gas, the vacuum is incapable of permanently absorbing light that propagates through it.

Thus the quantum vacuum does not absorb any light over macroscopic distance scales. This statement seems trivial, but it is never-the-less important when considering the quantum vacuum as a medium. On microscopic scales real photons are absorbed and re-emitted by individual virtual particles, in accordance with QED. Photon energy is lost in some collisions and regained in others so that on the average the energy loss is zero. This is because during the brief existence time of a virtual fermion particle, the virtual particle ***does*** possess energy, which is paid back almost immediately. This quantum process happens an enormous number of times as light travels through macroscopic distance scales. The energy balances out to zero over sufficiently large distance scales.

Furthermore unlike the electron gas, there can be no dispersion of light in the quantum vacuum. In other words all frequencies of electromagnetic radiation move at the same speed through the quantum vacuum in spite of the incredible numbers of virtual particle interactions that occur for any particular frequency of photon. Zero dispersion follows experimentally from many astronomical observations of distant supernova, where there is a dramatic change in light and electromagnetic radiation with time. Observations have been made of specific events in the light curves of supernovae light curves that range from the radio band frequencies to the X-ray / Gamma Ray frequencies. All the different frequencies are observed to arrive on the earth at the same time.

With distances of thousands or millions of light years away, any discrepancy in the photon velocity of supernovae at different frequencies would be very apparent. For example with the relatively nearby supernova 1987A (which exploded about 160,000 years ago in the Large Magellanic cloud) all the different frequencies of EM waves has reached us very much at the same time. If there had been a dispersion of only 0.01 m/sec in light velocity (i.e. 3 parts in $10^{-11}$) between two different frequencies, then the light of one frequency would arrive on the earth:

$160000 \times 365.25 \times 24 \times 60 \times 60 \times 10^{-2} / (3 \times 10^{-8})$ =170 seconds or 2.8 minutes later!

A result like this obviously disagrees with observations made of the spectrum of supernova 1987A. Spectra have been obtained for very distant supernovae up to a few billion light years away in other galaxies. One study places the maximum allowed dispersion to be on the order of 1 part in $10^{-21}$. Thus we conclude that there is no dispersion of light in the vacuum.

Is there a possibility for an index of refraction in the vacuum, as we have in an electron gas? Remember that an index of refraction requires two different media in which to compare the relative velocities of light. However the vacuum particle density must nearly uniform, with no transitions in density. Let us imagine a situation where somehow we have removed all the virtual particles in half of an empty box in vacuum, and the other half has the normal population of virtual particles in the normal quantum vacuum state. Would there be an index of refraction as light traveled from one side of the box to the other?

This is a very important question because the validity of special relativity at the sub-microscopic distance scales comes into question here. You might think that if the vacuum has no energy, there should no effect on the propagation speed of photons. However we believe that the virtual particles in the quantum vacuum ***do*** indeed delay the progress of photons through electrically charged vacuum particle scattering effects. Thus we believe that photon scattering reduces the light velocity on the half of the box with electrically charged virtual particles. How can we justify this belief, in spite of the contradiction to special relativity? Special relativity is a classical theory, and was developed in the macroscopic domain of physics. It is almost impossible to measure light velocities over the extremely short distance scales that we are talking about.

The electrically charged virtual particles in the quantum vacuum all have random velocities and move in random directions. They also have random energies $\Delta E$ during their brief life time $\Delta t$, which satisfies the uncertainty principle: $\Delta E \ \Delta t > h/(2\pi)$. Imagine a real photon propagating in a straight path through the electrically charged virtual particles in a given direction. The real photon will encounter an equal number of virtual particles moving towards it as it does moving away from it. The end result is that the electrically charged quantum vacuum particles do not contribute anything different than the situation where ***all*** the virtual particles in the it's path were at relative rest. Thus we can consider the vacuum as some sort of stationary crystal medium of virtual particles with a very high density, where each virtual particle is short-lived and constantly replaced (and carry no net average energy as discussed above).

The progress of the real photon is delayed as it travels through this quantum vacuum 'crystal', where it meets uncountable numbers of electrically charged virtual particles. Light travels through this with no absorption or dispersion. Based on our general arguments above, we ***postulate*** that the photon is delayed as it travels through the quantum vacuum. We can definitely say that the uncertainty principle places a lower limit on the emission and absorption time delay, and forbids the time delay from being *exactly* equal to zero.

***Therefore we conclude that the electrically charged virtual particles of the quantum vacuum frequently absorb and re-emit the real photons moving through the vacuum by introducing small delays during absorption and subsequent re-emission of the photon, thus reducing the <u>average</u> propagation speed of the photons in the vacuum (compared to the light speed of photons between absorption/re-emission events).***

Our examination of the physics literature has not revealed any previous work on a quantum time delay analysis of photon propagation through the quantum vacuum, presumably because of the precedent set by Einstein's postulate of light speed constancy in the vacuum under all circumstances. We will take the position that the delays due to photon scattering through the quantum vacuum are real. These delays reduce the much faster and absolutely fixed 'low-level light velocity $c_l$' (defined as the photon velocity between vacuum particle scattering events) to the average observed light velocity 'c' in the vacuum (300,000 km/sec) that we observe in our actual experiments.

Furthermore, we propose that the quantum vacuum introduces a sort of Vacuum Index of Refraction '$n_{vac}$' (compared to a vacuum without all virtual particles) such that $c = c_l / n_{vac}$. If this is true, what is the low-level light velocity? It is unknown at this time, but it must be significantly larger than 300,000 km/sec. In fact we believe that the vacuum index of refraction '$n_{vac}$' ***must be very large*** because of the ***high density*** of virtual particles in the vacuum. This concept is required in EMQG theory, and has become central to understanding the equivalence principle and 4D space-time curvature in accelerated frames and in gravitational fields.

## 9.6 THE FIZEAU EFFECT: LIGHT VELOCITY IN A MOVING MEDIA

It also has been known for over a century that the velocity of light in a moving medium differs from its value in the same stationary medium. Fizeau demonstrated this experimentally in 1851 (ref. 41). For example, with a current of water (with refractive index of the medium of n=4/3) flowing with a velocity V of about 5 m/sec, the relative variation in the light velocity is $10^{-8}$ (which he measured by use of interferometry). Fresnel first derived the formula (ref. 41) in 1810 with his ether dragging theory. The resulting formula relates the longitudinal light velocity '$v_c$' moving in the same direction as a transparent medium of an index of refraction 'n' defined such that 'c/n' is the light velocity in the stationary medium, which is moving with velocity 'V' (with respect to the laboratory frame), where c is the velocity of light in the vacuum:

Fresnel Formula: $$v_c = c/n + (1 - 1/n^2) V \qquad (9.61)$$

Why does the velocity of light vary in a moving (and non-moving) transparent medium? According to the principles of special relativity, the velocity of light is a constant in the vacuum with respect to all inertial observers. When Einstein proposed this postulate, he was not aware that the vacuum is not empty. However he was aware of Fresnel's formula and derived it by the special relativistic velocity addition formula for parallel velocities (to first order). According to special relativity, the velocity of light relative to the proper frame of the transparent medium depends only on the medium. The velocity of light in the stationary medium is defined as 'c/n'. Recall that velocities u and v add according to the formula: $(u + v) / (1 + uv/c^2)$
Therefore:

$v_c = [ c/n + V ] / [ 1 + (c/n) (V)/c^2 ] = (c/n + V) / ( 1 + V/(nc) ) \approx c/n + (1 - 1/n^2) V$ (9.62)

The special relativistic approach to deriving the Fresnel formula does not say much about the actual quantum processes going on at the atomic level. At this scale, there are several explanations for the detailed scattering process in conventional physics. We investigate these different approaches in more detail below.

## 9.7 LORENTZ SEMI-CLASSICAL PHOTON SCATTERING

The microscopic theory of the light propagation in matter was developed as a consequence of Lorentz's non-relativistic, semi-classical electromagnetic theory. We will review and summarize this approach to photon scattering, which will not only prove useful for our analysis of the Fizeau effect, but has become the basis of the 'Fizeau-like' scattering of photons in the accelerated quantum vacuum near large gravitational fields.

To understand what happens in photon scattering inside a moving medium, imagine a simplified one-dimensional quantum model of the propagation of light in a refractive medium. The medium consisting of an idealized moving crystal of velocity 'V', which is composed of evenly spaced, point-like atoms of spacing 'l'. When a photon traveling between atoms at a speed 'c' (vacuum light speed) encounters an atom, that atom absorbs it and another photon of the same wavelength is emitted after a time lag '$\tau$'. In the classical wave interpretation, the scattered photon is out of phase with the incident photon. We can thus consider the propagation of the photon through the crystal is a composite signal. As the photon propagates, part of the time it exists in the atom (technically, existing as an electron bound elastically to some atom), and part of the time as a photon propagating with the undisturbed low-level light velocity 'c'. When the photon changes existence to being a bound electron, the velocity is 'V'. From this, it can be shown (ref. 41, an exercise in algebra and geometry) that the velocity of the composite signal '$v_c$' (ignoring atom recoil, which is shown to be negligible) is:

$v_c = c [1 + (V\tau/l) (1 - V/c)] / [1 + (c\tau/l) (1 - V/c)]$ (9.71)

If we set V=0, then $v_c = c / (1 + c\tau/l) = c/n$. Therefore, $\tau/l = (n - 1)/c$. Inserting this in the above equations give:

$v_c = [(c/n) + (1 - 1/n) V (1 - V/c)] /[1 - (1 - 1/n)(V/c)] \approx c/n + (1 - 1/n^2) V$
(to first order in V/c). (9.72)

Again, this is Fresnel's formula. Thus the simplified non-relativistic atomic model of the propagation of light through matter explains the Fresnel formula to the first order in V/c through the simple introduction of a scattering delay between photon absorption and subsequent re-emission. This analysis is based on a semi-classical approach. What does quantum theory say about this scattering process? The best theory we have to answer this question is QED.

## 9.8 PHOTON SCATTERING IN THE ACCELERATED VACUUM

Anyone who believes in the existence of virtual fermion particles in the quantum vacuum that carry mass, will acknowledge the existence of a coordinated general downward acceleration of these virtual particles near any large gravitational field. In EMQG gravitons from the real fermions on the earth exchange gravitons with the virtual fermions of the vacuum (which carry electric charge), causing a downward acceleration. The virtual particles of the quantum vacuum (now accelerated by a large mass) acts on light (and matter) in a similar manner as a stream of moving water acts on light (and matter) in the Fizeau effect. How does this work mathematically?

Again, it is impossible to compute the interaction of an accelerated collection of virtual particles of the quantum vacuum with light exactly. However, a simplified model can yield useful results. We will proceed using the semi-classical model proposed by Lorentz, above. We have defined the raw light velocity '$c_r$' (EMQG, ref. 1) as the photon velocity in between virtual particle scattering. Recall that raw light velocity is the shifting of the photon information pattern by one cell at every clock cycle on the CA, so that in fundamental units it is an absolute constant. Again, we assume that the photon delay between absorption and subsequent re-emission by a virtual particle is '$\tau$', and the average distance between virtual particle scattering is 'l'. The scattered light velocity $v_c(t)$ is now a function of time, because we assume that it is constantly varying as it moves downwards towards the surface in the same direction of the virtual particles. The virtual particles move according to: a =gt, where $g = GM/R^2$.

Therefore we can write the velocity of light after scattering with the accelerated quantum vacuum:

$$v_c(t) = c_r \; [1 + (gt\tau/l)(1 - gt/c_r)] \,/\, [1 + (c_r\tau/l)(1 - gt/c_r)] \qquad (9.81)$$

If we set the acceleration to zero, or gt = 0, then $v_c(t) = c_r \,/\, (1 + c_r\tau/l) = c_r/n$. Therefore, $\tau/l = (n - 1)/c_r$. Inserting this in the above equation gives:

$$v_c(t) = [(c_r/n) + (1 - 1/n)\, gt\, (1 - gt/c_r)] \,/\, [1 - (1 - 1/n)(gt/c_r)] \approx c_r/n + (1 - 1/n^2)\, gt$$
…. to first order in $gt/c_r$. (9.82)

Since the average distance between virtual charged particles is very small, the photons (which are always created at velocity $c_r$) spend most of the time existing as some virtual charged particle undergoing downward acceleration. Because the electrically charged virtual particles of the quantum vacuum are falling in their brief existence, the photon *effectively* takes on the ***same downward acceleration*** *as the virtual vacuum particles* (as an average acceleration over macroscopic distances). In other words, because the *index of* *refraction of the quantum vacuum 'n' is so large* (compared to no vacuum particles), and because $c = c_r/n$ and we can write in equation 9.61:

$$v_c(t) = c_r/n + (1 - 1/n^2)\, gt = c + gt = c\,(1 + gt/c) \quad \text{if } n >> 1. \qquad (9.83)$$

Therefore for photons going with the flow (downwards): $\mathbf{v_c(t) = c\,(1 + gt/c)}$ (9.84)
Similarly, for photons going against the flow (upwards): $\mathbf{v_c(t) = c\,(1 - gt/c)}$ (9.85)
**as the refractive index of the quantum vacuum n** ® ¥**.**

Remember that this is a semi-classical derivation, and does not constitute an actual proof of the scattering effects on photons. An important limitation of these results is the issue of space-time, which we cover in the next section. This issue limits the accuracy of these expressions to small local regions of distance and time.

These formulas are used by EMQG as a starting point for the expression for the variation of light velocity near a large gravitational field at a point, and we show that this leads to the correct amount of general relativistic 4D space-time curvature, taking into account some additional assumptions. To see the connection between variable light velocity and space-time effects consider the following thought experiment. Imagine that two clocks that are identically constructed, and each calibrated with a highly stable monochromatic light source in the same reference frame. These clocks keep time by using a high-speed electronic divider circuit that divides the light output frequency by "n" such that an electrical voltage pulse is produced every second. For example, the light frequency used as the clock is precisely calibrated to $10^{15}$ Hz; this light frequency is converted in to an electronic pulse train of the same frequency, where it is divided by $10^{15}$ to give one electronic pulse every second. Another counter in this clock increments every time a pulse is sent, thus displaying the total time elapsed in seconds on the clock display.

Now, let us place these two clocks in a gravitational field on earth with one of them on the surface, and the other at a height "h" above the surface. Imagine that the clocks are compared every second to see if they are still running in unison in the gravitational field by exchanging light signals. We would find that as time progresses, the clocks loose synchronism, where the lower clock appears to run slower than the higher. According to general relativity, light always maintains a constant speed, and 4D space-time curvature is responsible for the difference in the timing of the two clocks, where the lower clock runs slower. We argue that the accelerated Fizeau-like quantum vacuum fluid affects the light velocity of the exchanged light signals, and it also affects the atomic photon emission process. In the next section we derive the same time dilation effect predicted by general relativity using eq. 9.85, which assumes that the light velocity has <u>exactly</u> the same downward acceleration component of the falling electrically charged virtual particles of the quantum vacuum (and a very high vacuum index of refraction).

## 9.9 SPACE-TIME CURVATURE OBTAINED FROM SCATTERING THEORY

In this section we are in a position to mathematically formulate EMQG 4D space-time curvature for the specific example of spherically symmetric mass. We derive the expected amount of space-time curvature effects near a spherical, non-rotating, massive object using our results from scattering theory in the previous section. We compare our results to those obtained from the Schwarzchild metric in section 9.1, which is the standard textbook method of general relativity.

We now take a bold step (that we justify later, with certain qualifiers to do with space and time that is inherent in these formulas) and assume that for the case on the surface of the earth, equation 9.85 holds. Therefore:

$c_h = c(1 - gt/c) = c(1 - gh/c^2)$, where t=h/c.

This describes the propagation of photons moving upwards for a distance h, but we assume it holds only for ***very short distances, where h® 0***. Technically this is true only at a point, which means that this equation must be written in differential form. ***The reason that we cannot say that it holds for large distances is because light velocity happens to also depend on the very nature of space and time,*** because the expression '$c(1 - gh/c^2)$' involves both space and time, implicitly.

Einstein himself derived a similar expression in 1911 for the velocity of light in a gravitational field (ref. 55), before developing his general theory of relativity (ref. 56):

$$c_0 = c\left(1 + \frac{\Delta\Phi_{so}}{c^2}\right) \quad \text{… Einstein's Formula} \qquad (9.90)$$

where $\Delta\Phi_{SO}$ is the difference of the gravitational potential of the source and the observation points S and O. This, however, leads to incorrect experimental values. Later this was corrected by himself in 1915. However, Einstein was not aware of the existence of the quantum vacuum at that time, and attributed this result to the degradation of the photon energy as it climbs out of a gravitational potential.

In what follows, we ignore the special relativistic postulate of the constancy of light velocity for now, and return to this issue later. We will take the position that photons continuously vary their velocity (remember that on the CA, the velocity of light is still an absolute constant in between vacuum particle scattering events) by scattering with the falling, electrically charged virtual particles.

In the scattering picture of light propagation, we found that the velocity of light takes on a ***different*** value when traveling straight upwards, compared to straight downwards, an effect that was independently proposed by V. Petkov (ref. 47, 1999). According to Petkov:

*"Up to now little attention has been paid to an expression for the velocity of light in a gravitational field derived by Einstein in 1911. … it is shown that the proper velocity of light is anisotropic in non-inertial frames …"*

Petkov believes that this can actually be measured experimentally (ref. 47). However, we disagree! Light plays a central role in defining the nature of space and time.

If light velocity varies on the earth, then why is it that we do not actually observe this variation in light velocity in real experiments? Part of the answer, of course, lies in the incredibly small variation of light velocity on the earth (on the order of one part in $10^{-15}$). More importantly, it involves the deep connection between the nature of space-time and the propagation of light itself, first discovered by Einstein himself in his work on special relativity (1905)! Einstein showed that many concepts related to space and time are traceable to the behavior of light.

For example, the concept of simultaneous events is deeply related to the motion of light (ref. 19, 20, 21). The derivation of the Lorentz transformations and the resulting time dilation and Lorentz contraction effects are solely based on the behavior of light. These considerations, and the results of section 9.2 lead us to believe that variations of light velocity will result in space-time variations on actual clocks and rulers.

EMQG provides a plausible mechanism for light velocity variation: the scattering of photons with the falling vacuum near a large gravitational mass. This concept is based on the concept that the vacuum acts like Fizeau-like quantum vacuum fluid. We will assume that light velocity varies, (but only for short distances) as it moves upward from the surface of the earth. As the photon moves upward from point r to point r+Δr it decelerates at -1g according to equation 16.54, and obtains a new velocity a short time Δt later:

$$c(r+\Delta r) = c(r)\left(1 - \frac{g\Delta t}{c}\right) \qquad (9.91)$$

Since $\Delta t = \Delta r / c$ for small distances (in the limit as $\Delta r \to 0$), we can then write:

$$c(r+\Delta r) = c(r)\left(1 - \frac{g\Delta r}{c^2}\right) \qquad (9.92)$$

Since, $g = GM/r^2$ at point r above the center of the earth, we can write this as:

$$c(r+\Delta r) = c(r)\left(1 - \frac{GM\Delta r}{r^2 c^2}\right) \qquad (9.93)$$

Since, the only observable property of light that we can be *sure* about is the red shift, as discussed in section 9.2, and with $c = \nu\,\lambda$, it follows that the change in frequency is:

$$n(r+\Delta r) = n(r)\left(1 - \frac{GM\Delta r}{r^2c^2}\right) \tag{9.94}$$

from which the corresponding wavelength appears longer by the same factor, or

$$l(r+\Delta r) = \ l(r)\left(1 + \frac{GM\Delta r}{r^2c^2}\right) \tag{9.95}$$

To find the total change in frequency from point r on the earth's surface to infinity (no gravity), we integrate for the change in frequency GM $\Delta$r / $r^2$ $c^2$, with respect to r as follows:

$$\int_r^\infty \frac{GM}{r^2c^2}dr = 0 + \left[\frac{GM}{rc^2}\right] \tag{9.96}$$

$$n(\infty) = n(r)\left(1 - \frac{GM}{rc^2}\right) \tag{9.97}$$

But, since $\nu$ = 1/t by definition, therefore time must be affected as follows:

$$\frac{1}{t(\infty)} = \frac{1}{t(r)}\left(1 - \frac{GM}{rc^2}\right) \tag{9.98}$$

Finally, we have:

$$t(r) = t(\infty)\left(1 - \frac{GM}{rc^2}\right) \tag{9.99}$$

***which is the same expression for time dilation in a gravitational field we obtained from the Schwarzchild metric from equation 9.14, i.e.*** $t(r) \approx (1 - \frac{GM}{rc^2})t(\infty)$**.**

Similarly, wavelength received at infinity is increased by the following expression:

$$l(\infty) = \ l(r)\left(1 - \frac{GM}{rc^2}\right) \tag{9.991}$$

Now, an observer at infinity can use the light signal from the surface of the earth to make measurements of distance in his reference frame at infinity. For example, suppose that in his own reference frame, a reference laser light source is used to measure a given reference length, and say that this corresponds to 1,000,000 wavelengths or $10^6$ $\lambda_r$, where $\lambda_r$ is the reference wavelength. Subsequently, he uses the light received from the surface of the earth from an identically constructed reference laser light source ($\lambda_r$) to measure the same length, and finds that when he counts the standard 1,000,000 wavelengths the reference length has shortened (because of the wavelength increase). In general he concludes that the distances at infinity s( ∞ ) are contracted by the amount:

$$s(\infty) = s(r)\left(1 - \frac{GM}{rc^2}\right) \qquad (9.992)$$

compared to distances s( r ) on the surface of the earth. Finally, we can write:

$$s(r) = \frac{s(\infty)}{1 - \frac{GM}{rc^2}} \qquad (9.993)$$

***which is again, exactly the same expression for length that we obtained from the Schwarzchild metric in eq. 16.15.***

This equation specifies the amount of distortion for rulers on the earth "s( r )" compared to rulers positioned at infinity "s( ∞ )". Thus by postulating that it is the light velocity that is actually varying (and not space-time curvature), we are led to the same amount of red shift, and the same amount of space-time curvature.

## DISCUSSION

So why insist that it is light velocity rather than space-time that is responsible for these effects on earth, especially when it becomes impossible experimentally to tell the difference between the two results? With the space-time approach we have to assume, without any prior reason, that a large mass curves space-time. We have to accept that it just does this, without understanding the action principle! With the variable light velocity approach the physical action that exists between matter and space-time becomes well understood! Matter acts on the vacuum particles, and the vacuum particles act on matter and light, to give a curved space-time.

The other advantage with the EMQG approach is that the equivalence of space-time in accelerated and gravitational fields also becomes clear. Therefore, the principle of equivalence is seen to hold from a ***common*** cause, the action of the quantum vacuum which appears to be the same to a light particle or a mass in accelerated frames and in gravitational frames. In short, we have closure as to why the principle of equivalence holds for space-time!

We can now see that in order to formulate a theory of gravity involving observers with measuring instruments (such as clocks and rulers) we must take into account how these measurements are affected by the local conditions of the quantum vacuum. Our analysis above shows that quantum vacuum can be viewed as a Fizeau-like fluid undergoing downward acceleration near a massive object, which affects the velocity of light. Indeed, not only is the velocity of light affected, it is *all* the particle exchange processes including graviton exchanges. Therefore, we find that the accelerated Fizeau-like 'quantum vacuum fluid' effects all forces.

This has consequences for the behavior of clocks, which are constructed with matter and forces. After all, nobody questions the fact that a mechanical clock submerged in moving water cannot keep proper time with respect to an external clock. Similarly, a clock near a gravitational field (with a Fizeau-like, quantum vacuum fluid flow through the clock) also cannot be expected to keep proper time with respect to an observer outside the gravitational field. The accelerated Fizeau-like 'quantum vacuum fluid' moves along radius vectors directed towards the center of the earth, and thus has a specific direction of action. Therefore, the associated space-time effects should also work along the radius vectors (and not parallel to the earth).

For the case of light moving parallel to the earth's surface, the light path is the result of a tremendous number of photon to virtual particle scattering interactions (figure 10). Again in between virtual particle scattering, the light velocity is constant and 'straight'. The total path is curved as shown in figure 10. The path the light takes is called a geodesic in general relativity. In EMQG, this path simply represents the natural path that light takes through the accelerated vacuum, which affects the motion of light. For the case of light moving parallel to the floor of the accelerated rocket (figure 11), the path for light is also the result of virtual particle scattering, but now the quantum vacuum is not in a state of relative acceleration. Therefore, the path is straight for the observer outside the rocket. The observer inside the rocket sees a curved path simply because he is accelerating upwards.

We now see why Einstein's gravitational theory takes the form that it does. Because of the continuously varying frequency and wavelength of the light with height, Einstein interpreted this as a variation of space and time with height. We postulated that the scattering of light with the falling vacuum changes the light velocity in absolute CA units, which cause the *measurements of space and time* to be affected. As we have already seen, these two alternative explanations ***cannot*** be distinguished by direct experimentation. This is why the principle of the constancy of light velocity is still a postulate in general relativity (through the acceptance of special relativity).

We are now in a position to understand the concept of the geodesic proposed by Einstein. ***The downward acceleration of the virtual electrically charged masseons of the quantum vacuum serves as an effective 'electrical guide' for the motion of light (and for test masses) through space and time***. This 'electrical guide' concept replaces the 4D space-time geodesics that guide matter in motion in relativity. For light, this guiding action is through the electromagnetic scattering process of section 9.4. For matter, the electrically charged virtual particles guide the particles of a mass by the electrical force interaction that results from the relative acceleration. Because the quantum vacuum virtual particle density is quite high, but not infinite (at least about $10^{90}$ particles/$m^3$), the quantum vacuum acts as a very effective reservoir of energy to guide the motion of light or matter.

The *relative nature* of 4D space-time can now be easily seen. ***Whenever the background virtual particles of the quantum vacuum are in a state of relative acceleration with respect to an observer, the observer lives in curved 4D space-time.*** Why should the reader accept this new approach, when both approaches give the same result? The reason for accepting EMQG is that the action between a large mass and 4D space-time curvature becomes quite clear. The reason that 4D space-time is curved in an accelerated reference is also clear, and very much related to the gravitational case.

The relative nature of curved 4D space-time also becomes very obvious. An observer inside a gravitational field would normally live in a curved 4D space-time. If he decides to free-fall, he cancels his relative acceleration with respect to the quantum vacuum, and 4D space-time is restored to flat 4D space-time for the observer. The principle of general covariance no longer becomes a principle, but merely results for the deep connection between the quantum vacuum state for accelerated frames and gravitational frames. Last, but not least, the principle of equivalence is completely understood as a reversal of the (net statistical) relative acceleration vectors of the charged virtual masseons of the quantum vacuum, and real masseons that make up a test mass.

## 10. EXPERIMENTAL VERIFICATION OF THE EQUIVALENCE MODEL

We have suggested that the strong principle of equivalence does not hold at all in section 8. Mass Equivalence is also not perfect (section 8). Imagine that a very large test mass and a very small test mass are dropped simultaneously on the earth (in a vacuum), We predict that there will be an *extremely small* difference in the arrival time of the masses on the surface of the earth, which is in slight violation of the principle of equivalence. This occurs because on the earth there are direct graviton exchanges between the test mass and the earth, which tend to unbalance perfect equivalence. The larger test mass has a much larger excess of gravitons exchanged compared to the tiny mass, resulting in a greater pure gravitational force of attraction. However, the electrically charge quantum vacuum dominates over these pure graviton forces, and stabilizes the fall rate and causes the masses to reach the ground almost at the same time. This is because the electrical force from the vacuum is on the order of $10^{40}$ times greater than the pure graviton generated force for these test masses.

We propose several new experimental tests of the principle of equivalence that gives results that are different from the conventional general relativistic physics. These experiments are designed to show that the equivalence principle is not a perfect symmetry of nature, and contains has a few flaws.

(1) ***ANTI-MATTER GRAVITATIONAL PHYSICS*** - EMQG opens up a new field of investigation, which we call anti-matter gravitational physics. We propose that if two sufficiently large pieces of anti-matter are manufactured to allow measurement of the mutual gravitational interaction, then the gravitational force will be found to be repulsive! The force will be equal in magnitude to $-GM^2/r^2$ where M is the mass of each of the equal anti-matter masses, r is their mutual separation, and G is Newton's gravitational constant). This is in clear violation of the principle of equivalence, since in this case $M_i = - M_g$ , instead of masses $M_i = M_g$. Antimatter that is accelerated in far space has the same inertial mass '$M_i$' as ordinary matter, but when interacting gravitationally with another antimatter mass it is repelled ($M_g$). (**Note:** The earth will *attract* bulk anti-matter because of the large abundance of gravitons originating from the earth of the type that induce attraction). This means that no violation of equivalence is expected for anti-matter dropped on the earth, where anti-matter falls normally (recall that virtual masseons and anti-masseons are both attracted to the earth). However, an antimatter earth will repel an antimatter mass dropped on the earth. Recent attempts at measuring earth's gravitational force on anti-matter (e.g. anti-protons will not reveal any deviation from the principle of equivalence).

(2) ***VIOLATION OF THE WEAK EQUIVALENCE PRINCIPLE*** - For an extremely large test mass and a very small test mass dropped simultaneously on the earth (in a vacuum free of air resistance), there will be an extremely small difference in the arrival time of the masses, in slight violation of the principle of equivalence. This effect is on the order of $\approx \Delta N \times \delta$, where $\Delta N$ is the difference in the number of masseon particles in the two masses, and $\delta$ is the ratio of the gravitational to electric forces for one masseon. This experiment is very difficult to perform on the earth, because $\delta$ is extremely small ($\approx 10^{-40}$), and $\Delta N$ cannot be made sufficiently large. To achieve a difference of $\Delta N = 10^{30}$ particles between the small and large mass requires dropping a molecular-sized atomic cluster and a large military tank simultaneously in the vacuum in order to give a measurable deviation. Note: For ordinary objects that might seem to have a large enough difference in mass (like dropping a feather and a tank), the difference in arrival time may be obscured by background interference, or by quantum effects like the Heisenberg uncertainty principle which restrict the accuracy of time measurements.

(3) ***DETECTION OF THE VIOLATION OF THE STRONG PRINCIPLE OF EQUIVALENCE USING A GRAVITON DETECTOR*** - The strong equivalence principle does not hold at all, as we suggested in section 8! To see this, we suggest a thought experiment involving a hypothetical graviton detector. If gravitons can be detected by the invention of a graviton detector/counter in the distant future, then there will be easy experimental proof for the violation of the strong principle of equivalence. The strong equivalence principle states that all the laws of physics are the same for an observer situated on the surface of the earth as it is for an accelerated observer on a rocket (1 g). The graviton detector will find a tremendous difference in the graviton count in these two cases, because gravitons are vastly more numerous here on the earth due to the vast numbers of masseons in the earth. On the rocket, the graviton count would be negligible. Therefore a visual indicator could be placed on the graviton detector box that would easily distinguish between accelerated frames and gravitational frames. This is a ***gross violation*** of the strong equivalence principle.

*(4)* ***REDUCTION OF GRAVITATIONAL MASS ELECTROMAGNETICALY*** - Since mass has a strong electromagnetic force component, a sensitive gravitational mass measurement near the earth might be disrupted by experimentally manipulating the electrically charged virtual particles of the nearby quantum vacuum through ***electromagnetic means***. If a rapidly fluctuating magnetic field (or rotating magnetic field) is produced under a mass it might effect the instantaneous virtual charged particle spectrum, and disrupt the tiny electrical forces contributed by each electrically charged masseon of the mass. This may reduce the measured gravitational mass of an object in the vicinity (this would also affect the inertial mass). In a sense, this device would act like a primitive weak "anti-gravity" device. The virtual particles are constantly being "turned-over" in the vacuum at different rates depending on the energy, with the high frequency particles (and therefore, high-energy particles) being replaced the quickest. If a magnetic field is made to fluctuate fast enough so that it does not allow the new virtual particle pairs to replace the old and smooth out the disruption, the spectrum of the vacuum will be altered. According to conventional physics, the energy density of virtual particles is infinite, which means that all frequencies of virtual particles are present. In EMQG there is a definite upper cut-off to the frequency, and therefore the highest energy according to the Plank's law: $E=h\upsilon$, where $\upsilon$ is the frequency that a virtual particle can have. This frequency cutoff is very roughly on the order of the plank distance scale. We can therefore state that the smallest wavelength that a virtual particle can have is on the order of about $10^{-35}$ meters, e.g. the plank wavelength (or a corresponding maximum Plank frequency of about $10^{43}$ hertz for very high velocity ($\approx c$) virtual particles). Unfortunately for our "anti-gravity" device, it is technologically impossible to disrupt the highest frequencies. According to the uncertainty principle, the relationship between energy and time is: $\Delta E \times \Delta t < h$. This means that the high frequency end of the spectrum consists of virtual particles that "turns-over" the fastest. To give measurable mass change the higher frequencies of the vacuum must be disrupted, which requires magnetic fluctuations on the order of at least $10^{20}$ cycles per seconds. Therefore, only lower frequencies virtual particles of the vacuum can be practically affected, and only small changes in the measured mass can be expected with today's technology. As a result of this, a relationship should exist between the amount of gravitational (or inertial) mass loss and the frequency of electromagnetic fluctuation or disruption. The higher the frequency the greater the mass loss. Work on the Quantum Hall Effect (ref. 29) by Laughlin has suggested that the electron density in a two-dimensional sheet under the influence of a strong magnetic field causes the electrons to move in concert, with very high speed swirling vortices created in the resulting 2D electron gas. In ordinary magnetic fields, electrons are merely 'pushed' around, while a strong magnetic field causes the electrons to swirl in high-speed 'whirlpools'. There is also a possibility that this 'whirlpool' phenomena holds for the virtual particles of the quantum vacuum under the influence of a strongly fluctuating magnetic field. These high-speed whirlpools might disrupt the high frequency end of the spectral distribution of electrically charged virtual particles in small pockets. Therefore, there might be a greater mass loss under these circumstances. Recent experiments on mass reduction with rapidly rotating magnetic fields are inconclusive at this time. Reference 30 gives an excellent and detailed review of the various experiments on reducing the gravitational force with superconducting magnets.

## 11. CONCLUSIONS

Using a newly developed quantum field theory of gravity theory called EMQG, we have illustrated the hidden quantum processes involved in Einstein's principle of equivalence. We found that almost the same quantum processes occurring in inertial mass are also happening in gravitational mass. We found that gravity involves *two* pure, quantum force particle exchange processes. *Both* the photon and graviton particle exchanges occur *simultaneously* for a test mass in a large gravitational field, whereas for inertial mass only photon exchanges are involved.

We modified a new theory of inertia first introduced by Haisch, Rueda, and Puthoff, which we call HRP inertia. In HRP inertia, inertia is the resulting electromagnetic force interaction of the charged 'parton' particles making up a mass with the background virtual photon field, which is called the zero point fluctuations (or ZPF). We modified HRP inertia, (which we call Quantum Inertia, or QI), which involves the introduction of a new particle of nature called the masseon. The masseon contains the smallest possible quanta of electric charge as well as the smallest possible quanta of 'mass-charge'.

Masseons combine with other masseons to produce all the known fermion mass particles of the standard model. The masseon is electrically charged, as well as possessing a new form of charge called 'mass-charge'. Mass-charge is analogous to electric charge, where gravitons take the place of photons as the exchange particle for the pure gravitational force, and masseons take the place of electrons as the charge source and destination. The physics of graviton exchanges between masseons is virtually the same as photon exchanges for electrons in QED. Gravitons have spin 1, just as the photon (not spin 2).

Quantum Inertia is based on the idea that inertial force is due to the tiny electromagnetic force interactions originating from each charged masseon particle of real matter undergoing relative acceleration with respect to the vast swarm of virtual, electrically charged masseon particles of the nearby quantum vacuum. These tiny forces add up to the total resistance force opposing the accelerated motion in Newton's law 'F = MA', where the sum of each of the tiny masseon forces equals the total inertial force. When masseons move through the vacuum at a constant velocity (i.e., an inertial frame), the sum of the total vacuum forces is zero, and the vacuum does not oppose the motion. Therefore, the virtual masseons of the quantum do not act as a form of an ether for inertial observers.

We proposed that gravity also involves this very *same* ‘inertial’ electromagnetic force component found inside an accelerated mass above. This is the source of the deep connection between inertia and gravity, which is at the heart of Newtonian mass equivalence. Since virtual masseons possesses mass-charge, and the earth (composed of real masseons) possesses mass-charge, the result is that virtual masseons of the quantum vacuum fall during their very brief life-times, which has profound effects on test masses.

Newtonian mass equivalence has been explained as a consequence of the falling quantum vacuum. Inside a test mass subjected to a large gravitational fields, there exists a similar quantum vacuum process that occurs for an inertial mass, where the roles of the real charged masseon particles of the mass and the virtual, electrically charged masseons of the quantum vacuum are reversed. Now it is the electrically charged virtual masseon particles of the quantum vacuum that are accelerating (falling), while the mass particles are at relative rest. The reason why the virtual particles of the quantum vacuum fall in a large gravitational field is the huge numbers of graviton particles that are exchanged between the earth and the surrounding virtual masseon particles of the quantum vacuum.

Furthermore, the general relativistic Weak Equivalence Principle (WEP) also results from this common physical process existing at the quantum level in both gravitational mass and inertial mass. The falling, electrically charged, virtual masseons of the quantum vacuum affect both the motion of real test masses, and also the motion of ***light***. The action of the falling quantum vacuum on light is very much reminiscent of the action of flowing water on the motion of light in the Fizeau experiment. The path that light takes while moving parallel to the surface of the earth, through the falling quantum vacuum, is ***curved***. This implies space-time curvature. The action of falling ‘Fizeau-like’ quantum vacuum on clocks and rulers is responsible for the origin Riemannian curved 4D space-time geometry near the earth, and is the basis of a quantum theory of general relativity.

Therefore based on a new theory of quantum gravity called EMQG, we have discovered the hidden quantum interactions that occur in Newtonian inertia and the quantum machinery responsible for Einstein’s Weak Equivalence Principle.

## 13. ILLUSTRATIONS

The captions for the figures are shown below:

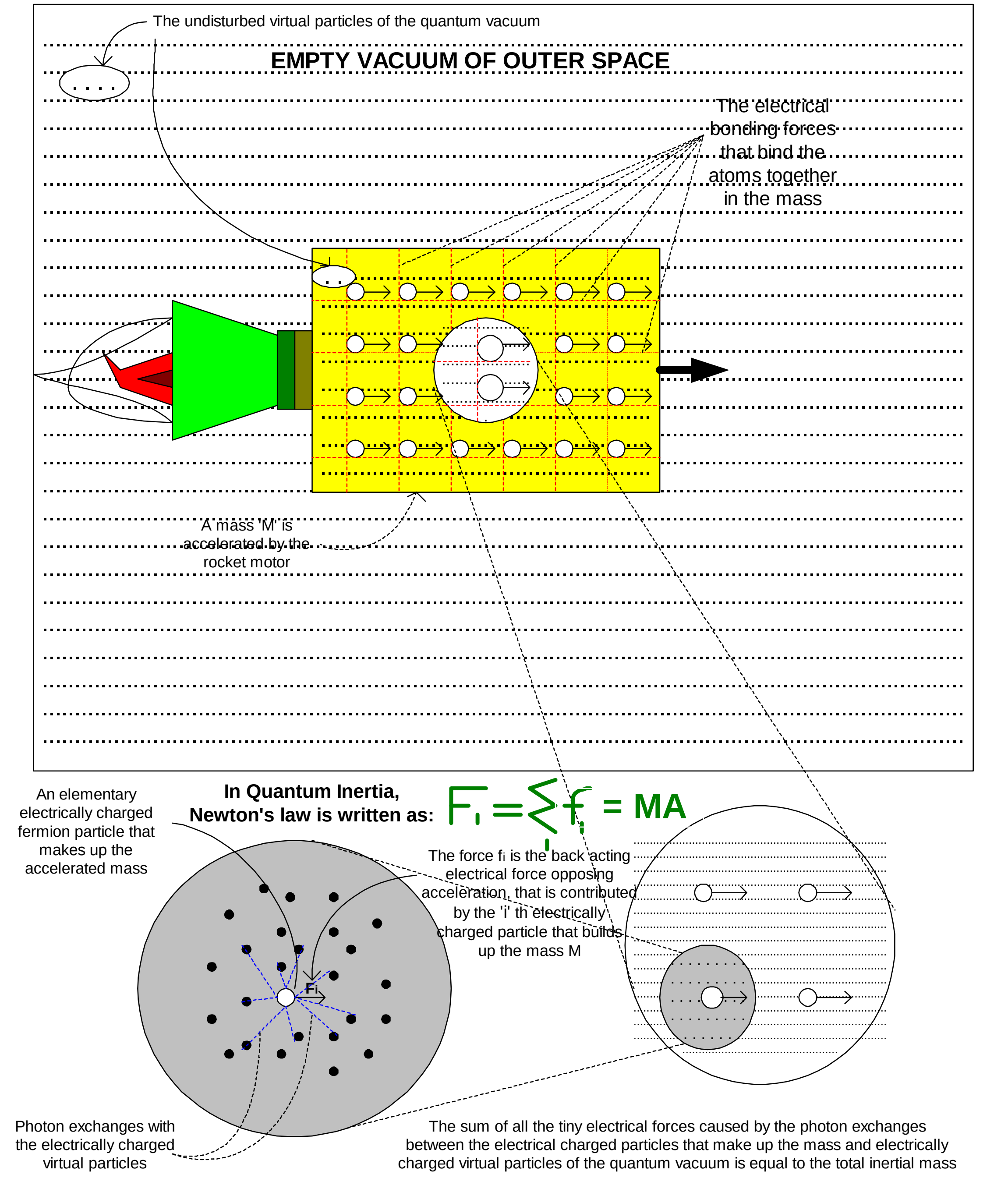

**FIGURE 1 - The Quantum Mechanism behind Newtonian Inertia**

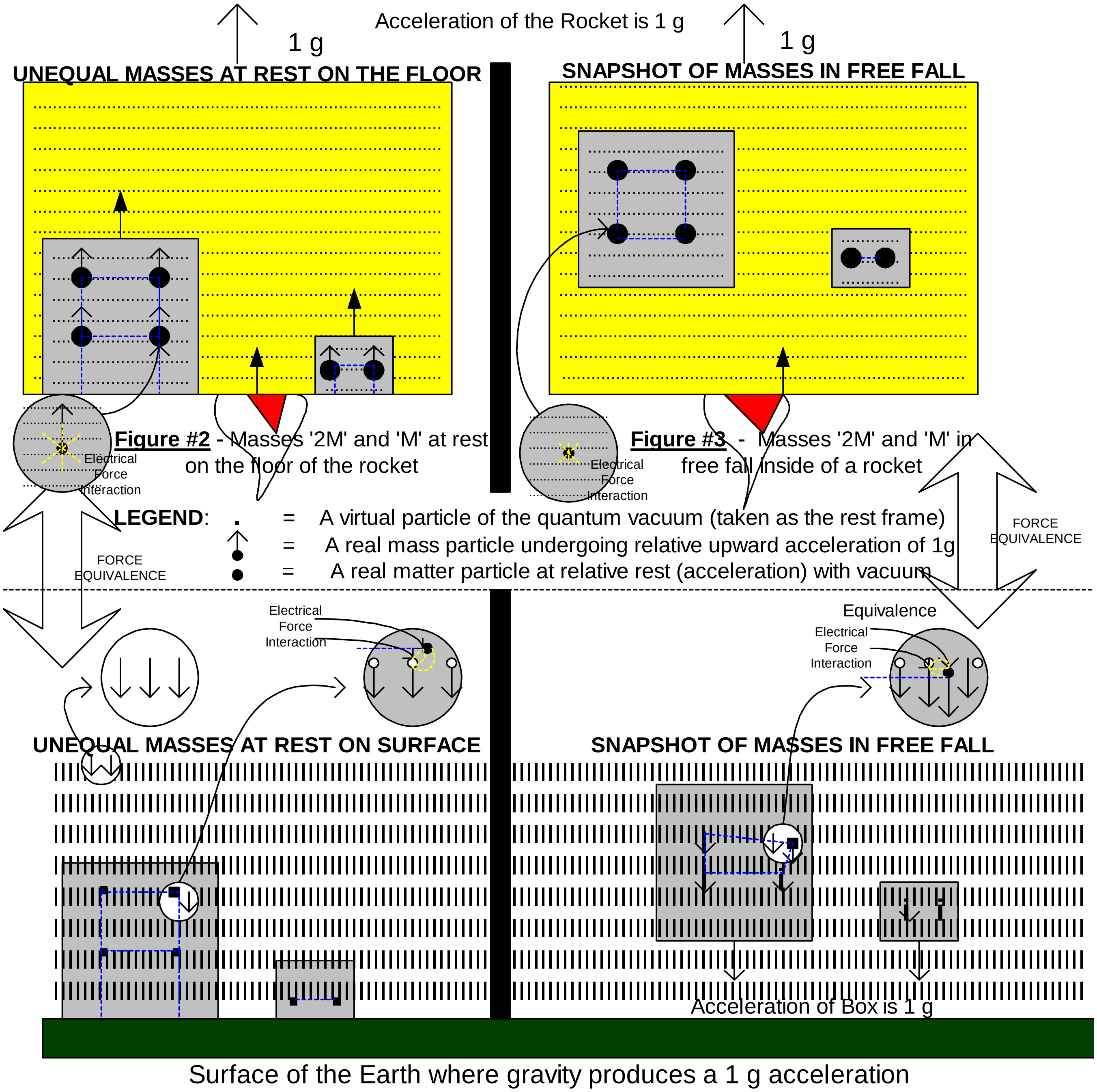

**Figure #2** - Masses '2M' and 'M' at rest on the floor of the rocket

**Figure #3** - Masses '2M' and 'M' in free fall inside of a rocket

**LEGEND**:
- . = A virtual particle of the quantum vacuum (taken as the rest frame)
- ↑● = A real mass particle undergoing relative upward acceleration of 1g
- ● = A real matter particle at relative rest (acceleration) with vacuum

**Figure #4** - Masses '2M' and 'M' at rest on Earth's surface

**Figure #5** - Masses '2M' and 'M' in free fall above the Earth

**LEGEND:**
- **I** = Relative downward acceleration (1g) of a virtual particle
- **i** = Relative downward acceleration (1g) of a real matter particle
- **.** = A real stationary matter particle (with respect to the earth's center)

## FIGURES 2 TO 5 - THE PRINCIPLE OF EQUIVALENCE FOR A STATIONARY MASS ON THE EARTH AND INSIDE A ROCKET

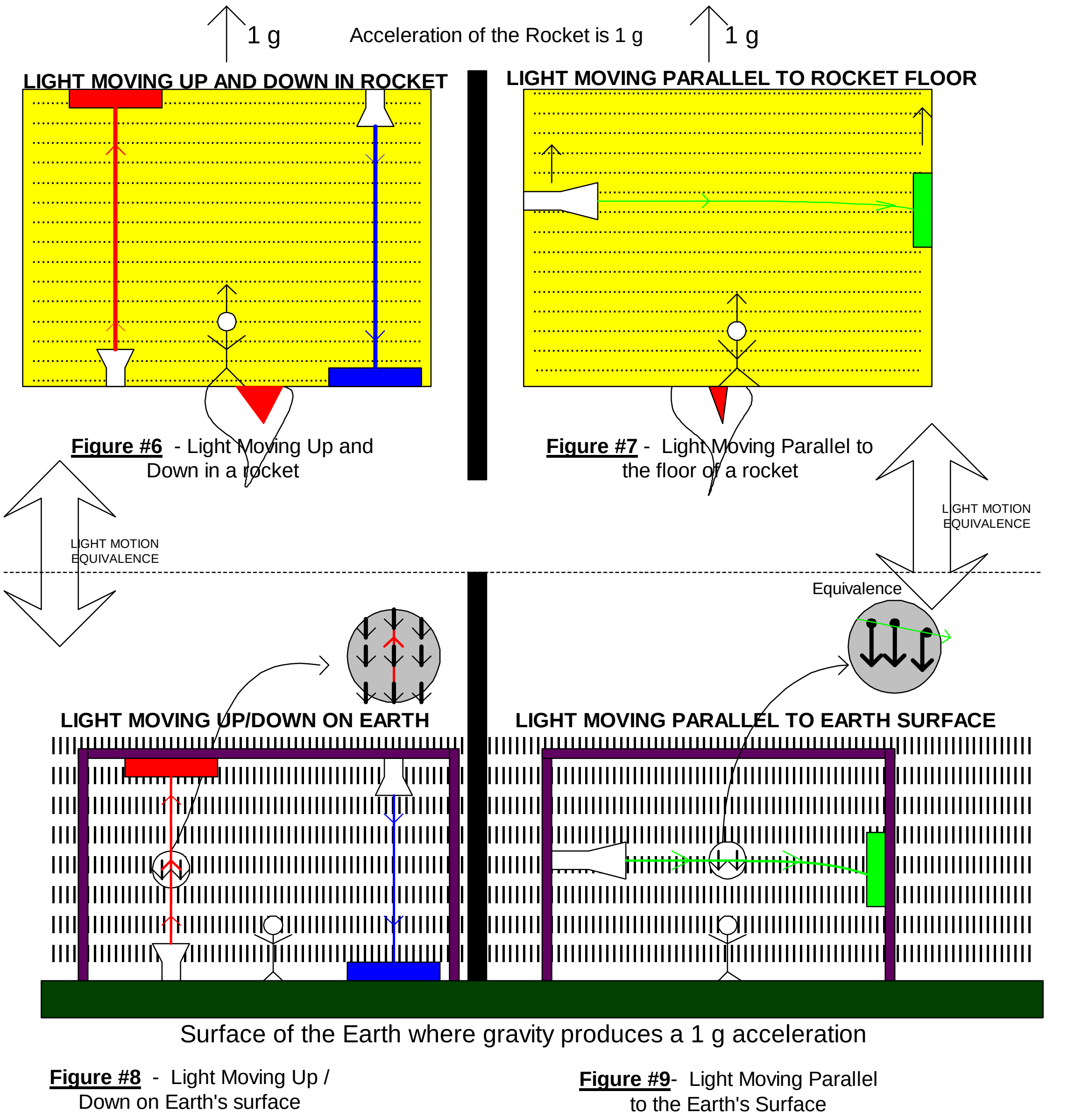

**FIGURES 6 TO 9 - THE PRINCIPLE OF EQUIVALENCE FOR LIGHT MOTION ON THE EARTH AND INSIDE A ROCKET**

FIGURE #10 - **MICROSCOPIC PRINCIPLE OF EQUIVALENCE FOR VIRTUAL PARTICL**

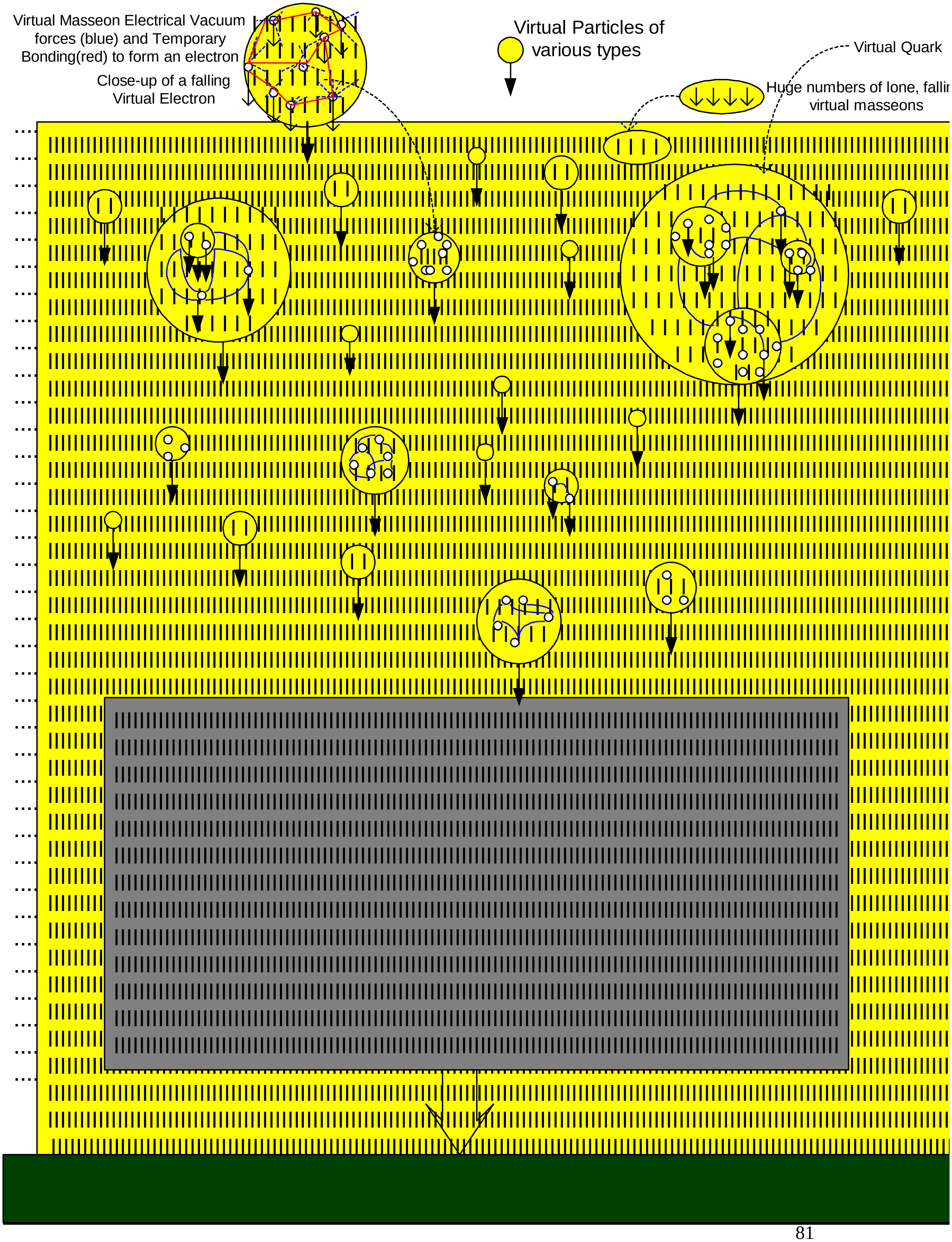

**Figure #11- VIRTUAL PARTICLE PATTERN FOR THE EARTH AND MOON IN FREE FALL NEAR THE SUN**

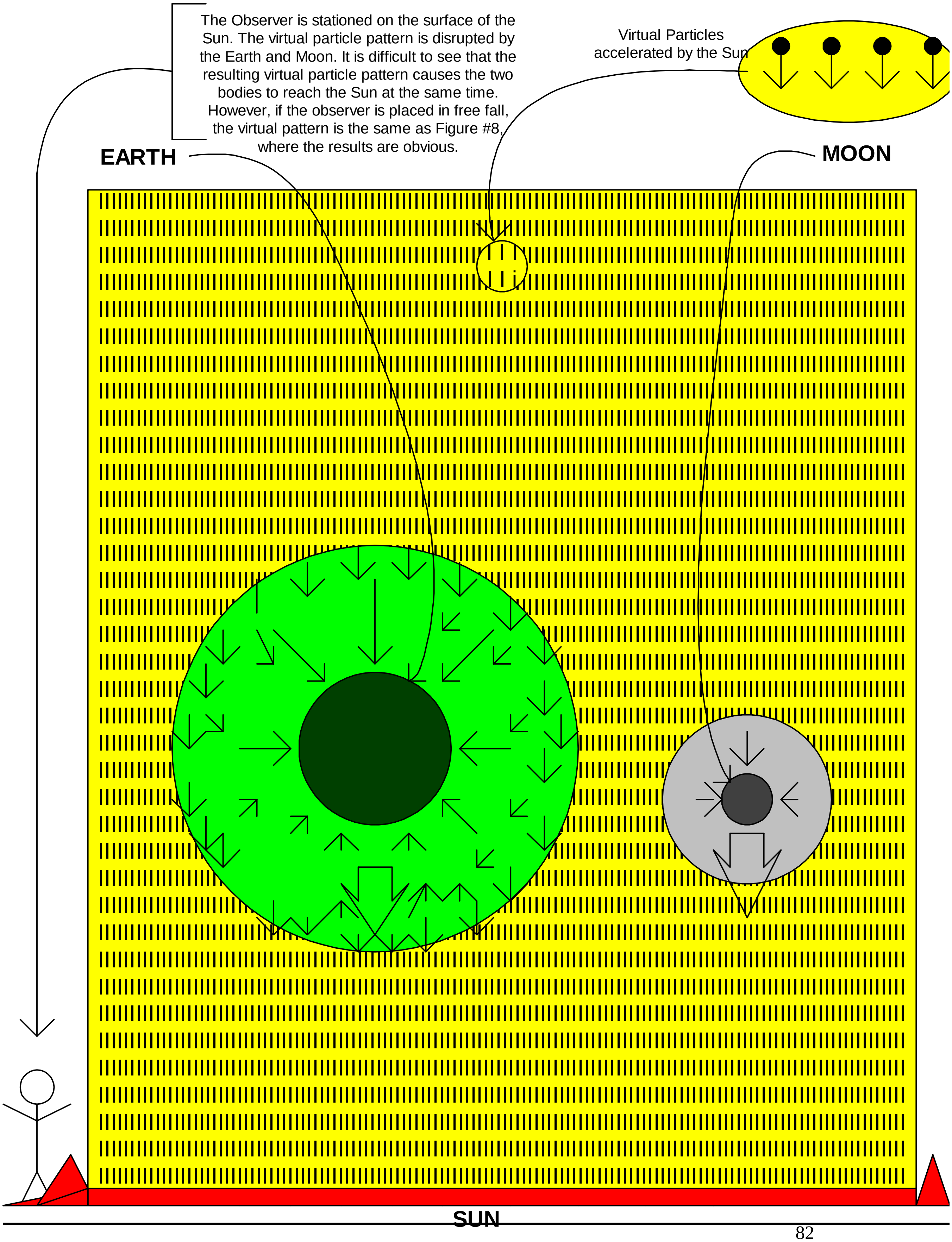

**Figure #12 - VIRTUAL PARTICLE PATTERN FOR THE EARTH AND MOON IN FREE FALL IN A ROCKET**

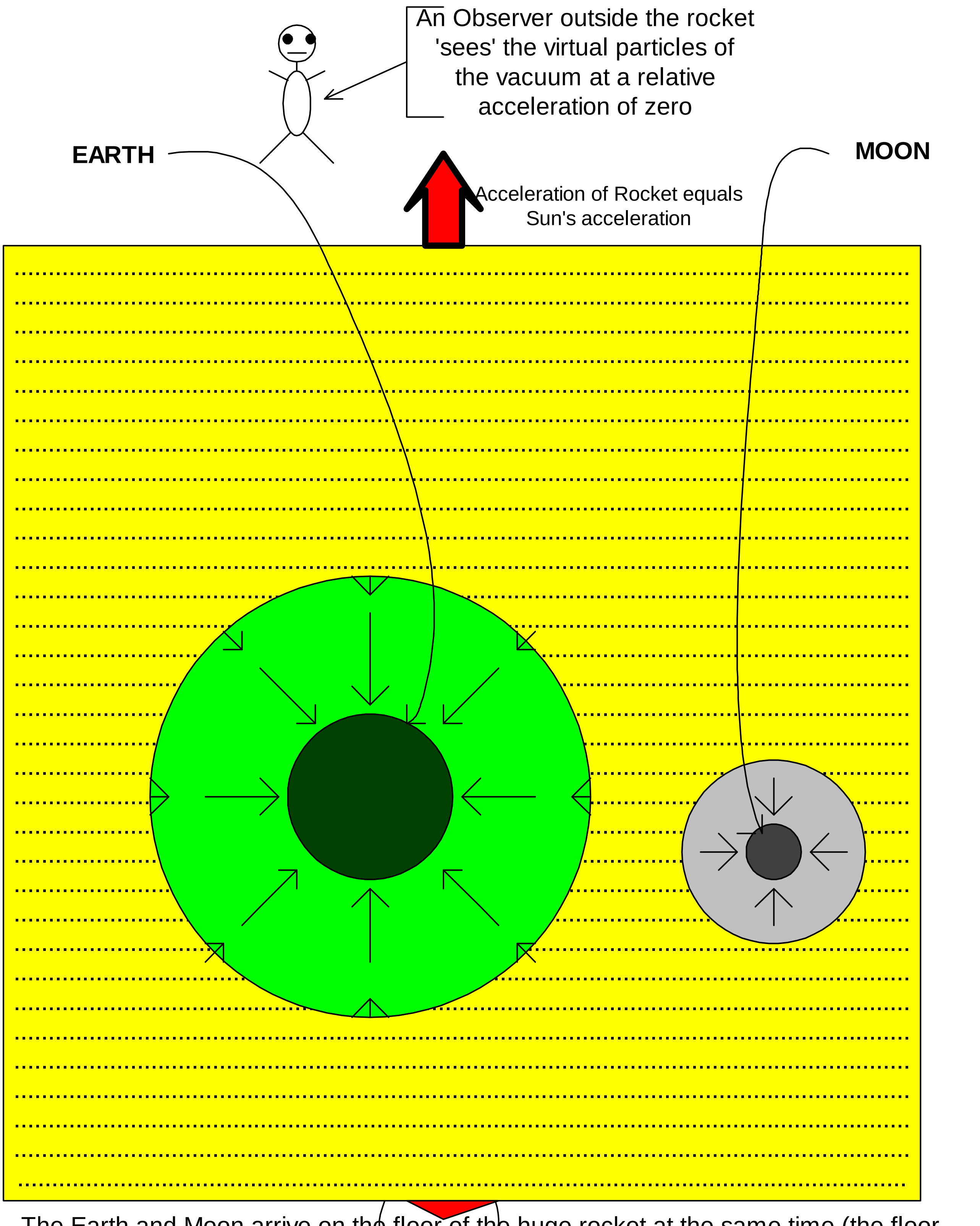

The Earth and Moon arrive on the floor of the huge rocket at the same time (the floor simply moves up to meet these bodies). But now, the virtual masseon particles near these two bodies are distorted and interacting with the real masseons in these two bodies.

**Figure #13 - MOTION OF REAL PHOTONS IN THE PRESENCE OF VIRTUAL PARTICLE NEAR EARTH**

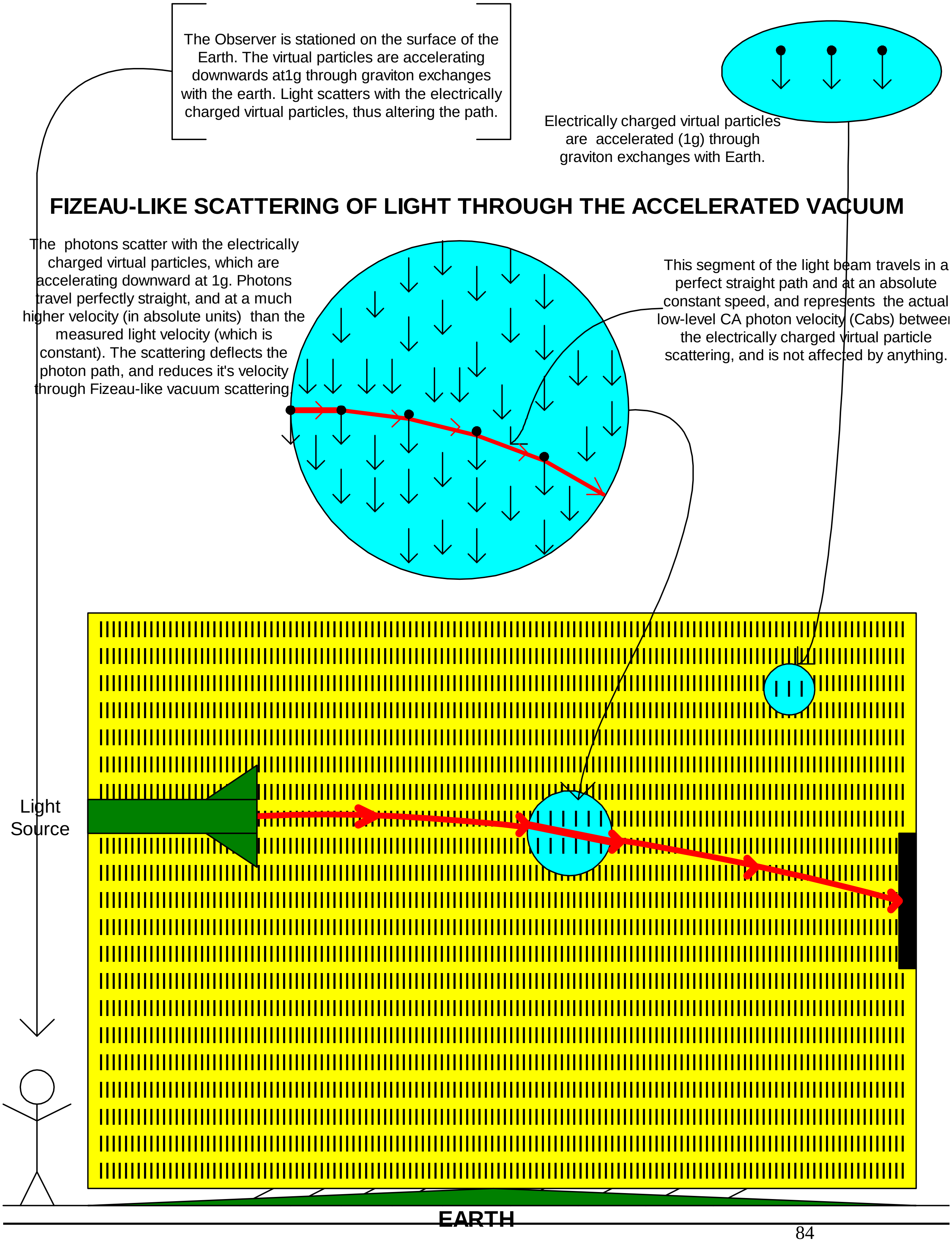

**Figure #14 - MOTION OF REAL PHOTONS IN A ROCKET ACCELERATING AT 1g**

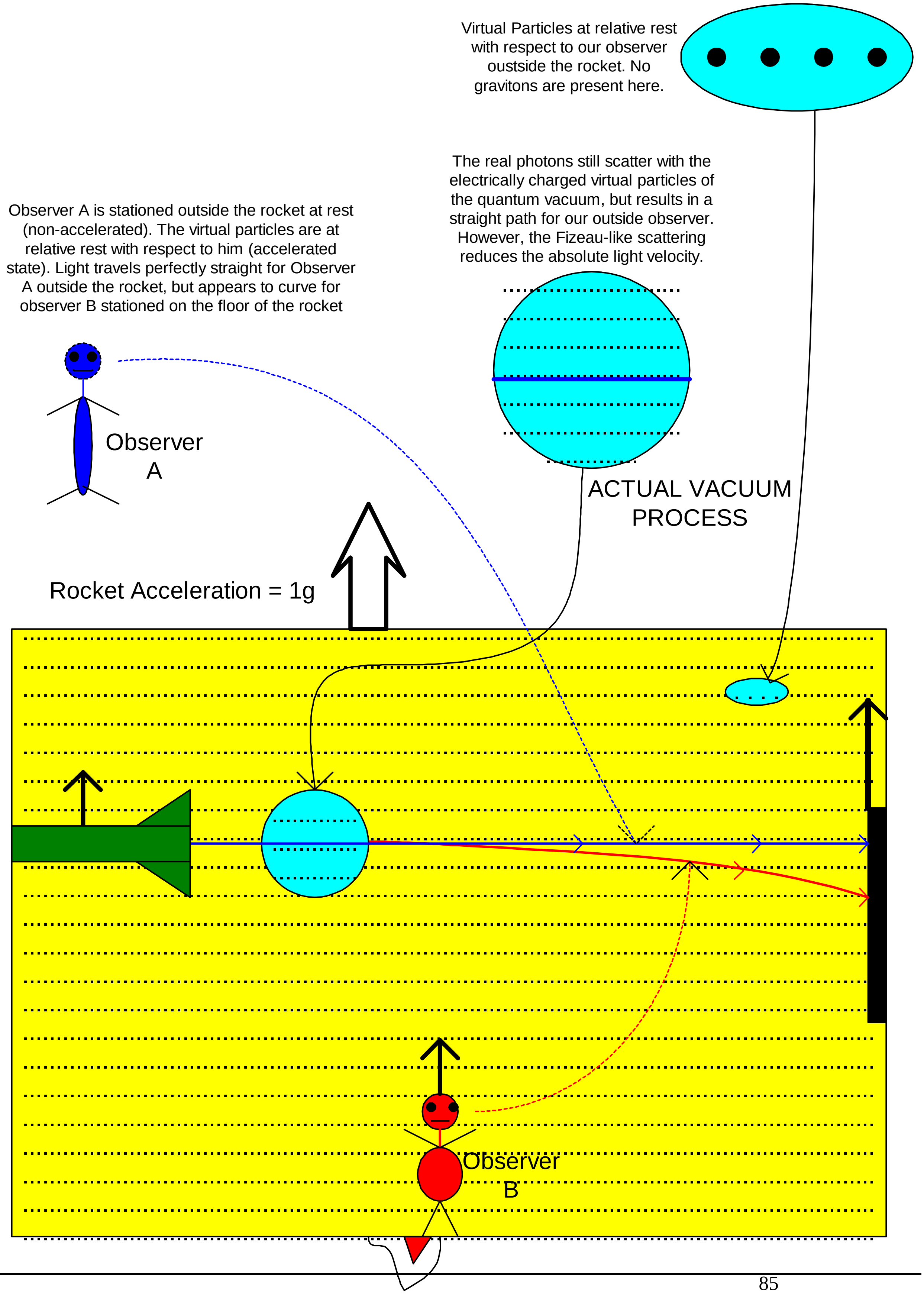

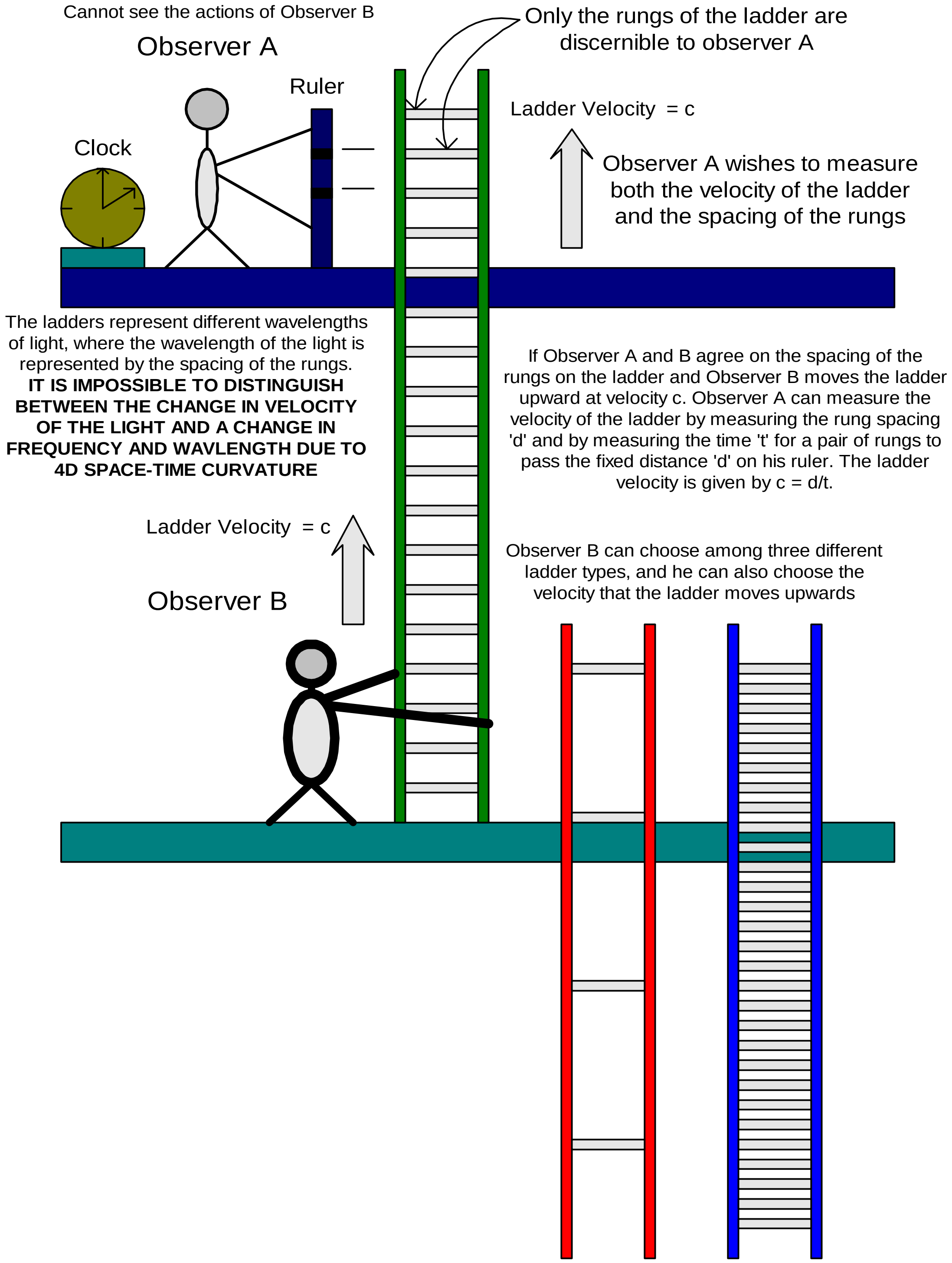

Figure #15 - It is impossible to distinguish between the change in the velocity of light versus space-time curvature when light propagates from a lower to a higher gravitational potential. Light is represented by a ladder where the peaks of the wave are replaced by rungs on a ladder.

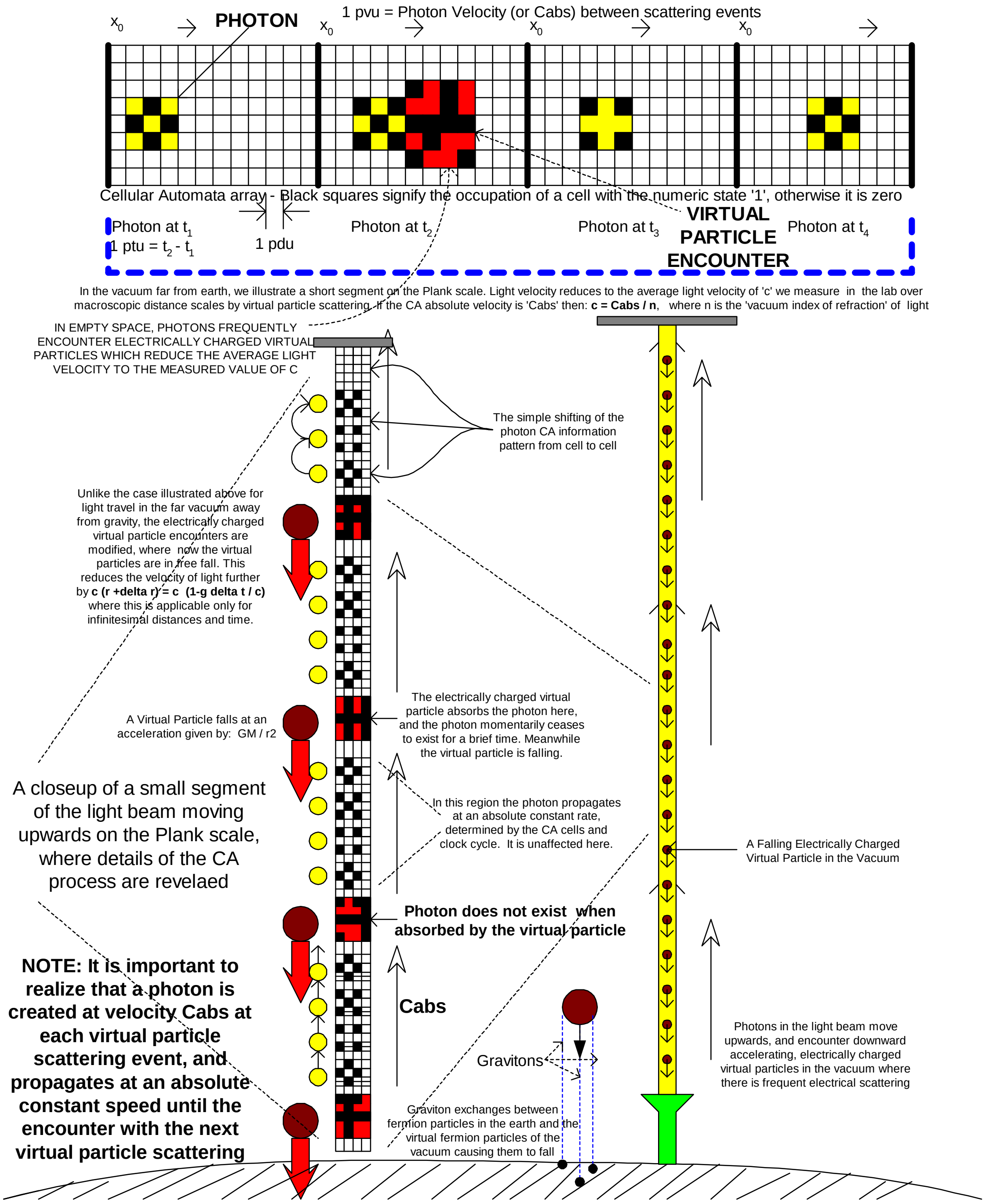

Figure 16 - EMQG and Space-Time effects from photon scattering by falling electrically charged virtual particles

**Figure #17 - BLOCK DIAGRAM OF RELATIONSHIP OF CA AND EMQG WITH PHYSICS**

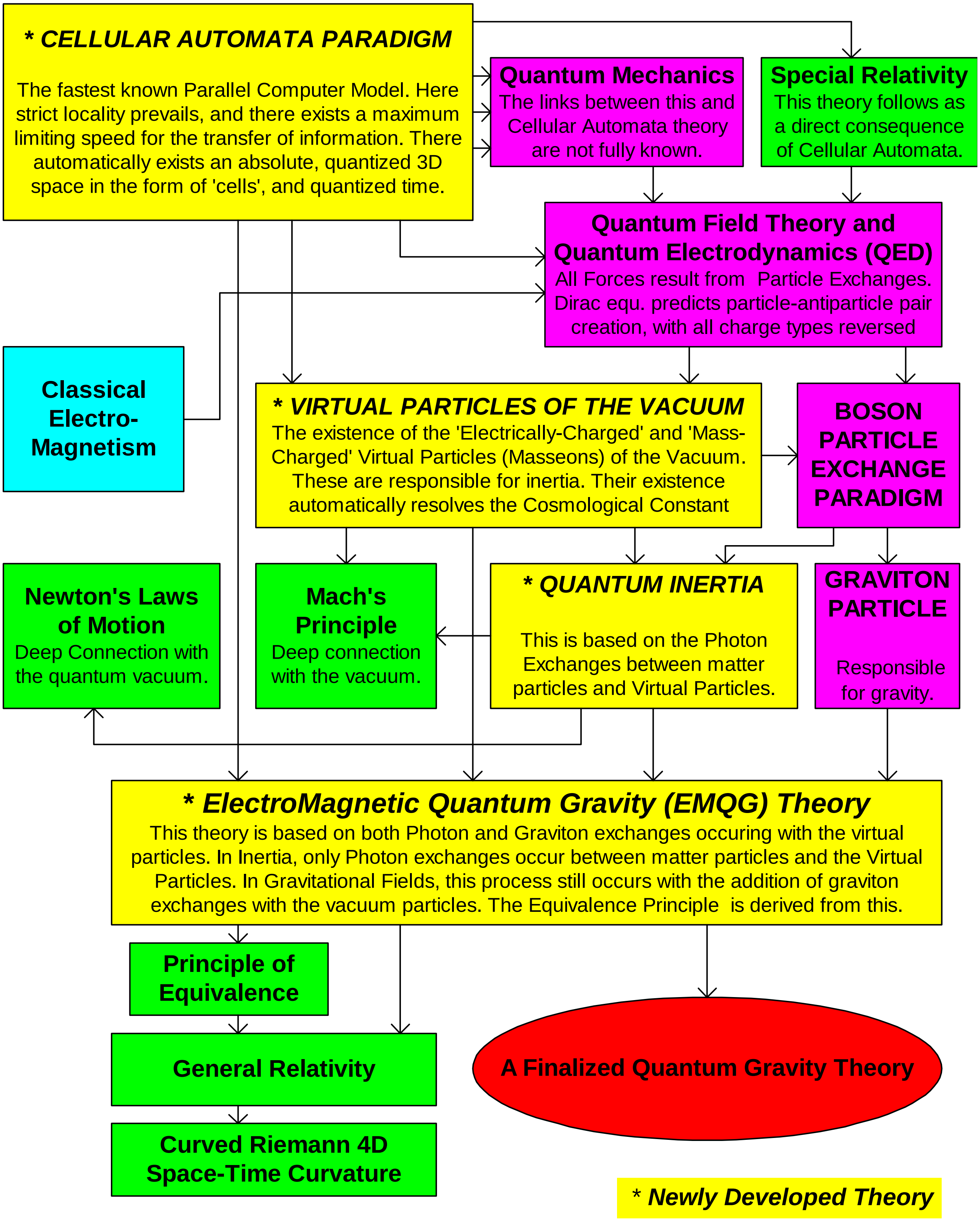

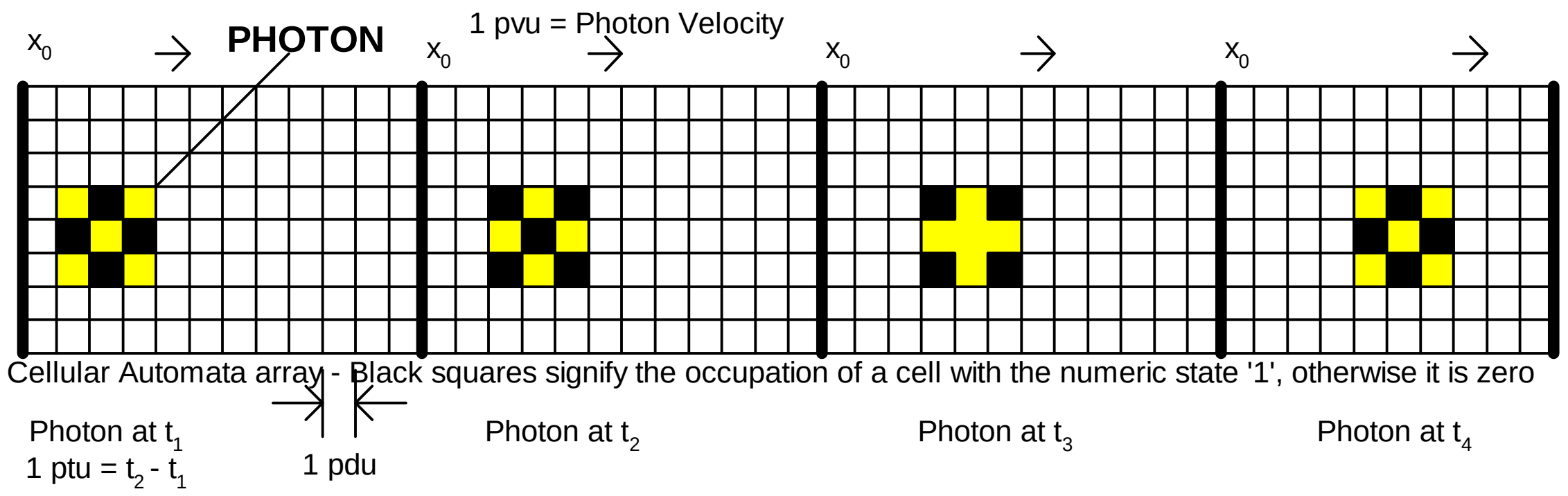

**Figure #18** - Simplified model of the motion of the photon information pattern on the CA. The photon information pattern moves 1 plank unit to the right at every plank 'clock cycle' (Note: The photon is actually an oscillating wavepattern shown highly simplified

**Absolute CA units:** 1 pdu is the shifting of information by 1 cell; 1 ptu is the time to shift 1 cell; 1 pvu = photon velocity

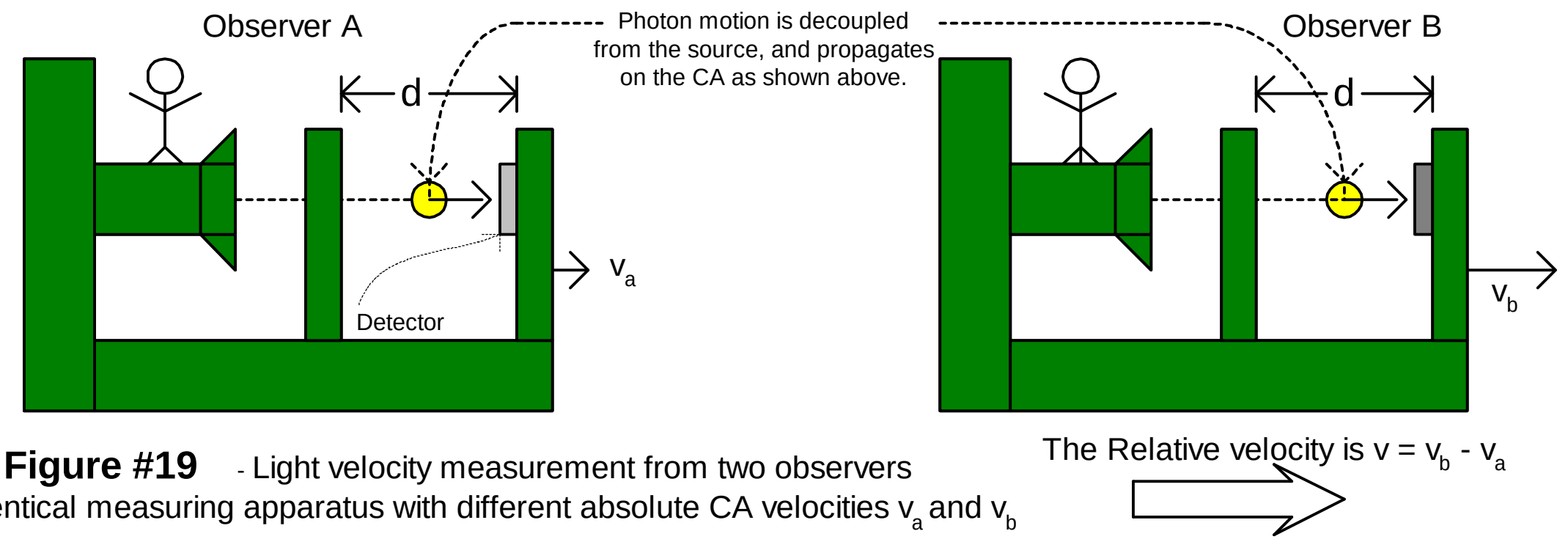

**Figure #19** - Light velocity measurement from two observers with identical measuring apparatus with different absolute CA velocities $v_a$ and $v_b$

***MEASUREMENTS ARE MADE USING ABSOLUTE CA UNITS OF DISTANCE AND TIME***

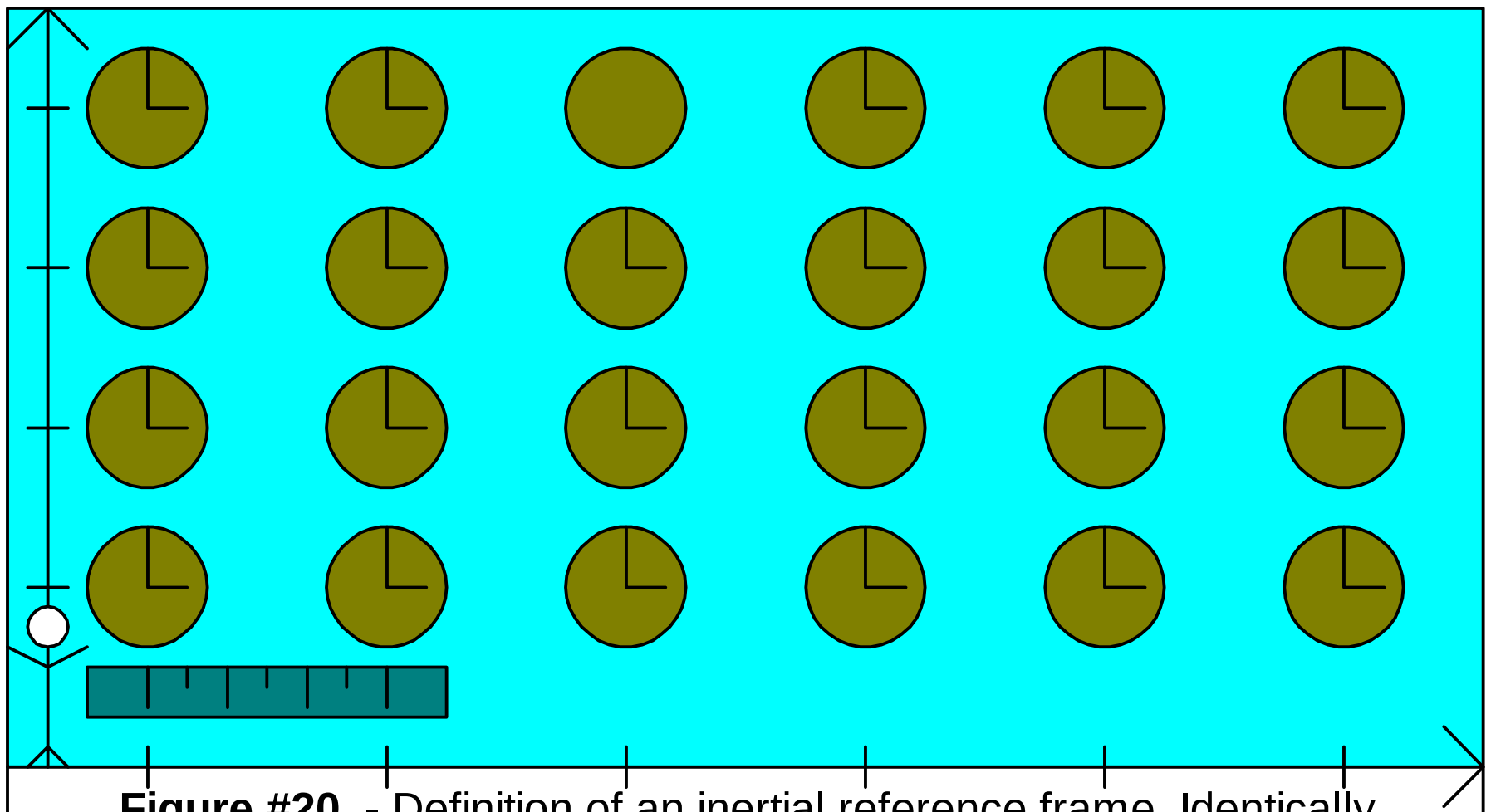

**Figure #20** - Definition of an inertial reference frame. Identically constructed clocks are spaced apart at regular intervals by a ruler

(A 2 dimenional arrangement is shown for simplicity)

## RELATIONSHIP BETWEEN THE CA MODEL, LIGHT MOTION ON THE CA, AND SPECIAL RELATIVITY

In a 3D Geometric Cellular Automata, the numeric content of Cell $C_{i,j,k}$ is uniquely determined by the numeric contents of each of the surrounding 26 neighbouring cells (and possibly with it's own numeric state). On the next CA 'clock cycle' the contents of cell $C_{i,j,k}$ is determined by a function (or algorithm) $F_{i,j,k}$ such that the contents of the cell $C_{i,j,k} = F(C_{i+x,j+y,k+z})$ where x,y,z take on all the following values: -1,0,1. This same function F is programmed into each and every cell in the entire CA. In the figure below, the binary number system is chosen for illustration purposes (any number system can be used). The dotted lines indicate what cells affect cell $C_{i,j,k}$.

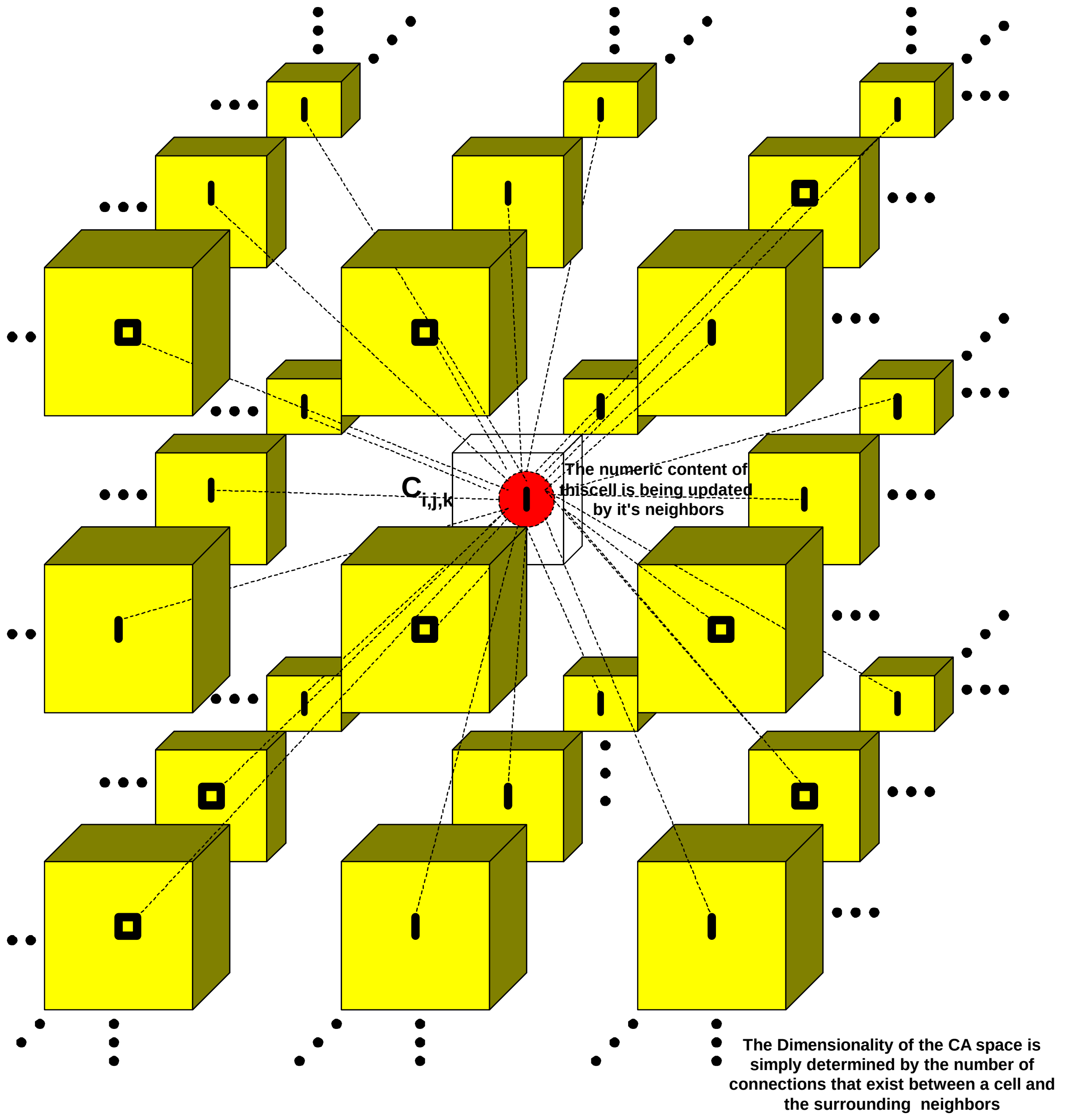

**Figure #21 - Schematic Diagram of what space looks look on the Cellular Automata**